\documentclass[12pt]{article}
\usepackage{amssymb}
\usepackage{epsfig}
\usepackage{bbm}

\usepackage{graphicx}

\def\L{\mathcal L}
\def\e{\varepsilon}

\newcommand{\wt}{\widetilde}

\begin{document}

\def\a{\alpha}
\def\b{\beta}
\def\c{\chi}
\def\d{\delta}
\def\e{\epsilon}
\def\f{\phi}
\def\g{\gamma}
\def\h{\eta}
\def\i{\iota}
\def\j{\psi}
\def\k{\kappa}
\def\l{\lambda}
\def\m{\mu}
\def\n{\nu}
\def\o{\omega}
\def\p{\pi}
\def\q{\theta}
\def\r{\rho}
\def\s{\sigma}
\def\t{\tau}
\def\u{\upsilon}
\def\x{\xi}
\def\z{\zeta}
\def\D{\Delta}
\def\F{\Phi}
\def\G{\Gamma}
\def\J{\Psi}
\def\L{\Lambda}
\def\O{\Omega}
\def\P{\Pi}
\def\Q{\Theta}
\def\S{\Sigma}
\def\U{\Upsilon}
\def\X{\Xi}

\def\ve{\varepsilon}
\def\vf{\varphi}
\def\vr{\varrho}
\def\vs{\varsigma}
\def\vq{\vartheta}

\def\dg{\dagger}                                     
\def\ddg{\ddagger}                                   
\def\wt#1{\widetilde{#1}}                    
\def\mt{\widetilde{m}_1}
\def\mti{\widetilde{m}_i}
\def\rt{\widetilde{r}_1}
\def\mtt{\widetilde{m}_2}
\def\mttt{\widetilde{m}_3}
\def\rtt{\widetilde{r}_2}
\def\mb{\overline{m}}
\def\VEV#1{\left\langle #1\right\rangle}        
\def\be{\begin{equation}}
\def\ee{\end{equation}}
\def\ds{\displaystyle}
\def\ra{\rightarrow}

\def\bea{\begin{eqnarray}}
\def\eea{\end{eqnarray}}
\def\NO{\nonumber}
\def\Bar#1{\overline{#1}}


\def\pl#1#2#3{Phys.~Lett.~{\bf B {#1}} ({#2}) #3}
\def\np#1#2#3{Nucl.~Phys.~{\bf B {#1}} ({#2}) #3}
\def\prl#1#2#3{Phys.~Rev.~Lett.~{\bf #1} ({#2}) #3}
\def\pr#1#2#3{Phys.~Rev.~{\bf D {#1}} ({#2}) #3}
\def\zp#1#2#3{Z.~Phys.~{\bf C {#1}} ({#2}) #3}
\def\cqg#1#2#3{Class.~and Quantum Grav.~{\bf {#1}} ({#2}) #3}
\def\cmp#1#2#3{Commun.~Math.~Phys.~{\bf {#1}} ({#2}) #3}
\def\jmp#1#2#3{J.~Math.~Phys.~{\bf {#1}} ({#2}) #3}
\def\ap#1#2#3{Ann.~of Phys.~{\bf {#1}} ({#2}) #3}
\def\prep#1#2#3{Phys.~Rep.~{\bf {#1}C} ({#2}) #3}
\def\ptp#1#2#3{Progr.~Theor.~Phys.~{\bf {#1}} ({#2}) #3}
\def\ijmp#1#2#3{Int.~J.~Mod.~Phys.~{\bf A {#1}} ({#2}) #3}
\def\mpl#1#2#3{Mod.~Phys.~Lett.~{\bf A {#1}} ({#2}) #3}
\def\nc#1#2#3{Nuovo Cim.~{\bf {#1}} ({#2}) #3}
\def\ibid#1#2#3{{\it ibid.}~{\bf {#1}} ({#2}) #3}

\title{
\vspace*{10mm}
\bf New aspects of leptogenesis bounds}
\author{{\Large Steve Blanchet$^a$ and Pasquale Di Bari$^b$}
\\
$^a${\it Max-Planck-Institut f\"{u}r Physik}
{\it (Werner-Heisenberg-Institut)} \\
{\it F\"{o}hringer Ring 6, 80805 M\"{u}nchen, Germany} \\
$^b$
{\it INFN, Sezione di Padova},
{\it Dipartimento di Fisica Galileo Galilei} \\
{\it  Via Marzolo 8, I-35131 Padua, Italy}
}

\maketitle \thispagestyle{empty}

\vspace{-5mm}

\begin{abstract}
We present a general analysis that reveals new aspects
of the leptogenesis bounds on neutrino masses and on the
reheat temperature of the Universe.
After revisiting a known effect coming from an unbounded term
in the total $C\!P$ asymmetry, we show that an unbounded term in the
flavored $C\!P$ asymmetries has a stronger impact. It
relaxes the lower bound on the reheat temperature down
to $10^8\,{\rm GeV}$ for $(M_2-M_1)/M_1={\cal O}(1-100)$ and for a mild
tuning of the parameters in the see-saw orthogonal matrix.
We also consider the effect of the Higgs asymmetry,
showing that it lowers the upper bound on the neutrino masses
in the so-called fully flavored regime where classic Boltzmann
equations can be used. Imposing independence of the initial
conditions contributes to lower the upper bound on neutrino masses as well.
We study the conditions for the validity of the usual $N_1$-dominated
scenario and for the applicability of the
lower bound on the lightest right-handed (RH) neutrino mass $M_1$. We find
that except for the two effective RH neutrino scenario, recovered for
$M_3\gg 10^{14}\,{\rm GeV}$, and for values $M_2 < {\cal O}(10^{11}\,{\rm GeV})$,
the final asymmetry is more naturally dominated by the contribution from $N_2$-decays.
Finally, we confirm in a general way that going beyond the hierarchical limit,
the effect of washout addition makes the lower bound on $M_1$ more stringent for
$(M_2-M_1)/M_1={\cal O}(0.1)$.
\end{abstract}

\newpage

\section{Introduction}

Leptogenesis \cite{fy} provides an elegant solution
to the problem of the non-observation of primordial antimatter
in the Universe. An appealing feature of this mechanism is that it
relies on a minimal extension of the Standard Model (SM) where
RH neutrinos are added to the Lagrangian. In the
see-saw limit \cite{seesaw}, small neutrino masses
can be naturally accommodated in agreement with the data.
Considering, for definiteness, the simplest and best motivated case
of three additional RH neutrinos, only few of the eighteen new parameters,
three neutrino masses, three mixing angles and three {\em CP}
violating phases, are or can hopefully be probed in neutrino experiments.
The remaining 9 `high-energy'
parameters are out of reach in low energy experiments.
In this respect, leptogenesis represents also an important cosmological tool
to access information on this `dark side' of the see-saw parameter space.

In an unflavored analysis and in a traditional $N_1$-dominated scenario,
where it is assumed that the final asymmetry is dominantly produced
from the lightest RH neutrino decays, one finds a lower bound
on the lightest RH neutrino mass $M_1$ \cite{di,cmb}.
This is given by  $M_1 \gtrsim 3 \times 10^9~{\rm GeV}$ \cite{geometry,flavorlep}
at the onset of the strong washout regime, where the
final asymmetry does not depend on the initial conditions.
This lower bound also implies an associated lower bound on the reheat
temperature, $T_{\rm reh}\gtrsim 1.5\times 10^9\,{\rm GeV}$
\cite{annals,giudice,flavorlep}. In addition an upper bound  on the neutrino
masses, $m_i\lesssim 0.1\,{\rm eV}$, holds as well \cite{window}.

A mild hierarchy in the spectrum of the RH neutrino masses,
such that $M_2\gtrsim 3\,M_1$, is a necessary condition for the
validity of the $N_1$-dominated scenario but is not sufficient.
Indeed, upon closer inspection, the observed asymmetry can still be
generated from the decays of the next-to-lightest RH neutrinos, realizing
a $N_2$-dominated scenario \cite{geometry}. The lower bound on $M_1$ is replaced by
a lower bound on $M_2$ but this still implies a lower bound on $T_{\rm reh}$.
It is therefore more correct to say that leptogenesis yields a
lower bound on $T_{\rm reh}$ rather than on $M_1$.

Even when flavor effects are considered \cite{nardi1,abada1} (see also \cite{bcst,seealso}),
the lower bound on $T_{\rm reh}$ has been found not to change \cite{flavorlep}.
On the other hand they induce other important modifications.
First, the final asymmetry receives an additional contribution
that also depends on neutrino mixing parameters \cite{nardi1,abada2}
and an interesting feature is that such an additional contribution
can originate solely by low-energy (Dirac or Majorana)
phases \cite{nardi1} and can even explain the whole observed asymmetry
\cite{flavorlep,pascoli1,pascoli2,branco,antusch,diraclep}.
Second, flavor effects can relax the
stringent upper bound on the neutrino masses, $m_i\lesssim 0.1~{\rm eV}$,
holding in the unflavored regime when a hierarchical heavy neutrino mass
spectrum is assumed \cite{window}.
 In \cite{abada1}, it was found that flavor effects
completely erase this upper bound. However, in \cite{zeno}, it
was pointed out that this conclusion relies on the use of classical Boltzmann
equations beyond their range of validity and has therefore to be
checked within a more general description making use of density matrix equations.
In \cite{desimone1} it was found that
using classic Boltzmann equations neutrino masses as large as
$2\,{\rm eV}$ are possible, a value larger than the upper bound holding
in the unflavored regime and than the current upper bound
from cosmological observations. A similar conclusion has been
recently obtained in \cite{review} as well.

Many of these results have been obtained employing
different assumptions or restrictions in the parameter space.
In this paper we revisit in detail the leptogenesis bounds
still assuming classical Boltzmann equations and the $N_1$-dominated scenario but
without further restrictions on the parameter space and
relaxing many assumptions that are usually made. For example, we
find more general conditions for the validity of the
$N_1$-dominated scenario. In general,
our analysis reveals various new aspects including effects
confirming the necessity to go beyond a classical
Boltzmann kinetic approach to solve the problem of
the upper bound on the neutrino masses.

In Section 2 we set up the general notation.
In Section 3 we show that the final asymmetry can be
written as the sum of different contributions acting
independently of each other.
The first step is to distinguish between a contribution
from the lightest RH neutrino decays and
a contribution from the heavier RH neutrino decays, just
two in our case. In a traditional $N_1$-dominated scenario
the first one is dominant. This can be still conveniently
re-cast as the sum of an unflavored term plus a flavored term.
Finally, the unflavored term can be further decomposed into a
piece proportional to a contribution to the total $C\!P$ asymmetry
respecting the usual upper bound \cite{CPbound,di} and into a term
that is not upper bounded but vanishes when $M_2=M_3$. If $M_2\neq M_3$,
this term is typically strongly suppressed when a mild hierarchy
is assumed and barring a fine-tuned choice of the parameters
\cite{hambye,kitano,geometry}. Analogously, we recast also the flavored $C\!P$
asymmetries as the sum of an upper-bounded term plus an unbounded
extra-term that has been neglected in previous works but that proves
to be important in interesting cases.

In the end of Section 3 we review, as a starting point for our analysis,
the neutrino mass bounds that arise in the minimal scenario,
that we call `vanilla leptogenesis', when all simplifications are made at once:
the heavier RH neutrino contribution,
the flavored terms and
the extra-term to the total $C\!P$ asymmetry are neglected and
the hierarchical limit is taken.
In this case the final asymmetry depends
only on 6 parameters \cite{annals,geometry}.

In Section 4 we show the role played by the unbounded term in
the total $C\!P$ asymmetry of the lightest RH neutrino, $\ve_1$,
that has to be considered when $M_2\neq M_3$ \cite{hambye,kitano,geometry}.
We study its effect on the bounds in a quite conservative mild
hierarchical limit $M_2\simeq 3\, M_1$, showing how
a rotation in the complex 2-3 plane is the most effective ingredient
in enhancing this term and showing that the bounds get modified
only for fine tuned parameter choices. Our analysis shows moreover that
this extra-term acts independently of flavor effects.

In Section 5 we study how the bounds change when flavor effects are
taken into account and, interestingly, we show that a so far neglected
term in the flavored $C\!P$ asymmetries is actually able to relax the
lower bound on the reheating temperature in presence of wash-out.
In the case of the upper bound on the absolute neutrino mass scale,
we show that this is sensitive to a variation of different assumptions and conditions.
We also show that, taking into account the Higgs asymmetry  then,
within the validity of classical Boltzmann equations,
one cannot say whether the bound $m_1\lesssim 0.1\,{\rm eV}$
holding in the unflavored regime is evaded. This conclusion
is strengthened when independence of the initial
conditions  is imposed.

In Section 6 we consider the contribution from the two heavier RH
neutrinos, including flavor effects and still assuming the hierarchical limit,
showing a (non-trivial) sufficient condition for the $N_1$-dominated scenario to hold.

Finally, in Section 7, we show how the neutrino mass bounds change going beyond
the limit of hierarchical RH neutrinos when flavor effects are included.
Contrarily to a naive expectation, we show that the allowed region,
instead of getting enlarged, actually shrinks when mild degeneracies for
the heavy neutrinos masses are considered. The bounds  get relaxed
only when the heavy neutrino mass degeneracies are much smaller than
those of the light neutrinos.

\section{General set-up}

Leptogenesis is based on a popular extension of the Standard Model,
\begin{equation}\label{lagrangian}
\mathcal{L}= \mathcal{L}_{\rm SM} +i \overline{N_{R i}}\g_{\m}\partial^{\m} N_{Ri} -
h_{\a i} \overline{\ell_{L\a}} N_{R i} \tilde{\F} -
{1\over 2}\,M_i \overline{N_{R i}^c}N_{R i} +h.c.\quad (i=1,2,3,\quad \a=e,\m,\t),
\end{equation}
where three RH neutrinos $N_{R i}$, with
a Majorana mass term $M$ and Yukawa couplings $h$,
are added. After spontaneous symmetry breaking,
a Dirac mass term $m_D=v\,h$, is generated
by the vev $v=174$ GeV of the Higgs boson.
In the see-saw limit, $M\gg m_D$,
the spectrum of neutrino mass eigenstates
splits in two sets: 3 very heavy neutrinos, $N_1,N_2$ and $N_3$
respectively with masses $M_1\leq M_2 \leq M_3$ almost coinciding with
the eigenvalues of $M$, and 3 light neutrinos with masses $m_1\leq m_2\leq m_3$,
the eigenvalues of the light neutrino mass matrix
given by the see-saw formula \cite{seesaw},
\be
m_{\nu}= - m_D\,{1\over M}\,m_D^T \, .
\ee
Neutrino oscillation experiments measure two neutrino mass-squared
differences. For normal schemes one has
$m^{\,2}_3-m_2^{\,2}=\Delta m^2_{\rm atm}$ and
$m^{\,2}_2-m_1^{\,2}=\Delta m^2_{\rm sol}$,
whereas for inverted schemes one has
$m^{\,2}_3-m_2^{\,2}=\Delta m^2_{\rm sol}$
and $m^{\,2}_2-m_1^{\,2}=\Delta m^2_{\rm atm}$.
For $m_1\gg m_{\rm atm} \equiv
\sqrt{\Delta m^2_{\rm atm}+\Delta m^2_{\rm sol}}=
(0.050\pm 0.001)\,{\rm eV}$ \cite{gonzalez}
the spectrum is quasi-degenerate, while for
$m_1\ll m_{\rm sol}\equiv \sqrt{\D m^2_{\rm sol}}
=(0.0088\pm 0.0001)\,{\rm eV}$ \cite{gonzalez}
it is fully hierarchical (normal or inverted).
The most stringent upper bound on the
absolute neutrino mass scale comes from
cosmological observations. Recently, quite a conservative
upper bound,
\be\label{bound}
m_1 < 0.2\,{\rm eV} \, \hspace{5mm} (95\%\, {\rm CL}) \, ,
\ee
has been obtained by the
WMAP collaboration combining CMB, baryon acoustic oscillations
and supernovae type Ia observations \cite{WMAP5}.

With leptogenesis, this simple extension
of the Standard Model is also able to explain
the observed baryon asymmetry of the Universe
\cite{WMAP5}
\be\label{etaBobs}
\eta_B^{\rm CMB} = (6.2 \pm 0.15)\times 10^{-10} \, .
\ee
It is widely known that, in order to generate a baryon asymmetry
in the early Universe, one needs to satisfy the three Sakharov
conditions \cite{sakharov}. At temperatures
$T\gtrsim 100\,{\rm GeV}$, baryon number is violated
by the non-perturbative sphaleron processes \cite{sphalerons}.
Moreover $C\!P$ is violated in the decays of the heavy RH neutrinos.
Indeed the Dirac mass matrix is
in general complex and this provides a natural source of $C\!P$ violation.
At the same time departure from thermal equilibrium occurs in the
decays of the heavy RH neutrinos as well. This can be conveniently quantified
in terms of the decay parameters, defined as
$K_i\equiv \widetilde{\G}_i/H_{T=M_i}$, given by the ratio of the
decay widths to the expansion rate when the RH neutrinos become non-relativistic.
The decay parameters can be expressed in terms of the Yukawa couplings by
\be
K_i={\mti\over m_{\star}} \, ,
\hspace{10mm}
{\rm where}
\hspace{10mm}
\mti\equiv{(m_{\rm D}^{\dagger}\,m_{\rm D})_{ii} \over M_i}
\ee
are the effective neutrino masses and
$m_{\star}$ is the equilibrium neutrino mass \cite{annals} given by
\begin{equation}\label{d}
m_{\star}\equiv
{16\, \pi^{5/2}\,\sqrt{g_*} \over 3\,\sqrt{5}}\,
{v^2 \over M_{\rm Pl}}
\simeq 1.08\times 10^{-3}\,{\rm eV}.
\end{equation}
There are two ways how $C\!P$ violation can manifest itself.
A first one is given by having a decay rate of $N_i$ into leptons,
$\G_{i}$, different from the decay rate into anti-leptons,
$\bar{\G}_{\rm i}$. This is parameterized by the total $C\!P$ asymmetries
\be
\ve_i\equiv -\,{\G_i-\bar{\G}_i\over \G_i+\bar{\G}_i} \, .
\ee
A perturbative calculation from the
interference of tree level with one
loop self-energy and vertex diagrams gives \cite{crv}
\be\label{CPas}
\ve_i =\, {3\over 16\pi}\, \sum_{j\neq i}\,{{\rm
Im}\,\left[(h^{\dagger}\,h)^2_{ij}\right] \over
(h^{\dagger}\,h)_{ii}} \,{\xi(x_j/x_i)\over \sqrt{x_j/x_i}}\, ,
\ee
having introduced \cite{window}
\be\label{xi}
\xi(x)= {2\over 3}\,x\,
\left[(1+x)\,\ln\left({1+x\over x}\right)-{2-x\over 1-x}\right] \, .
\ee
A  second way \cite{nardi1} is given by the possibility
that in general, indicating with $|\ell_i\rangle$ the final
lepton quantum state and with $|\bar{\ell}'_i\rangle $ the final
anti-lepton quantum state, one has $|\ell_i\rangle\neq CP |\ell'_i\rangle$.
This can be easily understood when the flavor composition of the final
lepton state is considered. Indeed, introducing the projectors
on the flavor eigenstates, they can be written like the sum of two terms,
\bea
P_{i\alpha} & \equiv  &
|\langle \ell_{i}|\ell_{\alpha}\rangle |^2  =
P_{i\alpha}^0 + {\Delta P_{i\alpha}\over 2} \\
\bar{P}_{i\alpha}& \equiv &
|\langle \bar{\ell}'_{i}|\bar{\ell}_{\alpha}\rangle |^2  =
P_{i\alpha}^0 - {\Delta P_{i\alpha}\over 2} \, .
\eea
The first term is the tree level contribution and is
common to both projectors but the second term, from loop
corrections, changes sign and gives rise to a
different flavor composition when $\D\,P_{i\a}\neq 0$.
These two $C\!P$ violating effects translate respectively in two
separate terms in the flavored $C\!P$ asymmetries,
\be
\ve_{i\a}\equiv -{\G_{i\alpha}-\overline{\G}_{i\alpha}
\over \G_{i}+\overline{\G}_{i}}= P^0_{i\a}\,\ve_{i\a}+{\D\,P_{i\a}\over 2} \, ,
\ee
where $\G_{i\alpha}\equiv P_{i\a}\,\G_i$ and
$\bar{\G}_{i\alpha}\equiv \bar{P}_{i\a}\,\bar{\G}_i$.
The flavored $C\!P$ asymmetries can be calculated using \cite{crv}
\be\label{eps1a}
\ve_{i\a}=
\frac{3}{16 \p (h^{\dag}h)_{ii}} \sum_{j\neq i} \left\{ {\rm Im}\left[h_{\a i}^{\star}
h_{\a j}(h^{\dag}h)_{i j}\right] \frac{\x(x_j/x_i)}{\sqrt{x_j/x_i}}+
\frac{2}{3(x_j/x_i-1)}{\rm Im}
\left[h_{\a i}^{\star}h_{\a j}(h^{\dag}h)_{j i}\right]\right\} \, .
\ee
The second $C\!P$ violating contribution  yields
an additional contribution to the final asymmetry only
if the flavor composition plays a role in the
determination of the final asymmetry.
This depends on the effectiveness of the charged lepton interactions
implied by the term $f_{\a}\bar{\ell}_{L\a}e_{R\a}\F$ in the Lagrangian,
which is diagonal in flavor space.
The latter implies that the processes
$\ell_{\a}\bar{e}_{\a} \leftrightarrow \F$ and $\ell_{\a}\bar{e}_{\a}
\leftrightarrow \F \, A^{\rm a}$ and the {\emph{CP} conjugated,
(where $A^{\rm a}~({\rm a}=1,2,3)$
 are the $SU(2)_L$ gauge bosons) occur at a
rate $\G_{\alpha}\simeq 5\times 10^{-3}\,T\,
f^2_{\alpha}\,(\a=e,\m,\t)$ \cite{Campbell:1992jd}.
If these processes are effective, then they measure the flavor
composition of the final leptons and this becomes a relevant ingredient
in the determination of the final asymmetry if a second condition is
fulfilled as well, as we will comment.

In the hierarchical limit the decays of just one species of RH neutrino,
the lightest \cite{fy} or the next-to-lightest \cite{geometry}, dominantly contribute to
the final asymmetry. If $\G_{\alpha}\ll \G_{\rm ID}^i \;(i=1,2)$
during the relevant period of the asymmetry generation,
where $\G_{\rm ID}^i$ denotes the inverse-decay rate,
then the coherence of the lepton states is preserved on average
between a decay and a subsequent inverse decays and the unflavored
regime, where flavor effects are negligible, holds.
This requirement implies \cite{zeno}
\be\label{unflavored}
M_i\gtrsim 5\times 10^{11}\,{\rm GeV} \, .
\ee
In this case an approximate set of Boltzmann equations is given by
\begin{eqnarray}
{dN_{N_i}\over dz} & = &
-D_i\,(N_{N_i}-N_{N_i}^{\rm eq}) \;,
\hspace{10mm} i=1,2,3 \label{dlg1} \\\label{unflke}
{dN_{B-L}\over dz} & = &
\sum_{i=1}^3\,\varepsilon_i\,D_i\,(N_{N_i}-N_{N_i}^{\rm eq})-
N_{B-L}\,[\D W(z)+\sum_i \,W_i^{\rm ID}(z)] \; ,
\label{dlg2}
\end{eqnarray}
where $z \equiv M_1/T$ and where we indicated with $N_X$
any particle number or asymmetry $X$ calculated in a portion of co-moving
volume containing one heavy neutrino in ultra-relativistic thermal equilibrium,
so that $N^{\rm eq}_{N_i}(T\gg M_i)=1$.
With this convention, the predicted baryon-to-photon ratio $\eta_B$ is
related to the final value of the final $B-L$ asymmetry by the relation
\be\label{etaB}
\eta_B=a_{\rm sph} {N_{B-L}^{\rm f}\over N_{\g}^{\rm rec}}\simeq 0.96\times
10^{-2} N_{B-L}^{\rm f}\, ,
\ee
where $N_{\g}^{\rm rec}\simeq 37$, and $a_{\rm sph}=28/79$.
Defining $x_i\equiv M_i^2/M_1^2$ and $z_i\equiv z\,\sqrt{x_i}$,
the decay factors are given by
\be
D_i \equiv {\G_{{\rm D},i}\over H\,z}=K_i\,x_i\,z\,
\left\langle {1\over\gamma_i} \right\rangle   \, ,
\ee
where $H$ is the expansion rate. The total decay rates,
$\G_{{\rm D},i} \equiv \G_i+\bar{\G}_i$,
are the product of the decay widths times the
thermally averaged dilation factors
$\langle 1/\gamma\rangle$, given by the ratio
${\cal K}_1(z)/ {\cal K}_2(z)$ of the modified Bessel functions.
The equilibrium abundance and its rate are also expressed through
the modified Bessel functions,
\be
N_{N_i}^{\rm eq}(z_i)= {1\over 2}\,z_i^2\,{\cal K}_2 (z_i) \;\; ,
\hspace{10mm}
{dN_{N_i}^{\rm eq}\over dz_i} =
-{1\over 2}\,z_i^2\,{\cal K}_1 (z_i) \, .
\ee
After proper subtraction of the resonant contribution from
$\Delta L=2$ processes \cite{dolgov}, the inverse decay
washout terms are simply given by
\be\label{WID}
W_i^{\rm ID}(z) =
{1\over 4}\,K_i\,\sqrt{x_i}\,{\cal K}_1(z_i)\,z_i^3 \, .
\ee
The washout term $\D W (z)$ is the non-resonant $\D L=2$ processes contribution.
It gives a non-negligible effect only at $z\gg 1$ and in this case
it can be approximated as \cite{annals}
\be
\Delta W(z) \simeq {\o \over z^2}\,\left(M_1\over 10^{10}\,{\rm GeV}\right)\,
\left({\overline{m}^{\, 2} \over {\rm eV^2}}\right) \, ,
\ee
where $\o\simeq 0.186$ and $\overline{m}^2\equiv m_1^2+m_2^2+m_3^2$.
Notice that we are neglecting $\D L=1$ scatterings \cite{luty,abada2},
giving a correction to a level less than $\sim 10 \% $  \cite{flavorlep},
thermal corrections \cite{giudice}, giving relevant
(though with big theoretical uncertainties) corrections only in the weak washout,
and spectator processes \cite{buchplum,nardi2}, that produce corrections to a
level less than $\sim 20\% $ \cite{nardi2}.

On the other hand, if the charged lepton Yukawa interactions are in equilibrium
($\G_{\a}> H$) and faster than inverse decays, i.e.
\be\label{condition1}
\G_{\a}\gtrsim \G_{\rm ID}^i \, ,
\ee
during the relevant period of the asymmetry generation,
then lepton quantum states lose coherence between the production
at decay and the subsequent absorption in inverse processes.
If the quantum state becomes completely incoherent
and is fully projected on one of the flavor eigenstates, each lepton flavor
eigenstate $\ell_{\a}$ can be treated as a statistically independent particle species
and a `fully flavored regime' is obtained.
Note that one has to distinguish a two-flavor regime, for
$M_i\gtrsim 10^{\rm 9}\,{\rm GeV}$, such that
the condition Eq.~(\ref{condition1}) is satisfied
only for $\a=\t$, and a three-flavor regime, where
it applies also to $\a=\m$.

In the fully flavored regime (two or three flavors), classical Boltzmann equations
can be still used like in the unflavored regime, with the difference, in general,
that now each single flavor asymmetry has to be tracked independently.
Since sphaleron processes conserve the quantities
$\D_{\a}\equiv B/3-L_{\a}$ ($\a=e,\m,\t $), these are the convenient
independent variables to be used in the set of
Boltzmann equations that can be written as
\bea
{dN_{N_i}\over dz} & = & -D_i\,(N_{N_i}-N_{N_i}^{\rm eq})
\hspace{52mm} (i=1,2,3) \\
{dN_{\D_{\a}}\over dz} & = &
\sum_i\,\ve_{i\a}\,D_i\,(N_{N_i}-N_{N_i}^{\rm eq})
-\sum_{i,\b}\,P_{i\a}^{0}\,(C^{\ell}_{\a\b}+C^H_{\b})\,W_i^{\rm ID}\,N_{\D_{\b}} \, ,
\label{flke}
\eea
where  we are using the same approximations as in
the unflavored case but neglecting the non-resonant
$\D L=2$ term, since this counts only for
$M_1\gtrsim 10^{14}\,{\rm GeV}\,(m_{\rm atm}^2/\sum_i\,m_i^2)$
like also a contribution from  $\D  L=0$ processes
that one has to consider in the flavored case.
Notice that the final $B-L$ asymmetry is now calculated as
$N_{B-L}^{\rm f}= \sum_{\a}\,N_{\D_{\a}}^{\rm f}$.

The $C^{\ell}$ matrix \cite{bcst}  relates the asymmetries
stored in lepton doublets, ${\ell}_{\a}$, to the asymmetries
$\D_{\a}\equiv B/3-L_{\a}$ asymmetries and is given, in a
two-flavor regime, by~\cite{abada2}
\be\label{Cl}
C^{\ell}= \, {1\over 316}\left(\begin{array}{cc}
270 & -32\\
-17 & 208\end{array}\right).
\ee
The $C^H$ matrix takes into account the washout due to
the asymmetry stored in the Higgs field~\cite{nardi1}
and is given by
\be
C^{H} ={1\over 158}(41,56) \, .
\ee
The Higgs asymmetry has been neglected so far
but, as we will see, it has a relevant
effect on the upper bound on the neutrino masses.
Indeed the sum of the two matrices gives
\be\label{C}
C \equiv C^{\ell}+C^{H}\simeq \left(\begin{array}{cc}
1.11 & 0.25\\
0.21 & 1.01\end{array}\right) .
\ee
The off-diagonal terms give a small effect
in the calculation of the final asymmetry \cite{abada2} and
therefore, in the end, one can safely use the approximation
$C\simeq I$ in the derivation of the bounds.
In Section 5 we will compare the results when the Higgs
asymmetry is neglected, when it is taken into account and when one uses the
approximation $C\simeq I$, showing that the latter works very well
justifying its use for the remainder of the paper.

Taking for simplicity the two flavor case,
it is instructive to sum over the flavor Eq.~(\ref{flke}),
obtaining
\be
{dN_{B-L}\over dz} \simeq
\sum_{i=1}^3\,\ve_i\,D\,(N_{N_i}-N_{N_i}^{\rm eq})
-{1\over 2}\,N_{B-L}\,\sum_{i=1}^3\,W_i^{\rm ID}\,
+{1\over 2}\,[N_{\D_{\a}}-N_{\D_{\b}}]\,
\sum_{i=1}^3\,(P^0_{i\a}-P^0_{i\b})\,W_i^{\rm ID} \, .
\ee
This equation clearly shows that when the washout vanishes
there must be no difference between the unflavored and the fully flavored
regime. Therefore, the lower bounds on $M_1$ and on $T_{\rm reh}$
obtained in the limit of no-washout,
rigorously for an initial thermal abundance
and approximately for an initial vanishing abundance,
do not change when flavor effects are taken into account \cite{flavorlep}.

A convenient parametrization of the Dirac mass matrix is obtained
in terms of the orthogonal matrix \cite{casas}
\be\label{h}
m_D=U\,D_m^{1/2}\,\O\, D_M^{1/2} \,  ,
\ee
where we defined
$D_m\equiv {\rm diag}(m_1,m_2,m_3)$ and
$D_M\equiv {\rm diag}(M_1,M_2,M_3)$. The matrix $U$ diagonalizes
the light neutrino mass matrix $m_{\nu}$, such that
$U^{\dagger}\,m_{\nu}\,U^{\star}=-D_m$, and it can be identified
with the lepton mixing matrix in a basis where
the charged lepton mass matrix is diagonal.
We will adopt the parametrization  \cite{PDG}
\begin{equation}\label{Umatrix}
U=\left( \begin{array}{ccc}
c_{12}\,c_{13} & s_{12}\,c_{13} & s_{13}\,e^{-{\rm i}\,\d} \\
-s_{12}\,c_{23}-c_{12}\,s_{23}\,s_{13}\,e^{{\rm i}\,\d} &
c_{12}\,c_{23}-s_{12}\,s_{23}\,s_{13}\,e^{{\rm i}\,\d} & s_{23}\,c_{13} \\
s_{12}\,s_{23}-c_{12}\,c_{23}\,s_{13}\,e^{{\rm i}\,\d}
& -c_{12}\,s_{23}-s_{12}\,c_{23}\,s_{13}\,e^{{\rm i}\,\d}  &
c_{23}\,c_{13}
\end{array}\right)
\times {\rm diag(e^{i\,{\Phi_1\over 2}}, e^{i\,{\Phi_2\over 2}}, 1)}
\, ,
\end{equation}
where $s_{ij}\equiv \sin\theta_{ij}$,
$c_{ij}\equiv\cos\theta_{ij}$ and, neglecting
the statistical errors, we will use
$\theta_{12}=\pi/5$ and $\theta_{23}=\pi/4$,
compatible with the results from neutrino oscillation experiments.
Moreover, we will adopt the $3\s$ range $s_{13}=0-0.20$,
allowed from a global $3\n$ analysis for unitary $U$ \cite{gonzalez},
an approximation that holds
with great precision in the see-saw limit with $M_i\gg 100\,{\rm GeV}$.
With the adopted convention for the light neutrino masses, $m_1<m_2<m_3$,
this parametrization is valid only for normal hierarchy, while for inverted
hierarchy one has to perform a column cyclic permutation.
In a general analysis, leptogenesis bounds are not depending
on the scheme, normal or inverted, but
in restricted scenarios, like in the effective two RH neutrino scenario
where the third is very heavy and decouples or in `Dirac phase leptogenesis' \cite{diraclep},
differences can arise and depend on flavor effects.
We will signal these differences in our analysis.

It will also prove useful to introduce the following
parametrization for the see-saw orthogonal matrix in terms
of complex rotations
\be\label{second}
\O({\o}_{21},{\o}_{31},{\o}_{32})
={\rm diag (\pm,\pm,\pm)}\,
 R_{12}(\o_{21})\,\,
 R_{13}(\o_{31})\,\,
 R_{23}(\o_{32})\,\, ,
\ee
where
\be\label{R}
\mbox{\tiny $
R_{12}=
\left(
\begin{array}{ccc}
 \pm \sqrt{1-{\o}^2_{21}}  &  -{\o}_{21}          & 0 \\
            {\o}_{21}  & \pm \sqrt{1-{\o}^2_{21}} & 0 \\
  0 & 0 & 1
\end{array}
\right) \,
         \,\, , \,\,
R_{13}=
\left(
\begin{array}{ccc}
\pm \sqrt{1-{\o}^2_{31}}  & 0 &  - {\o}_{31} \\
    0 & 1 & 0 \\
  {\o}_{31} & 0 & \pm \sqrt{1-{\o}^2_{31}}
\end{array}
\right)  \,\, ,
\,\,
R_{23}=
\left(
\begin{array}{ccc}
  1  &  0   & 0   \\
  0  &  \pm \sqrt{1-{\o}^2_{32}} & - {\o}_{32} \\
  0 & {\o}_{32}  & \pm \sqrt{1-{\o}^2_{32}}
\end{array}
\right)$ \,
}
\ee
and where the overall sign takes into account the possibility
of a  parity transformation as well.
Notice that, using the orthogonal parametrization, Eq. (\ref{h}), the
effective neutrino masses, and consequently the decay parameters,
can be expressed as linear combinations of the neutrino masses
\cite{fhy,annals}, such that $\mti = \sum_j\,m_j\,|\O_{ji}^2|$.
Notice that the orthogonality of $\O$ is equivalent to the see-saw relation
for the light neutrino masses and in particular one has that
${\rm Re}[\O_{ij}^2]$ is the contribution to $m_i$ from the term $\propto 1/M_j$.
Therefore, large absolute values of the $\O$ entries
imply a strong fine tuning not only because they require phase cancelations
but also because they imply that neutrino masses are much lighter than
terms $\propto m_D^2/M$ because of sign cancelations.
Therefore, such choices tend to transfer the explanation of neutrino lightness
from the see-saw mechanism to some other mechanism that has to explain
the fine-tuned cancelations. The interest
for considering models with very large $|\O_{ij}|$ is merely phenomenological since
they make possible to satisfy neutrino masses with the see-saw mechanism
and at the same time to have ${\rm TeV}$ RH neutrinos
with large Yukawa's, making possible to detect them in colliders \cite{colliders}.
This will not be our point of view in this paper and we will
conventionally consider orthogonal matrices to be `reasonable' if
$|\o_{ij}|\leq 1$, implying $|\O_{ij}|\lesssim 1$,
and `acceptable' if $|\o_{ij}|\leq 10$, implying $|\O_{ij}|\lesssim 10$.
These will be the two benchmark cases that we will adopt in the plots.

To conclude this Section we just want to recall the particularly relevant
case of two effective RH neutrinos, obtained in the limit
where $M_3 \gg 10^{14}\,{\rm GeV}$ \cite{2effRH,turzynski,seealso2}. In this limit
the orthogonal matrix necessarily collapses into
\be\label{2effRH}
\O=
\left(
\begin{array}{ccc}
     0            &   0                 &  1 \\
\pm \sqrt{1-\O^2_{31}}& -\O_{31}            &  0 \\
       \O_{31}    & \pm \sqrt{1-\O^2_{31}}  &  0
\end{array}
\right)
\, ,
\ee
corresponding to have $\o_{32}=1$ and $\o_{21}=1$  in the Eq.~(\ref{R}).

\section{Vanilla leptogenesis}

Using the approximation $C \simeq I$,
a general solution for the final asymmetry
can be written as \cite{flavorlep}
\be\label{NfB-L}
N_{B-L}^{\rm f}=
\sum_{\alpha}\,N_{\D\a}^{\rm in}\,
{\rm e}^{-[\D W(z)+\sum_i\,P_{i\a}^0\,\int_{z_{\rm in}}^z\,dz'\,W_i^{\rm ID}(z')]}
+\sum_{i,\a}\,\ve_{i\a}\,\k_{i{\a}}^{\rm f} \,  ,
\ee
with the final values of the 9 efficiency factors given by
\be\label{efial}
\k_{i\a}^{\rm f}(K_i,P^{0}_{i\a})=-\int_{z_{\rm in}}^\infty\,dz'\,{dN_{N_i}\over dz'}\,
{\rm e}^{-[\D W(z)+\sum_i\,P_{i\a}^0\,\int_{z'}^\infty\,dz''\,W_i^{\rm ID}(z'';K_i)]} \,.
\ee
This solution holds both in the fully flavored regime
and in the unflavored regime, adopting the convention that
when the condition (\ref{unflavored}) applies, all projectors $P_{i\a}^0=1$.
It is indeed easy to verify that in this case, summing over the flavor,
the set of equations~(\ref{flke}) reduces to Eq.~(\ref{unflke}).
On the other hand it should be also noticed that Eq.~(\ref{flke})
holds only if the condition~(\ref{condition1}) is respected and this, when
applied to the $N_1$ decays and inverse decays, translates into
\be\label{condition}
M_1\lesssim \,{10^{12}~{\rm GeV}\over 2\,W_1^{\rm ID}(z_{\rm B}(K_{1\a}))} .
\ee
For $W_1^{\rm ID}(z_{\rm B})\gtrsim 1$ there is an intermediate regime where
the unflavored regime does not hold but at the same time the validity of
the Eq.~(\ref{flke}), and therefore of the expression~(\ref{NfB-L}), is not
guaranteed. We will signal in the plots the results obtained in the
fully flavored regime but for which
the condition~(\ref{condition}) is not satisfied. These results should
be therefore checked within a more general kinetic description employing
density matrix equations able to describe the regime where coherence
(or decoherence) of the final lepton quantum state is only partial.

Notice that the $N_{\D_{\a}}^{\rm in}$'s are the values of
possible pre-existing flavored asymmetries.
In the unflavored regime $K_1$ is the only parameter that
determines whether
the final asymmetry depends or not on a possible pre-existing
asymmetry \cite{window}. Taking into account flavor effects
the problem is more involved and there are different issues
to be considered. Here we will not face this problem and we
will simply assume that the first term is negligible.

It is instructive to re-cast the Eq.~(\ref{NfB-L}) in an
approximate way that enlightens different effects and contributions.
For definiteness, we consider the
two-flavor regime holding for $M_1\gtrsim 10^{9}\,{\rm GeV}$.
The total $C\!P$ asymmetry $\ve_1$ can be recast as~\cite{geometry}
\be\label{ve1xi}
\ve_1=\xi(x_2)\,\bar{\ve}_1(M_1,m_1,\o_{21},\o_{31})+[\xi(x_3)-\xi(x_2)]\,
\Delta\ve_1(M_1,m_1,\O) \, ,
\ee
where, defining
\be
\bar{\ve}(M_1) \equiv
{3\over 16\pi}\,{M_1\,m_{\rm atm}\over v^2}
\hspace{7mm}
{\rm and}
\hspace{7mm}
\beta(m_1,\o_{21},\o_{31}) \equiv
{\sum_j\,m^2_j\,{\rm Im}
\,(\Omega^2_{j1})\over m_{\rm atm}
\,\sum_j\,m_j\,|\Omega_{j1}^2|} \, ,
\ee
one has
\be
\bar{\ve}_1(M_1,m_1,\o_{21},\o_{31})=
\bar{\ve}(M_1)\,\beta(m_1,\o_{21},\o_{31})
\ee
and
\be\label{extra}
\Delta\ve_1 (M_1,m_1,\O)=
\bar{\ve}(M_1)\,
{{\rm Im}[\sum_h\,m_h\,\O_{h1}^{\star}\O_{h3}]^2\over m_{\rm atm}\,\mt} \, .
\ee
With these definitions, the final asymmetry can be written approximately as
\bea\label{master}
N_{B-L}^{\rm f} & \simeq &
N_{\rm fl}\,\left\{\xi(x_2)\,\bar{\ve}_1(M_1,m_1,\o_{21},\o_{31})+
[\xi(x_3)-\xi(x_2)]\,\Delta\ve_1(M_1,m_1,\O)\right\}\,\kappa_1^{\rm f} \\ \nonumber
&  &
+{\Delta P_{1\alpha}(M_1,m_1,\O,U)\over 2}\,
[\kappa_{1\a}^{\rm f}-\k_{1\b}^{\rm f}]
+\sum_{\alpha}\,[\ve_{2\a}\k_{2\a}^{\rm f}+\ve_{3\a}\k_{3\a}^{\rm f}] \, ,
\eea
where $N_{\rm fl}$ is an effective number of flavors.
In the unflavored regime one has $N_{\rm fl}=1$, and
in the fully flavored regime, if both flavors experience a
strong washout ($P_{1\a,\b}^0 K_1 \gg 1$),
one has approximately $N_{\rm fl}\simeq 2$,
while in general $1\leq N_{\rm fl} \lesssim 2$.

The simplest scenario, that we call {\em vanilla leptogenesis}
borrowing the name from observational cosmology,
corresponds to taking all possible simplifying assumptions:
\begin{enumerate}\label{vanilla}
\item hierarchical limit, $M_2\gtrsim 3\,M_1$, so that $\xi(x_2)\simeq 1$;
\item negligible contribution from the heavier RH neutrinos;
\item negligible flavor effects ($N_{\rm fl}=1$ and $\D P_{1\a}=0$);
\item $M_2=M_3$, so that $\xi(x_2)-\xi(x_3)=0$.
\end{enumerate}
The calculation of the final asymmetry then simply reduces to
\be
N_{B-L}^{\rm f} \simeq
\bar{\ve}_1(m_1,M_1,\o_{21},\o_{31})\,\kappa_1^{\rm f}(m_1,M_1,\o_{21},\o_{31}) \, ,
\ee
depending only on 6 unknown parameters.
The efficiency factor is well approximated by~\cite{annals}
\be\label{ef}
\k_1^{\rm f} (M_1,m_1,K_1)=\k^{\rm f}_1(K_1)\,
\exp \left\{-{\o \over z_{\rm B}}\left(M_1\over 10^{10}~{\rm GeV}\right)
\left(\overline{m}\over {\rm eV}\right)^2\right\} \, .
\ee
For the case of a thermal initial $N_1$-abundance
($N_{N_1}^{\rm in}=1$), one has
\be\label{k}
\k_{1}^{\rm f}(K_1)\simeq \k(K_1) \equiv {2\over K_1\,z_{\rm B}(K_1)}\,
\left[1-\exp\left(-{1\over 2}{K_1\,z_{\rm B}(K_1)}\right)\right] \, ,
\ee
where  $z_{\rm B}$ is approximately given by the expression \cite{beyond}
\be\label{zB}
z_{\rm B}(K_1) \simeq 2+4\,{K_1}^{0.13}\,\exp\left(-{2.5\over K_1}\right) \, .
\ee
In the relevant range $5\lesssim K_1 \lesssim 100$
this expression is further well approximated by the simple
power-law $\kappa_1^{\rm f}(K_1)\simeq 0.5/K_1^{1.2}$ \cite{proc}.

In the case of a vanishing initial $N_1$-abundance ($N_{N_1}^{\rm in}=0$),
one has to take into account both a negative and a positive contribution,
such that
\be
\k_{1}^{\rm f} (K_1) = \k_{-}^{\rm f}(K_1)+ \k_{+}^{\rm f}(K_1) \, .
\ee
The analytic expressions for  $\k_{-}^{\rm f}(K_1)$ and
$\k_{+}^{\rm f}(K_1)$ can be found in \cite{annals}.
Imposing that the predicted final asymmetry, Eq.~(\ref{etaB}),
explains the observed one, Eq.~(\ref{etaBobs}), yields the condition
\be\label{M1unfl}
M_1={\overline{M}\over \kappa_1^{\rm f}(M_1,m_1,K_1)\,
                \beta(m_1,\o_{21},\o_{31})} \, ,
\ee
where we defined
\be\label{barM}
\overline{M}\equiv
{16\,\pi\over 3}\,
{N_{\g}^{\rm rec}\,v^2\over a_{\rm sph}}\,
{\eta_B^{\rm CMB}\over m_{\rm atm}}
=(6.6\pm 0.3)\times 10^8\,{\rm GeV} \gtrsim 5.7 \times 10^8{\rm GeV} \, .
\ee
The last inequality gives the $3\s$ value of $\overline{M}$ that we used in
the plots  to obtain the allowed region in the $(m_1,M_1)$ plane
scanning over all values of $\o_{21}$ and $\o_{31}$.
 The result is shown in the central panel of Fig.~\ref{fig:vanilla}.
\begin{figure}
\psfig{file=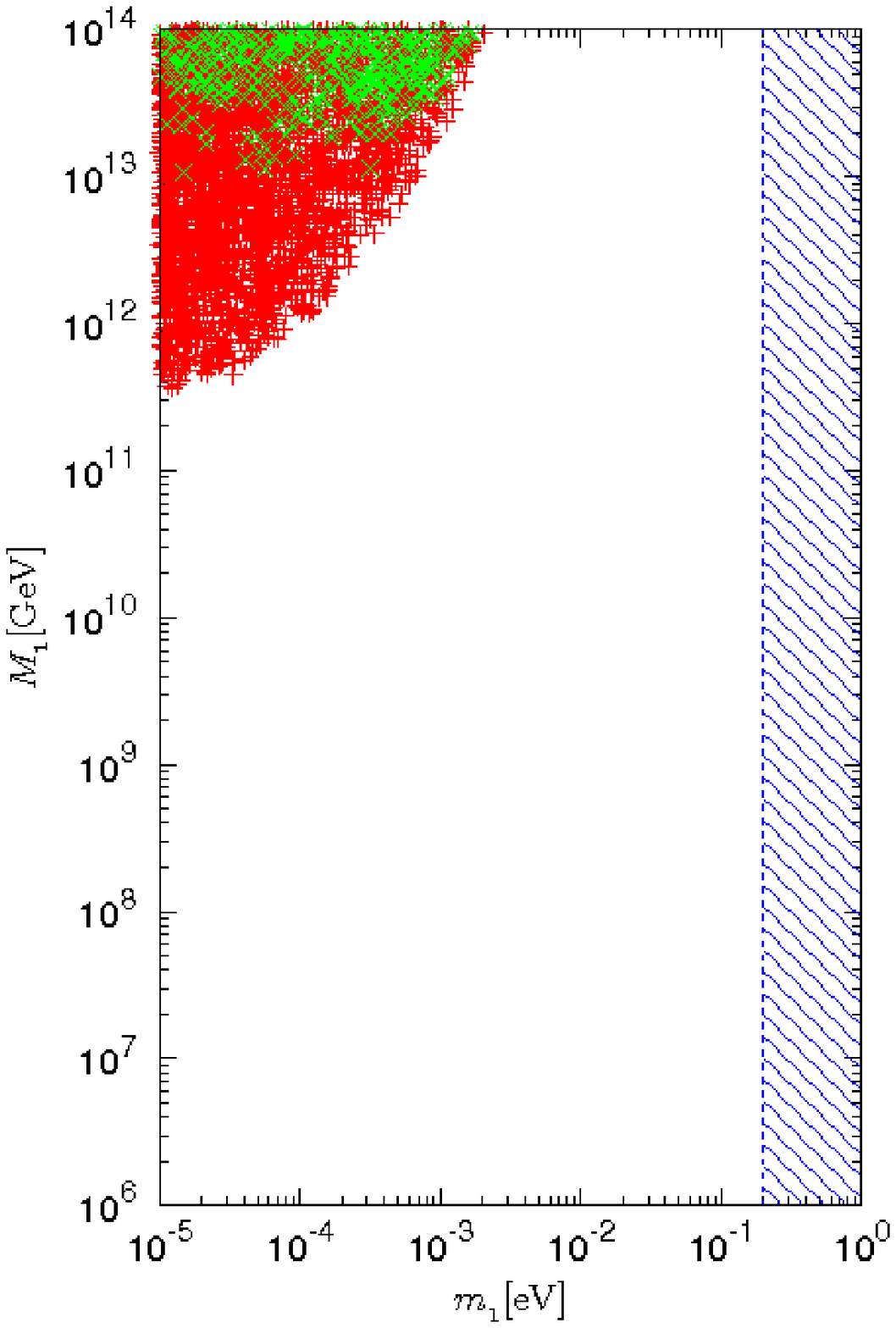,height=71mm,width=54mm}
\hspace*{-5mm}
\psfig{file=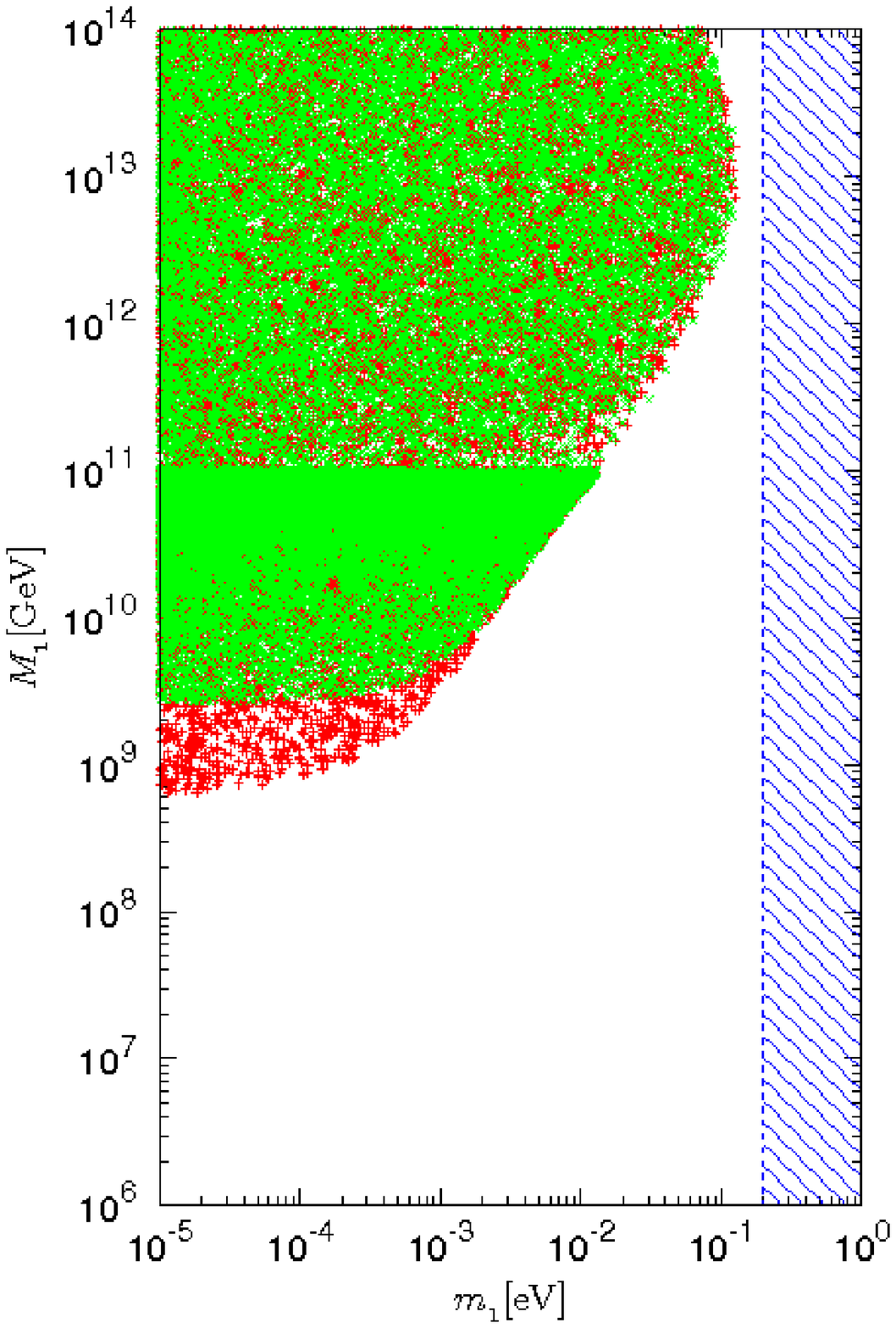,height=71mm,width=54mm}
\hspace{-1mm}
\psfig{file=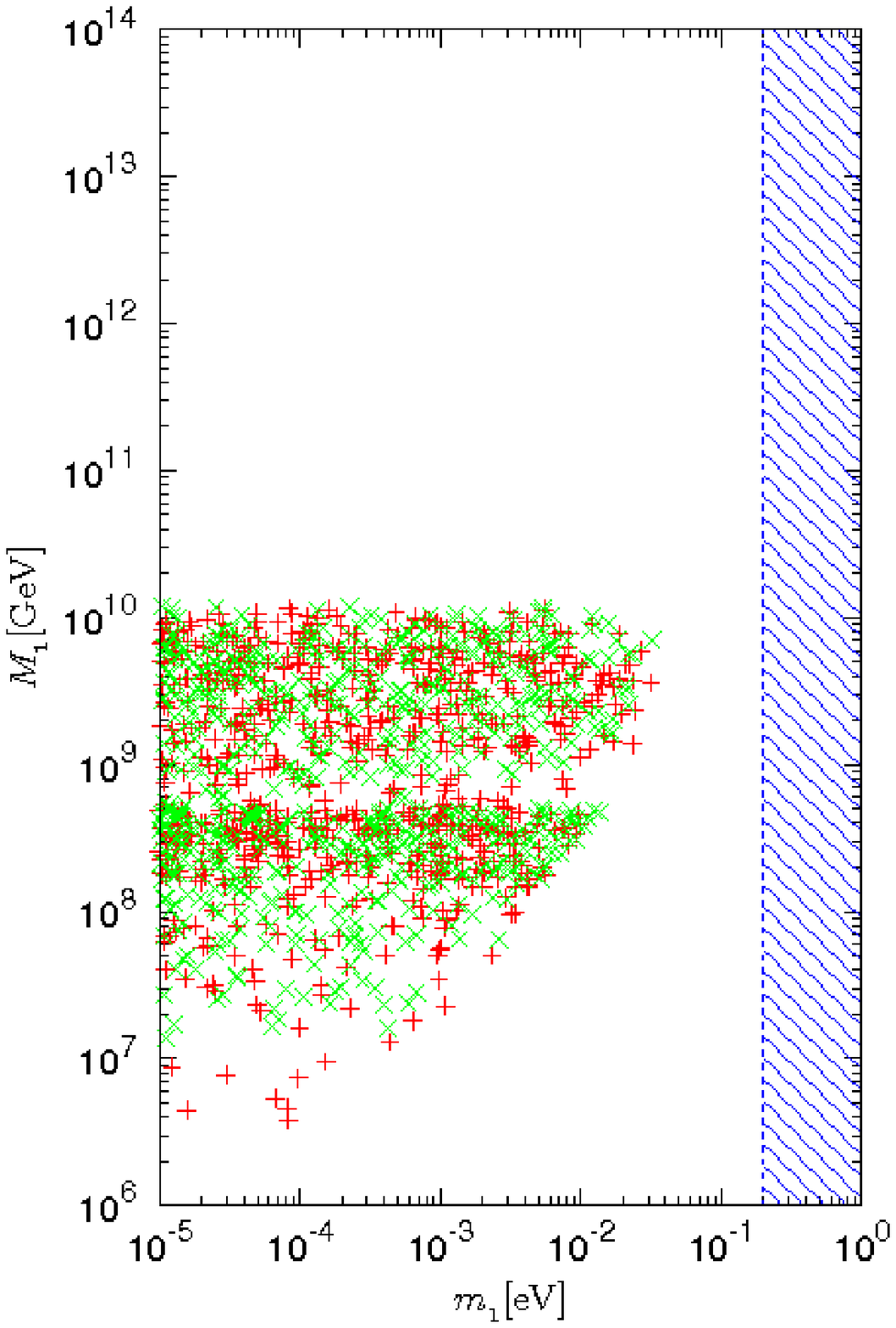,height=71mm,width=54mm}
\caption{Vanilla leptogenesis and leptogenesis conspiracy. The allowed
region in the plane $(m_1,M_1)$ is shown for different values of $m_{\rm atm}$
setting the value of $\overline{M}$ (cf. Eq.~(\ref{barM})).
In the central panel the true measured value is considered, $m_{\rm atm}\simeq 0.050\,{\rm eV}$,
while in the left and right panels two `wrong' values are considered,
$m_{\rm atm}=10^{-4}\,{\rm eV}$ (left) and $m_{\rm atm}=10\,{\rm eV}$ (right).
In all three panels, the red
crosses correspond to a thermal initial $N_1$-abundance while the green
ones to a vanishing initial $N_1$-abundance. The hatched area
indicates values of $m_1$ excluded by the upper bound Eq.~(\ref{bound}).}
\label{fig:vanilla}
\end{figure}
One can notice the presence of the usual lower bound
on the lightest RH neutrino mass, $M_1\gtrsim 2.3\times 10^{9}\,{\rm GeV}$
for initial vanishing abundance and $M_1\gtrsim 5.7\times 10^8\,{\rm GeV}$
for initial thermal abundance \cite{cmb}.
This can be easily inferred from the Eq.~(\ref{M1unfl})
neglecting the exponential factor in Eq.~(\ref{ef})
and using the well-known upper bound
\be\label{beta}
\beta(m_1,\o_{21},\o_{31}) \leq
{m_{\rm atm}\over m_1+m_3}\,f(m_1,\mt) \, ,
\ee
where the function $0\leq f(m_1,\mt) \leq 1$ is unity in the limit
$\mt/m_1\rightarrow \infty$, and vanishes for $\mt=m_1$.
This function can be derived analytically together with
simple analytic expressions valid in particular regimes \cite{geometry}.
For $m_1\ll \mt \ll m_{\star}$ and for initial thermal abundance one simply
finds $M_1\geq \overline{M} \gtrsim 5.7\times 10^{8}\,{\rm GeV}$,
in agreement with the numerical result.
From the central panel of Fig. 1 one can also notice that vanilla letogenesis
predicts $m_1\lesssim 0.12\,{\rm eV}$~\cite{window}, in
agreement with the current observational upper bound (cf. Eq.~(\ref{bound})).
In the plots the hatched region indicates the excluded values.
This upper bound can be derived analytically as well \cite{annals}.

We performed a simple exercise showing how the allowed region
in the $(m_1,M_1)$ plane would have been for values of $m_{\rm atm}$
and $m_{\rm sol}$ different from the true measured ones. We kept
the ratio $m_{\rm atm}/m_{\rm sol}$ constant.
In the left panel  $m_{\rm atm}=10^{-4}~{\rm eV}$,
while in the right panel  $m_{\rm atm}=10~{\rm eV}$.
One can see how the allowed region shrinks considerably, and almost disappears
for these extreme values. This is one way to show the `leptogenesis conspiracy' \cite{aspects},
that means how, order-of-magnitude-wise, the measured atmospheric and solar neutrino mass scales
are optimal for leptogenesis to be successful.
It should be noticed that for $m_{\rm atm}=10~{\rm eV}$, even though
the lower bound on $M_1$ is much lower, the density of points in the allowed region
is very low since they correspond to a very fine-tuned
situation where $\mt\ll m_{\rm atm}$.

\begin{figure}[t!]
\begin{center}
\psfig{file=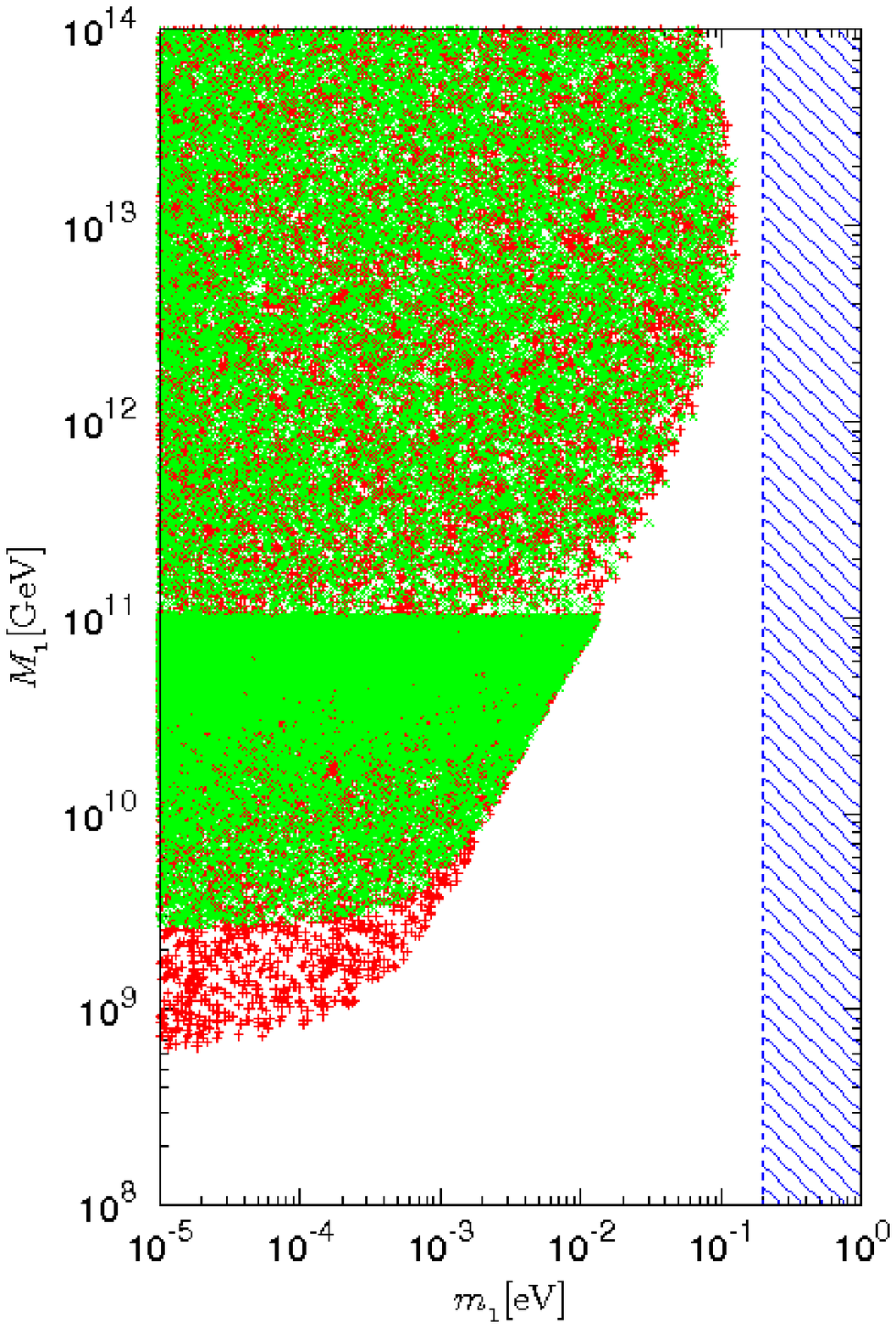,height=75mm,width=65mm}
\hspace{1cm}
\psfig{file=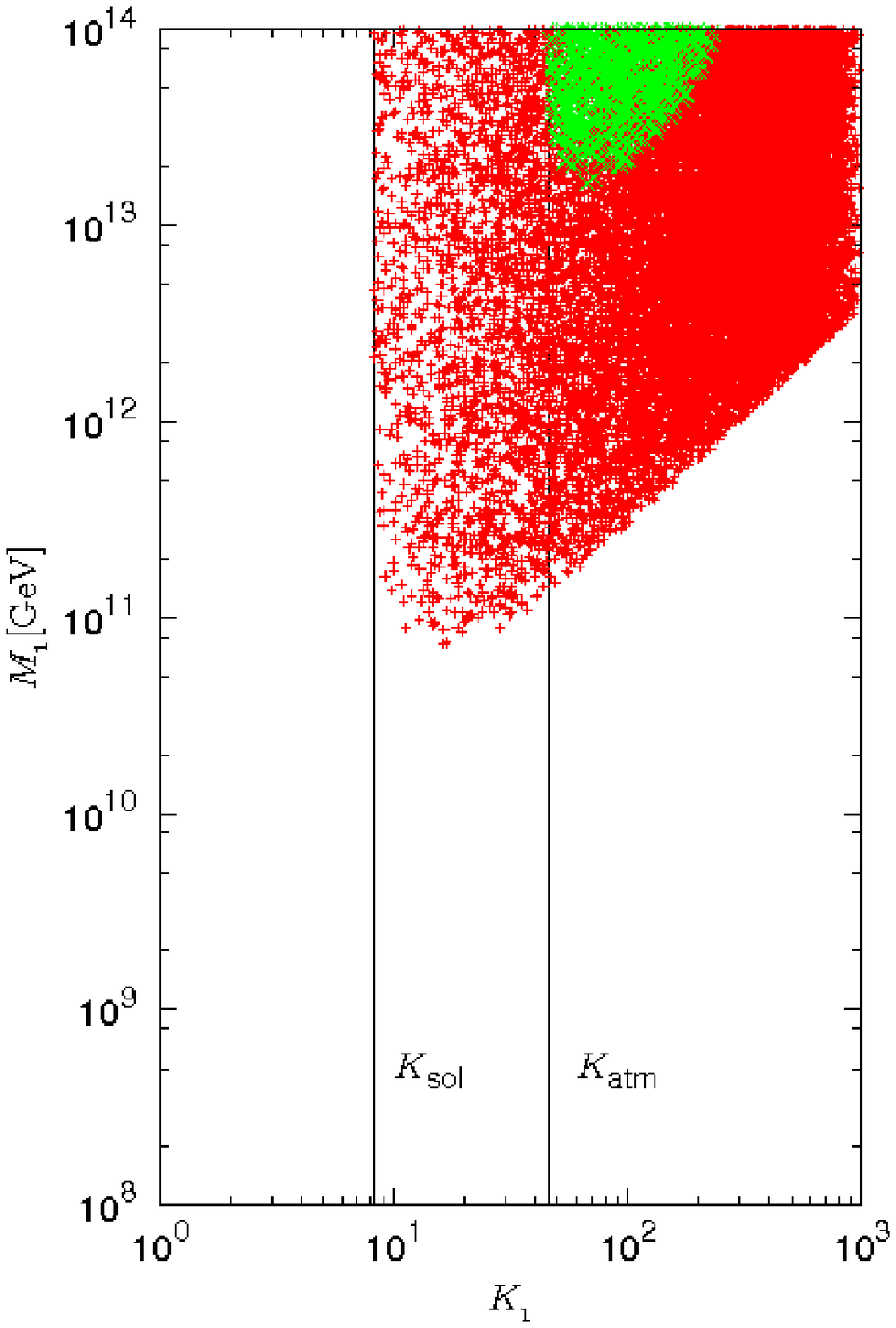,height=75mm,width=65mm}
\caption{Allowed region in the plane $(m_1,M_1)$ for $\O=R_{13}$ (left panel),
and $M_1$ lower bound versus $K_1$ for the effective 2 RH neutrino case obtained
for $M_3\to \infty$ and implying $m_1=0$ (right panel). In the left panel,
the color coding is the same as in the previous figure. In the right
panel, the red crosses correspond to normal hierarchy
whereas the green ones to inverted scheme.}\label{fig:R13vanilla}
\end{center}
\end{figure}

In Fig.~\ref{fig:R13vanilla} we show the bounds for particular choices of the orthogonal matrix:
in the left panel for $\O=R_{13}$, corresponding to $\o_{21}=\o_{32}=0$,
and in the right panel for the effective two RH neutrino case
obtained in the limit $M_3/10^{14}\,{\rm GeV} \gg 1$ and corresponding
to an orthogonal matrix with $\o_{21}=\o_{32}=1$. In the
first case, one can see how the bounds do not change compared to the
general case, showing that this is the choice saturating
the bounds, a well-known result~\cite{cmb}. Notice that for $\O=R_{13}$
there is no dependence  of the bounds on the light neutrino mass scheme, normal or inverted,
and therefore, in vanilla leptogenesis, normal or inverted schemes produce
the same bounds~\cite{cmb}. On the other hand, in the second case
the lower bound on $M_1$ becomes more stringent \cite{turzynski},
especially in the case of inverted hierarchy.
Indeed, in the vanilla case, for the same choice of
the orthogonal matrix, the final asymmetry in an inverted scheme
can be only less than in a normal scheme, or at most equal
in the special case $\O=R_{13}$, as discussed analytically in \cite{geometry}.
This result is easy to understand qualitatively:
the dominant term in the $C\!P$ asymmetry is suppressed
when the neutrino masses increase, either when $m_1$ increases,
or switching from normal to inverted hierarchy since in this case $m_2$ gets higher.

In the next sections we will relax the assumptions of vanilla leptogenesis,
studying how the leptogenesis bounds change accordingly.
In most cases the effect of the assumptions on the bounds is
independent of each other and therefore they can be studied individually.
However, in a few cases the interplay of different effects can yield interesting
interferences. For example,
going beyond the hierarchical limit one has also to consider the
contribution of the heavier RH neutrinos to the final asymmetry and
to the washout.

\section{Extra-term in the total \boldmath{$C\!P$} asymmetry}

In this Section we relax the assumption $M_2=M_3$ defining
vanilla leptogenesis, studying the effect on the bounds of
the term proportional to $\Delta\ve_1 (M_1,m_1,\O)$ in the Eq.~(\ref{master}),
and comparing our results with those obtained in~\cite{hambye}.
This effect clearly saturates to a maximum when $M_3/M_2 \gg 1$
and in the plots we fixed $M_3/M_2= 100$.

From the Eq.~(\ref{master}) it can be noticed that this effect acts
independently of flavor effects and can be even dominant when the
parameters are properly tuned. In particular
this extra-term is able, although with quite a strong fine-tuning,
to relax the lower bound on $M_1$.
A separate analysis is fully justified
since there is no interference between the two effects.
\begin{figure}
\hspace*{-5mm}
\psfig{file=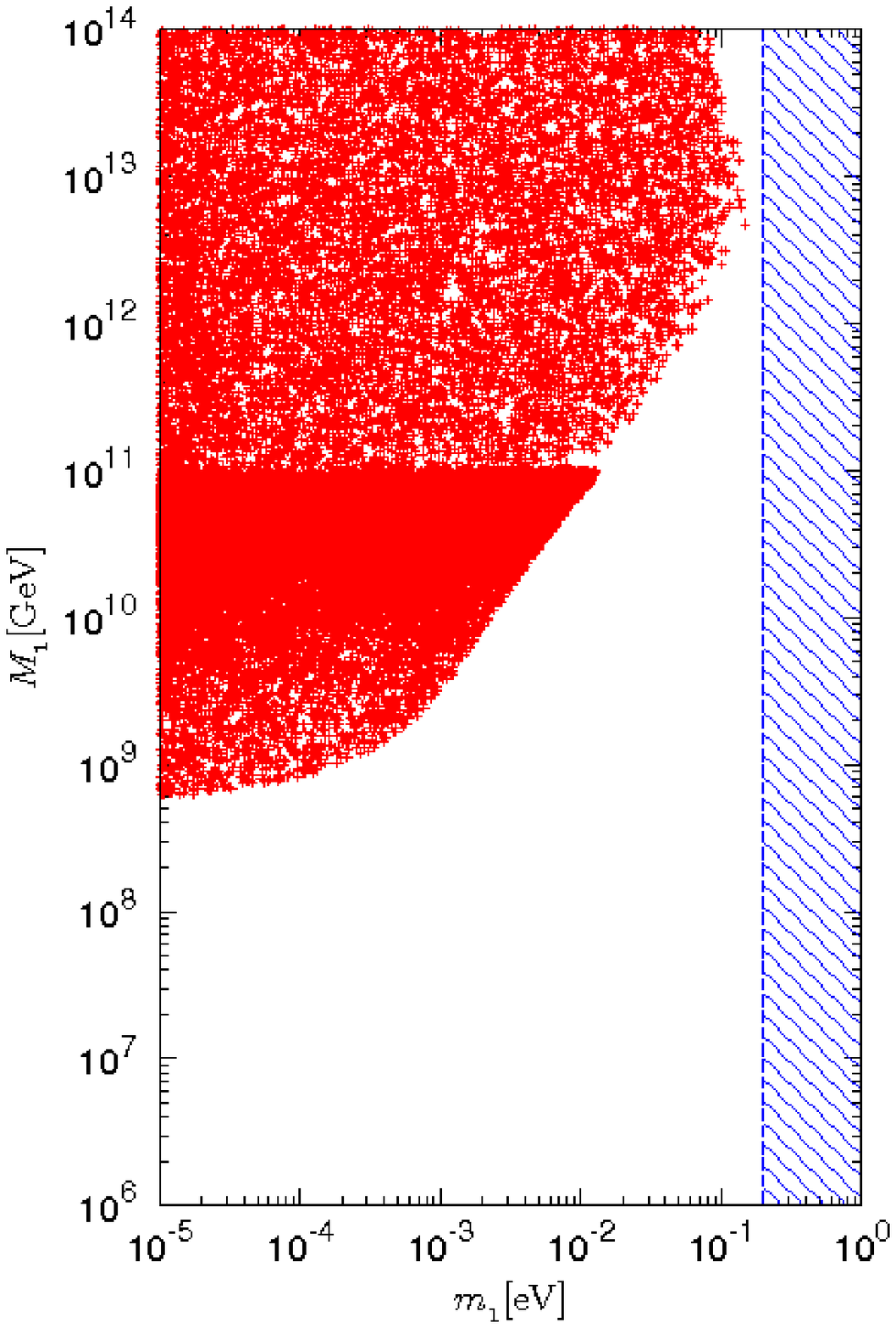,height=7cm,width=53mm}
\hspace{-1mm}
\psfig{file=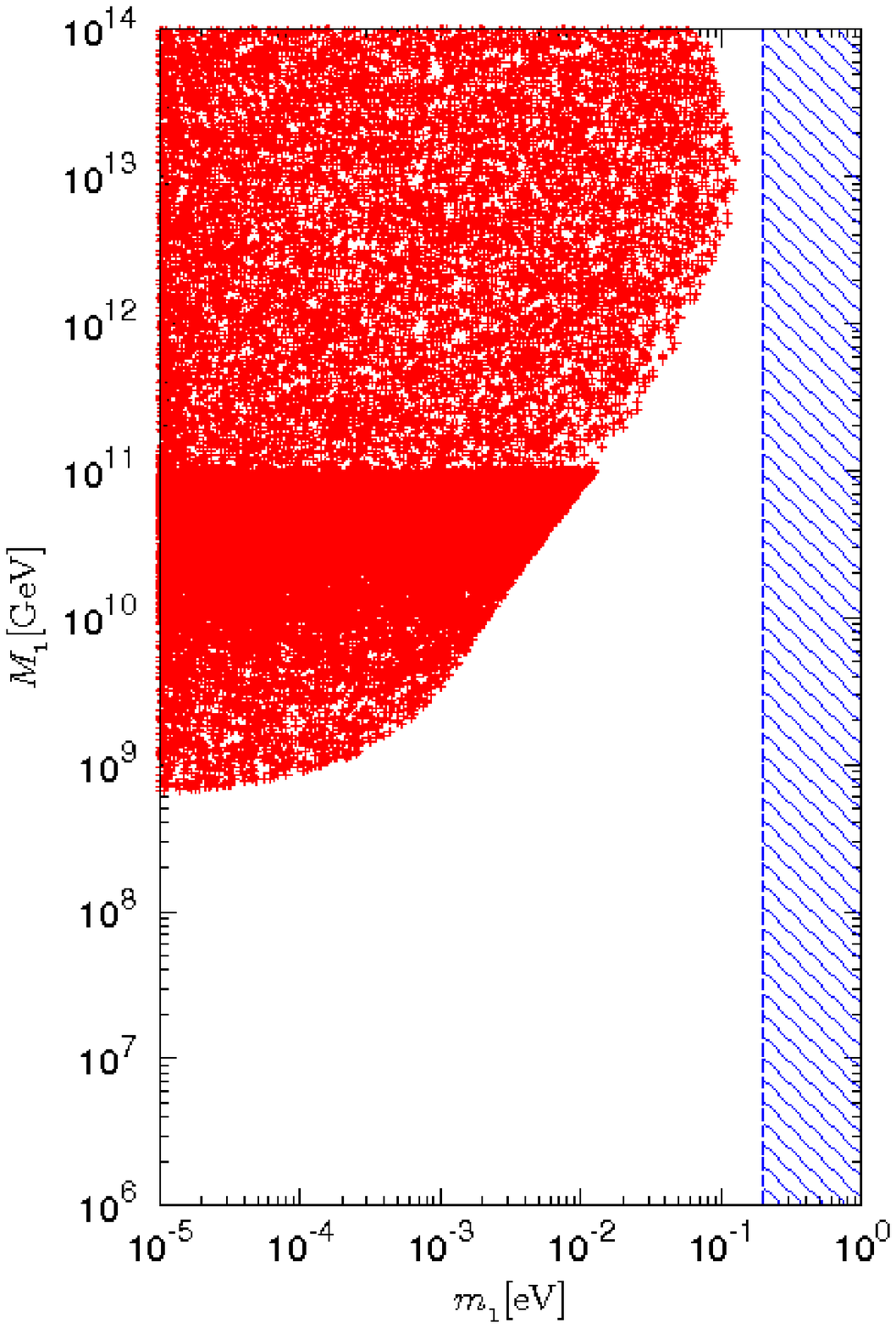,height=7cm,width=53mm}
\hspace{-1mm}
\psfig{file=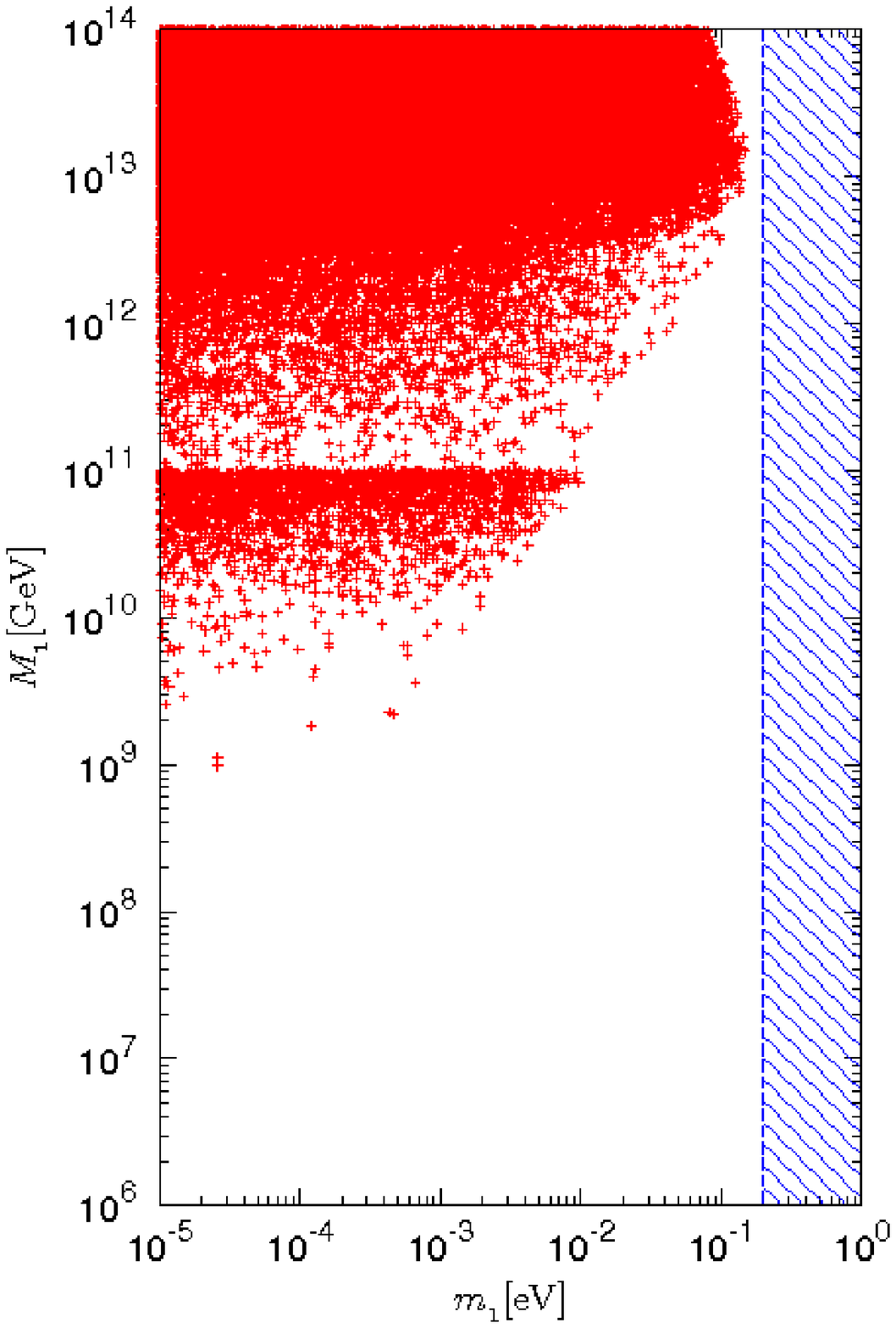,height=7cm,width=53mm} \\
\hspace*{-5mm}
\psfig{file=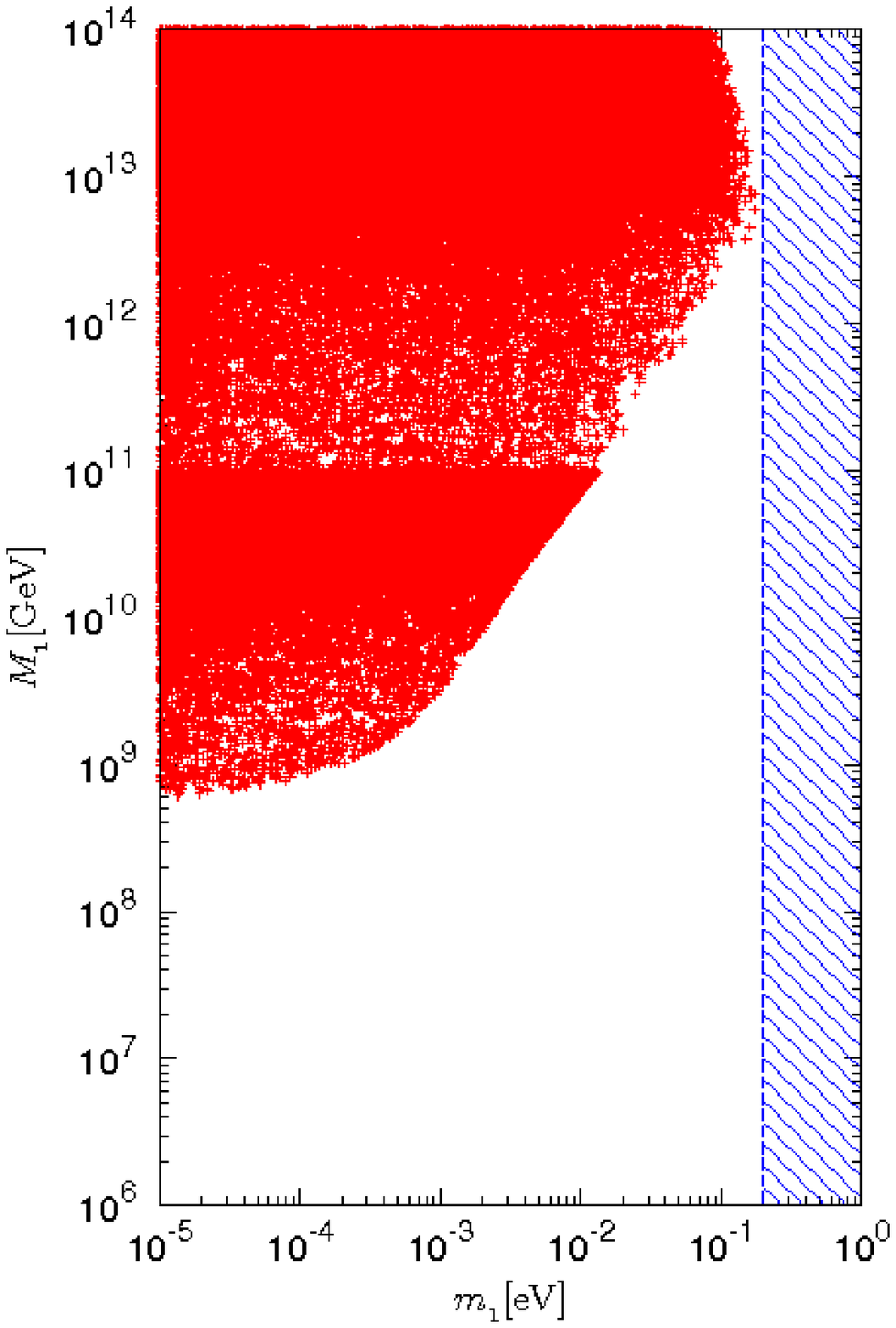,height=7cm,width=53mm}
\hspace{-1mm}
\psfig{file=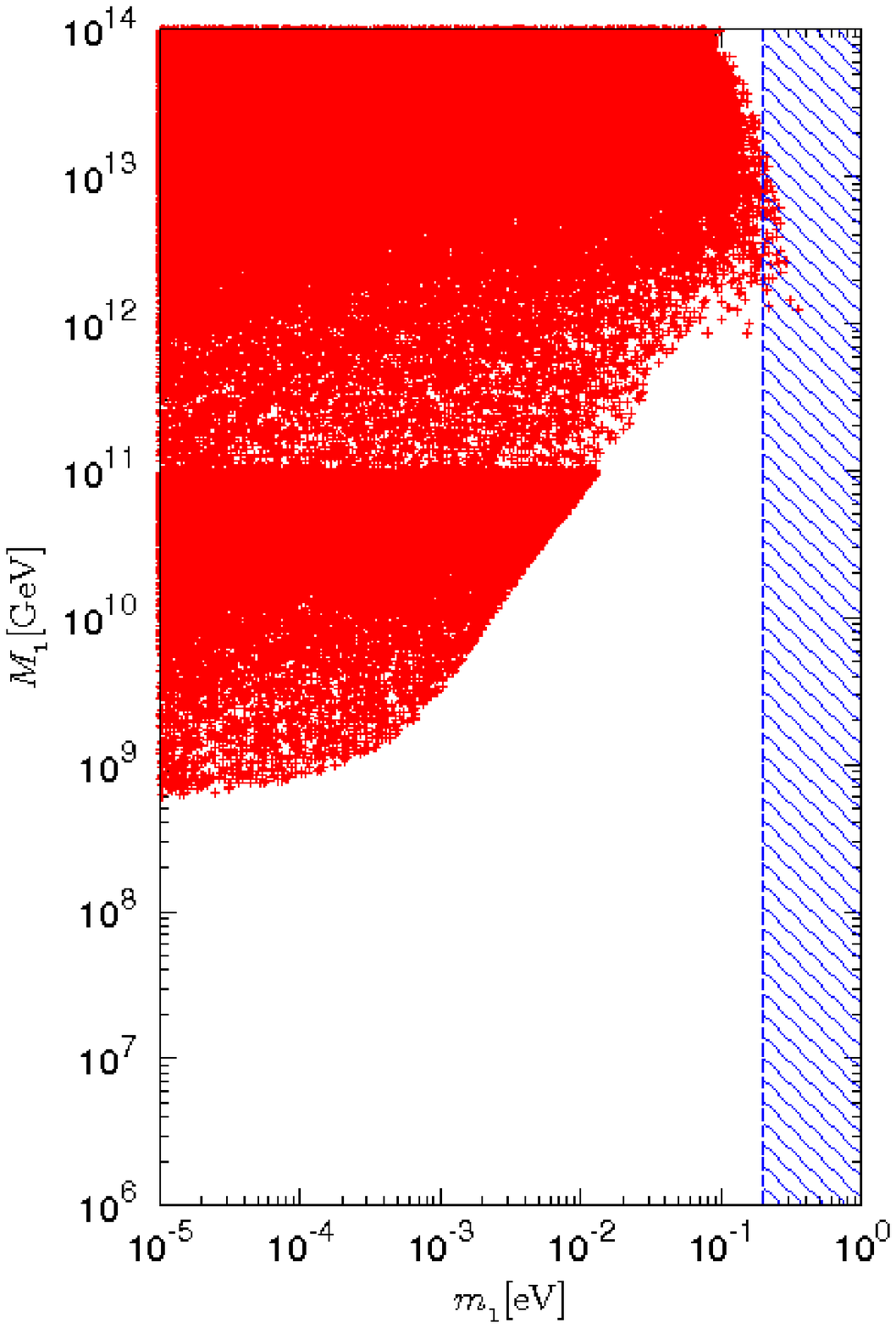,height=7cm,width=53mm}
\hspace{-1mm}
\psfig{file=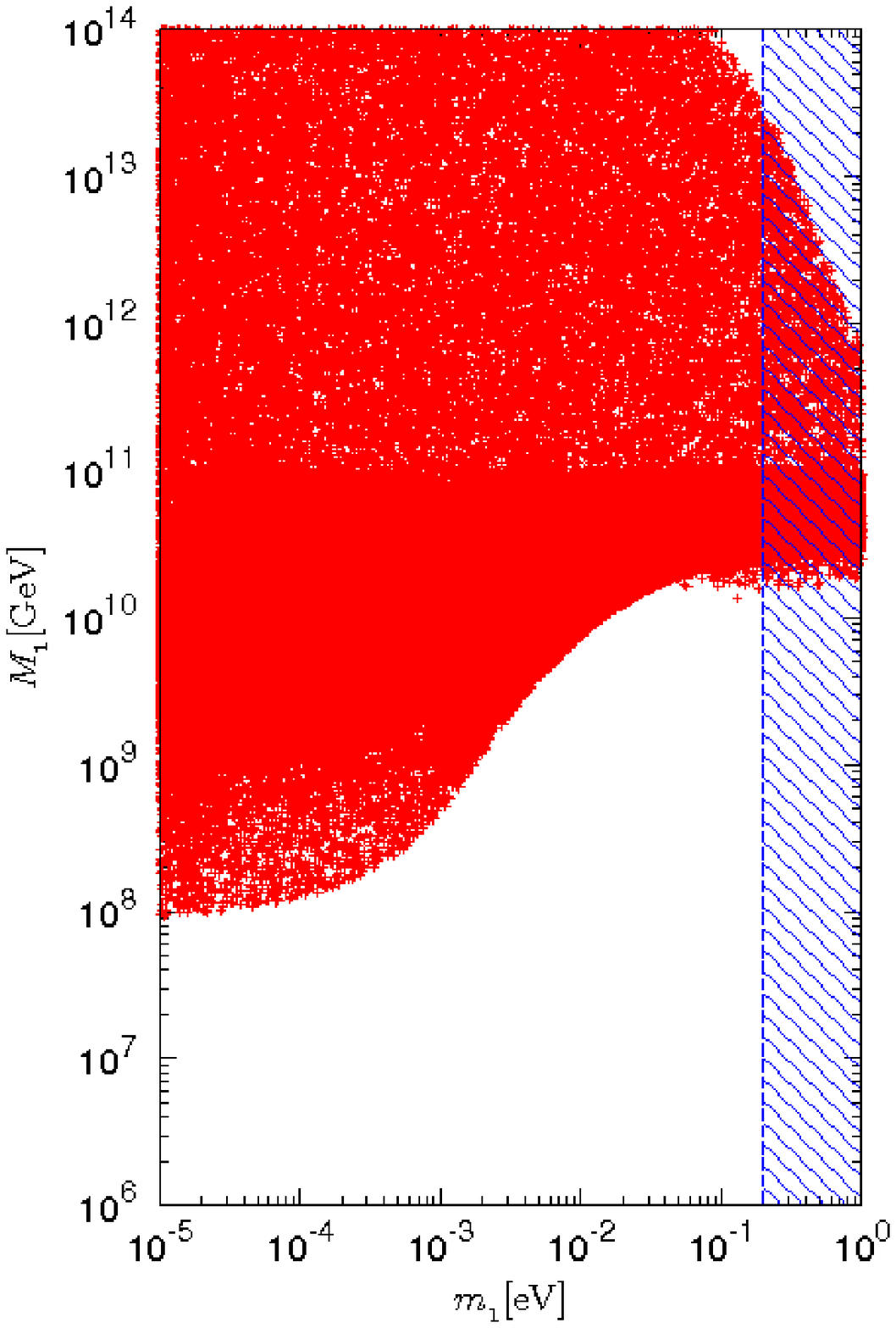,height=7cm,width=53mm}
\caption{Effect of the extra term [cf. Eqs. (\ref{extra}) and (\ref{master})]
in $\ve_1$ on the neutrino mass bounds. In all panels $M_3=100\,M_2$
and $M_2=3\,M_1$, except in the top center panel where $M_2=10\,M_1$.
Top left and top center panel: all three $|\o_{ij}|\leq 1$;
top right panel: $|\o_{32}|=0$ while $|\o_{21}|, |\o_{31}|\leq 10$;
bottom left panel: $|\o_{21}|, |\o_{32}|\leq 1$, $|\o_{31}|\leq 10$;
bottom center panel: $|\o_{31}|, |\o_{32}|\leq 1$, $|\o_{21}|\leq 10$;
bottom right panel: $|\o_{21}|, |\o_{31}|\leq 1$, $|\o_{32}|\leq 10$.}
\label{fig:extraterm}
\end{figure}
In the 6 panels of Fig.~\ref{fig:extraterm}, we show the crucial
role played by the parameter $\o_{32}$ \cite{geometry} in enhancing
the extra-term in the total $C\!P$ asymmetry (cf.~Eq.~(\ref{extra})).
This relaxes the mass bounds in a remarkable way if $|\o_{32}|\gg 1$ since
the $C\!P$ asymmetry enhancement is not counterbalanced by an increase of
the washout that is driven by $K_1$ and that is independent of $\o_{32}$.
It can be seen in the top right panel that when $\o_{32}=0$
the bounds are almost unchanged, even though we allowed
$|\o_{21}|, |\o_{31}|\leq  10$.

Notice that $\Delta\ve_1$ increases with the neutrino
masses, contrarily to $\bar{\ve}_1$.
This is the reason why it tends to relax the upper bound on $m_1$.
It also tends to be higher for inverted schemes compared
to normal schemes, even though the bounds are saturated for a choice of the
parameters  where there is no dependence on $m_2$ so that
inverted and normal schemes give the same results.

The results of the panels can be easily understood analytically
using Eqs.~(\ref{extra}) and~(\ref{master}). For example,
since $\xi(x)\simeq 1+5/(9x)$ when $x\gg 1$, the extra-term
is suppressed like $(M_1/M_2)^2$. It should be also noticed that the
extra-term vanishes exactly in the limit of two effective RH neutrinos,
obtained for $M_3/10^{14}\,{\rm GeV}\rightarrow\infty$.

In conclusion, the possibility to exploit the extra-term
$\propto\Delta\varepsilon_1(M_1,m_1,\O)$
to relax the lower bound on $M_1$ relies on models where
$|\o_{32}| \gtrsim 0.2\, (M_2/M_1)^2$.
Therefore, already for $M_2\gtrsim 3\,M_1$,
quite a high level of fine tuning is required.
Concerning the upper bound on $m_1$ the conditions are less stringent:
$|\o_{32}| \gtrsim 0.2\,(M_2/M_1)^2$ and/or
$|\o_{21}| \gtrsim \,(M_2/M_1)^2$, such that
one has to impose, more conservatively,
$M_2\gtrsim 10\,M_1$. We will see in the next Section
that flavor effects have a bigger impact in relaxing
the bounds compared to the vanilla scenario.

\section{Adding flavor to vanilla}
Relaxing only the second assumption defining vanilla leptogenesis, one has
\bea \label{Nf}
N_{B-L}^{\rm f} & = & \sum_{\a}\,\ve_{1\a}\,\kappa_{1\a}^{\rm f}
 \\ \nonumber
& \simeq &
N_{\rm fl}\,\bar{\ve}_1(M_1,m_1,\o_{21},\o_{31})\,\k_1^{\rm f}(M_1,m_1,K_1)
+{1\over 2}\,\Delta P_{1\alpha}(M_1,m_1,\O,U)\,
[\kappa_{1\a}^{\rm f}-\k_{1\b}^{\rm f}] \, .
\eea
From Eq. (\ref{eps1a}), in the HL, one finds \cite{flavorlep}
$\ve_{1\a}=\overline{\ve}_{1\a}+\D\ve_{1\alpha}$,
where
\be\label{bareps1a}
\overline{\ve}_{1\a}\equiv {3\over 16\,\pi\,(h^{\dagger}\,h)_{11}}
\,\sum_{j\neq 1}\,{1\over \,\sqrt{x_j}}\,
{\rm Im}\left[h^{\star}_{\a 1}\,h_{\a j}\,(h^{\dagger}h)_{1j}\right]
\ee
and
\be\label{Dveps1a}
\D\ve_{1\a} \equiv {1\over 8\,\pi\,(h^{\dagger}\,h)_{11}}
\,\sum_{j\neq 1}\,{1\over x_j}\,{\rm Im}
\left[h^{\star}_{\a 1}\,h_{\a j}\,(h^{\dagger}h)_{j1}\right] \, .
\ee
Taking advantage of the orthogonal parametrization (cf. Eq.~(\ref{h})) and defining
$r_{1\a}\equiv {\ve_{1\a}/ \overline{\ve}(M_1)}=\overline{r}_{1\a}+\D r_{1\a}$,
one has
\be\label{e1alOm}
\overline{r}_{1\a}=-\,\sum_{h,l}\,
{m_l\,\sqrt{m_l\,m_h}\over \mt\,m_{\rm atm}}
\,{\rm Im}[U_{\a h}\,U_{\a l}^{\star}\,\O_{h1}\,\O_{l1}]
\ee
and
\be
\Delta r_{1\alpha} = {2\over 3}\,\sum_{j,h,l,k} \,
{M_1\over M_j}\,{m_h\sqrt{m_l\,m_k}\over \mt\,m_{\rm atm}}\,
{\rm Im}[U^{\star}_{\a l}\,U_{\a k}\,\O^{\star}_{hj}\O^{\star}_{l1}\,\O_{h1}\,\O_{kj}] \, .
\ee
The second term has been neglected in previous analyses but,
as we will see, it can dominate under some conditions relaxing
the leptogenesis bounds holding in the vanilla scenario.
The most important difference between the two terms is that the
first is upper bounded \cite{abada1},
\be\label{CPbound}
\overline{r}_{1\a} < \sqrt{P^0_{1\a}}\,m_3/m_{\rm atm} \, ,
\ee
while the second is not. From this point of view the
$\Delta\ve_{1\a}$ term is analogous to the extra term in the total $C\!P$
asymmetry but, as we will see, it affects the bounds in a more relevant way.

The analytical expressions for the $\kappa_{1\a}^{\rm f}$, generalizing
those for $\k_1^{\rm f}$, can be found in~\cite{flavorlep}.
The generalization of the expression~(\ref{M1unfl}) becomes
\be\label{M1}
M_1={\overline{M}\over
\sum_{\a}\,\kappa_{1\a}^{\rm f}(M_1,m_1,\O,U)\,r_{1\a}(m_1,\O,U)} \, .
\ee
In the fully flavored regime this gives rise to a lower
bound on $M_1$ that always falls in the two-flavor regime
with negligible washout from $\D L=2$ processes. Therefore,
this can be expressed like
\be
M_1={\overline{M}\over
r_{1\t}\,\kappa_{1\t}^{\rm f}+r_{1,e+\m}\,\kappa_{1,e+\m}^{\rm f}}
\simeq \,
{\overline{M}\over N_{\rm fl}\,\k_1^{\rm f}(K_1)+{1\over 2}\,
[\D P_{1\t}/\bar{\ve}(M_1)]\,[\k_{1\t}^{\rm f}-\k_{1,e+\mu}^{\rm f}]} \, .
\ee
From the approximate expression, one can see once more that the lower bound,
can be relaxed compared to the unflavored case only if there is some
washout, otherwise $N_{\rm fl}=1$ and $\k_{1\t}^{\rm f}-\k_{1,e+\mu}^{\rm f}=0$.
This implies $K_1\gtrsim 1$. In the limit of no washout, for $K_1\ll 1$,
one recovers the usual lower bound $M_1>\overline{M}$ in the
case of initial thermal abundance. In the strong washout
a big relaxation is possible only in the one-flavor dominance case,
where one of the projectors $P_{1\a}^0\ll P_{1\b}^0$,
otherwise close to the democratic case, where $P_{1\a}^0\simeq P_{1\b}^0$,
the difference $[\k_{1\t}^{\rm f}-\k_{1,e+\mu}^{\rm f}]$
tends to suppress the final asymmetry.

\subsection{Lower bound on $M_1$}

In Fig.~\ref{lboundfl} we show the allowed region in the $(K_1,M_1)$ plane for $m_1=0$.
\begin{figure}
\hspace*{-5mm}
\psfig{file=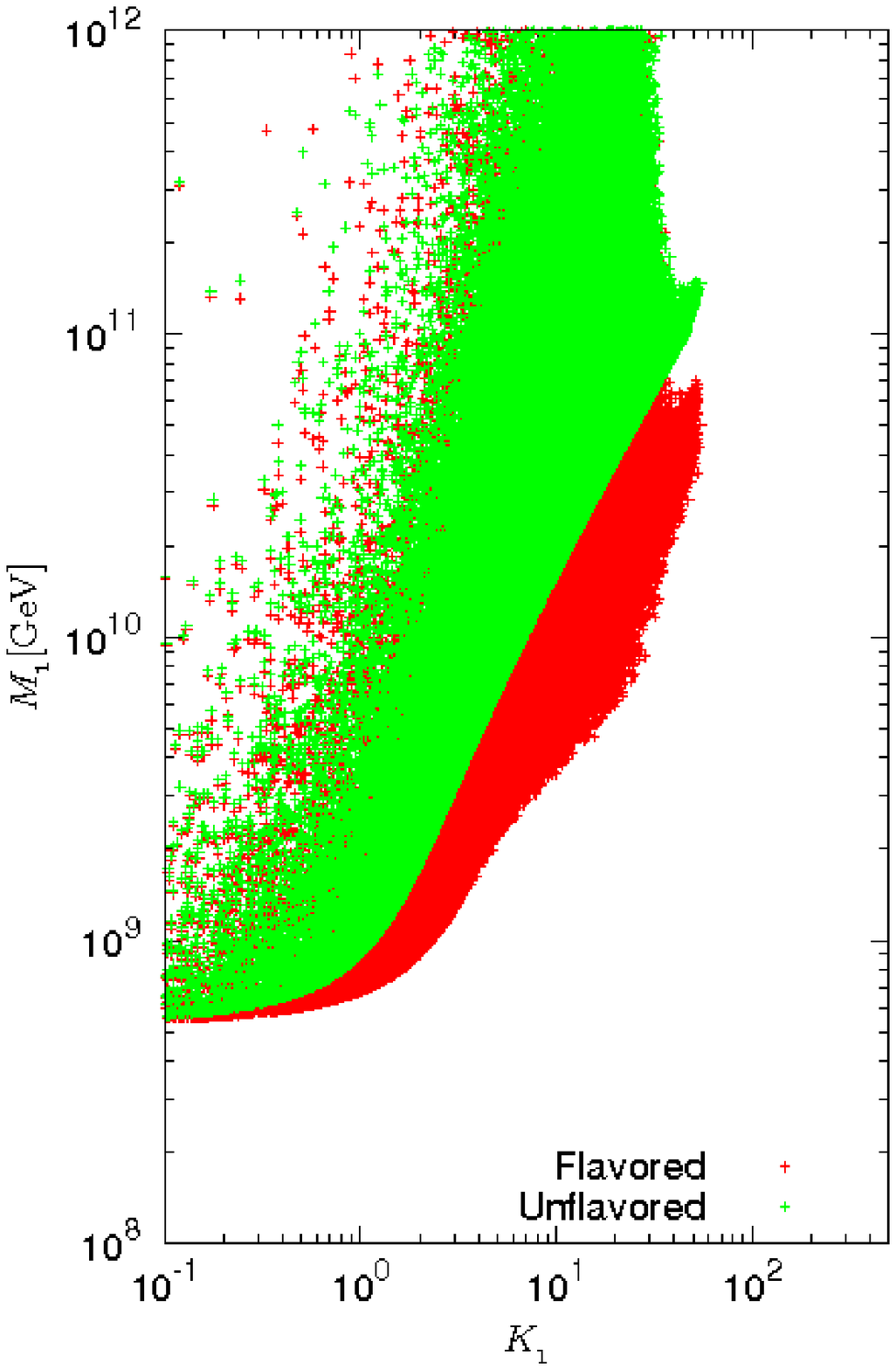,height=7cm,width=53mm}
\hspace{-1mm}
\psfig{file=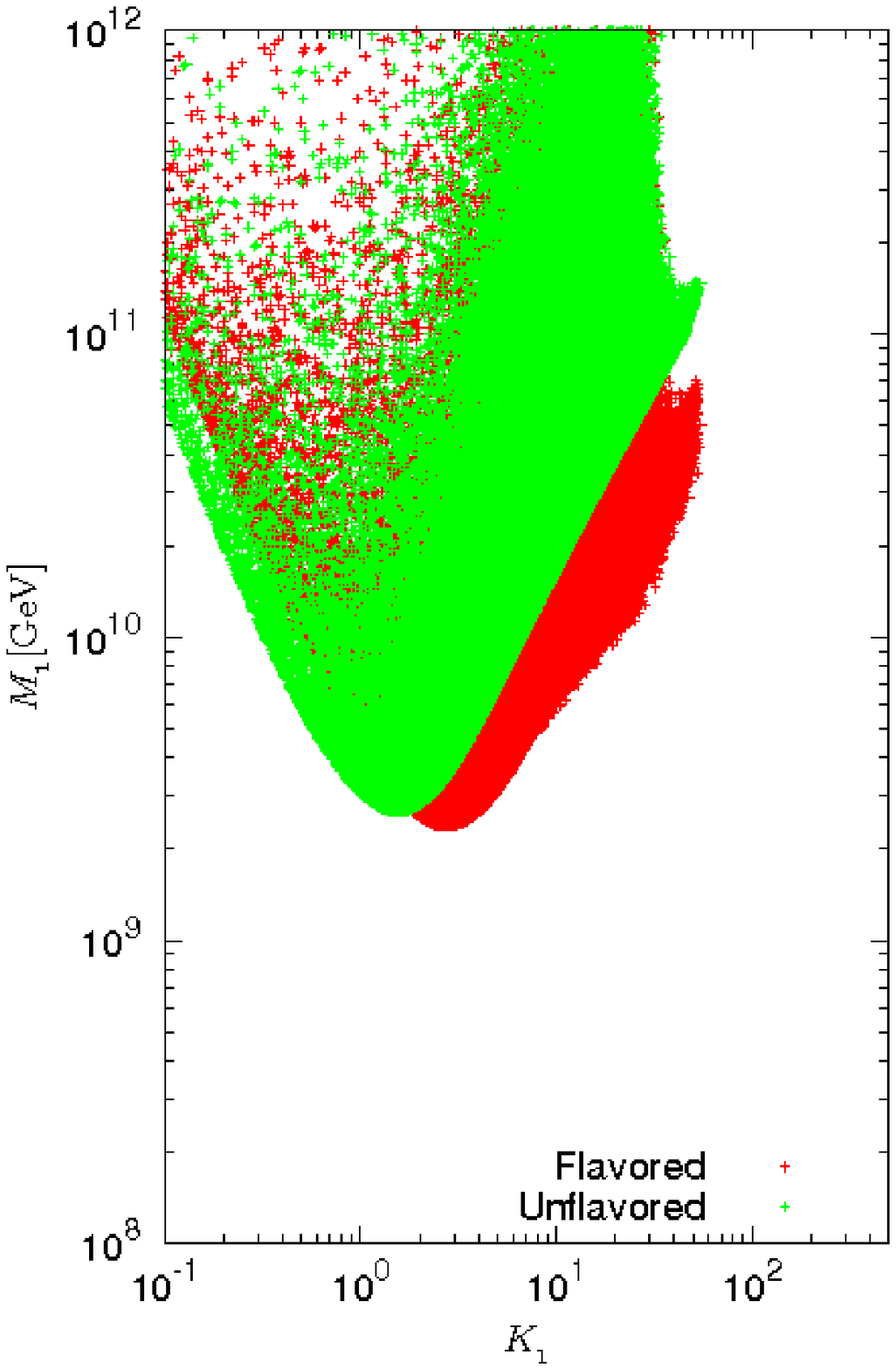,height=7cm,width=53mm}
\hspace{-1mm}
\psfig{file=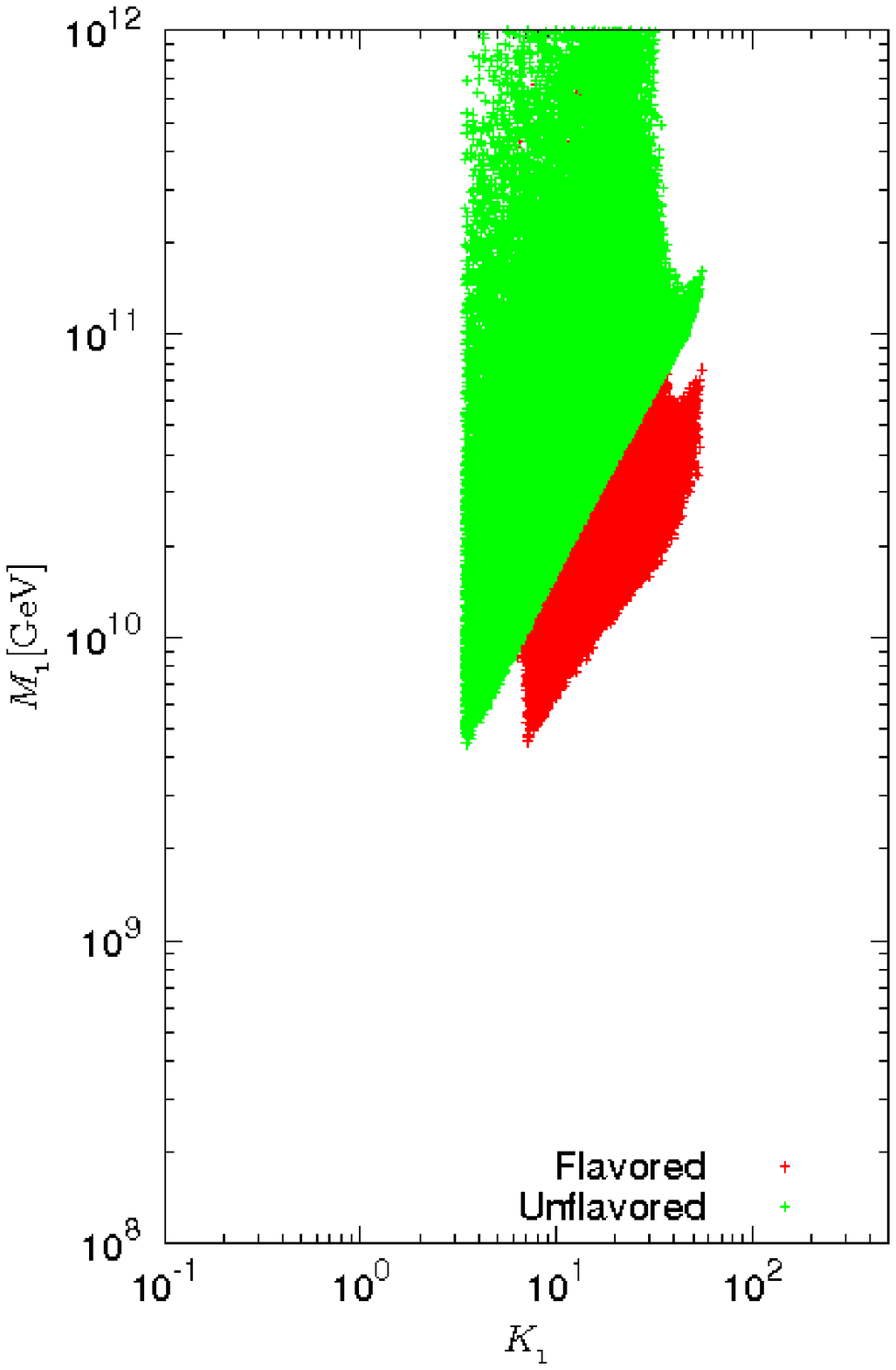,height=7cm,width=53mm}
\\
\psfig{file=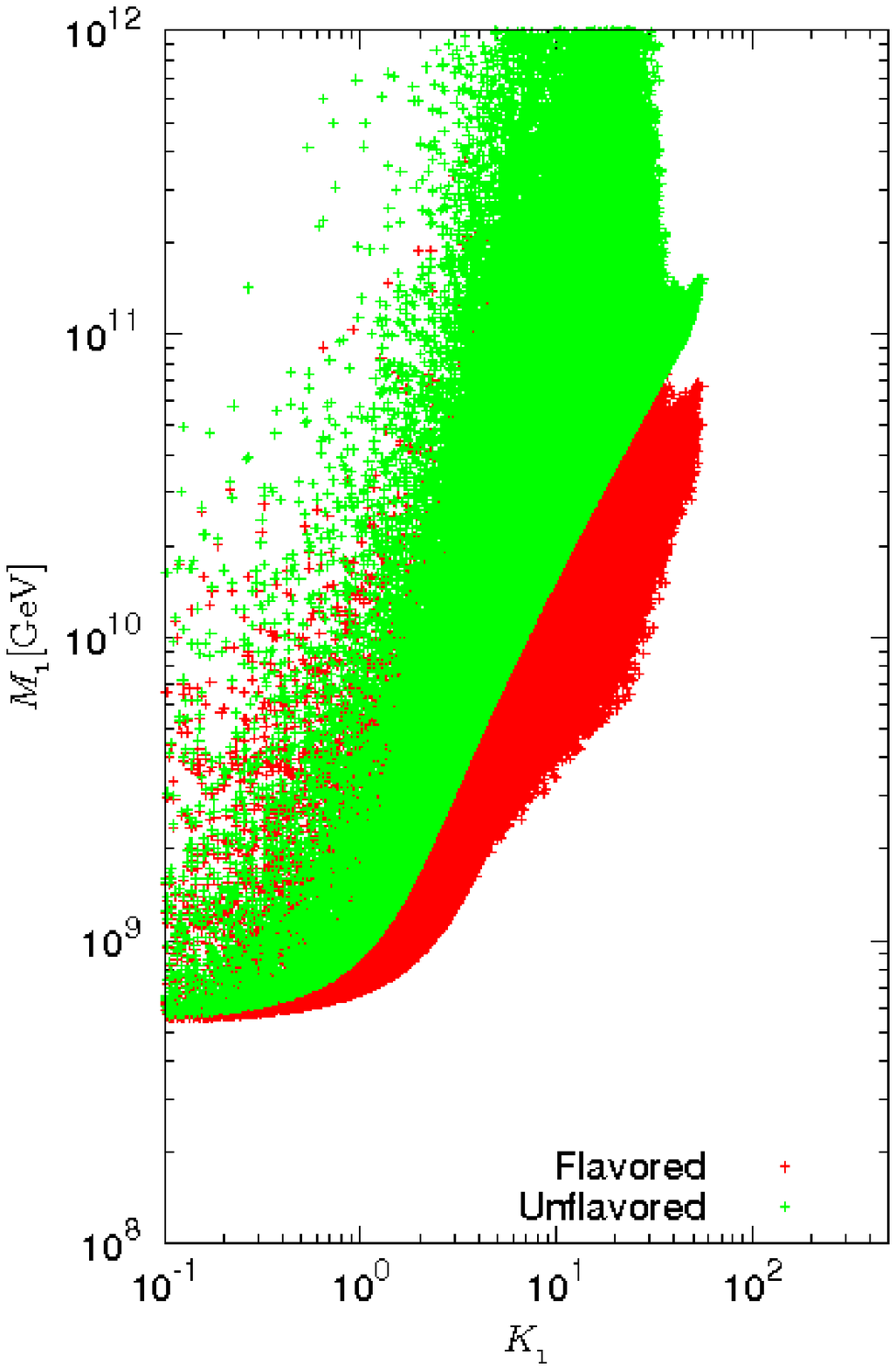,height=7cm,width=52mm}
\hspace{-1mm}
\psfig{file=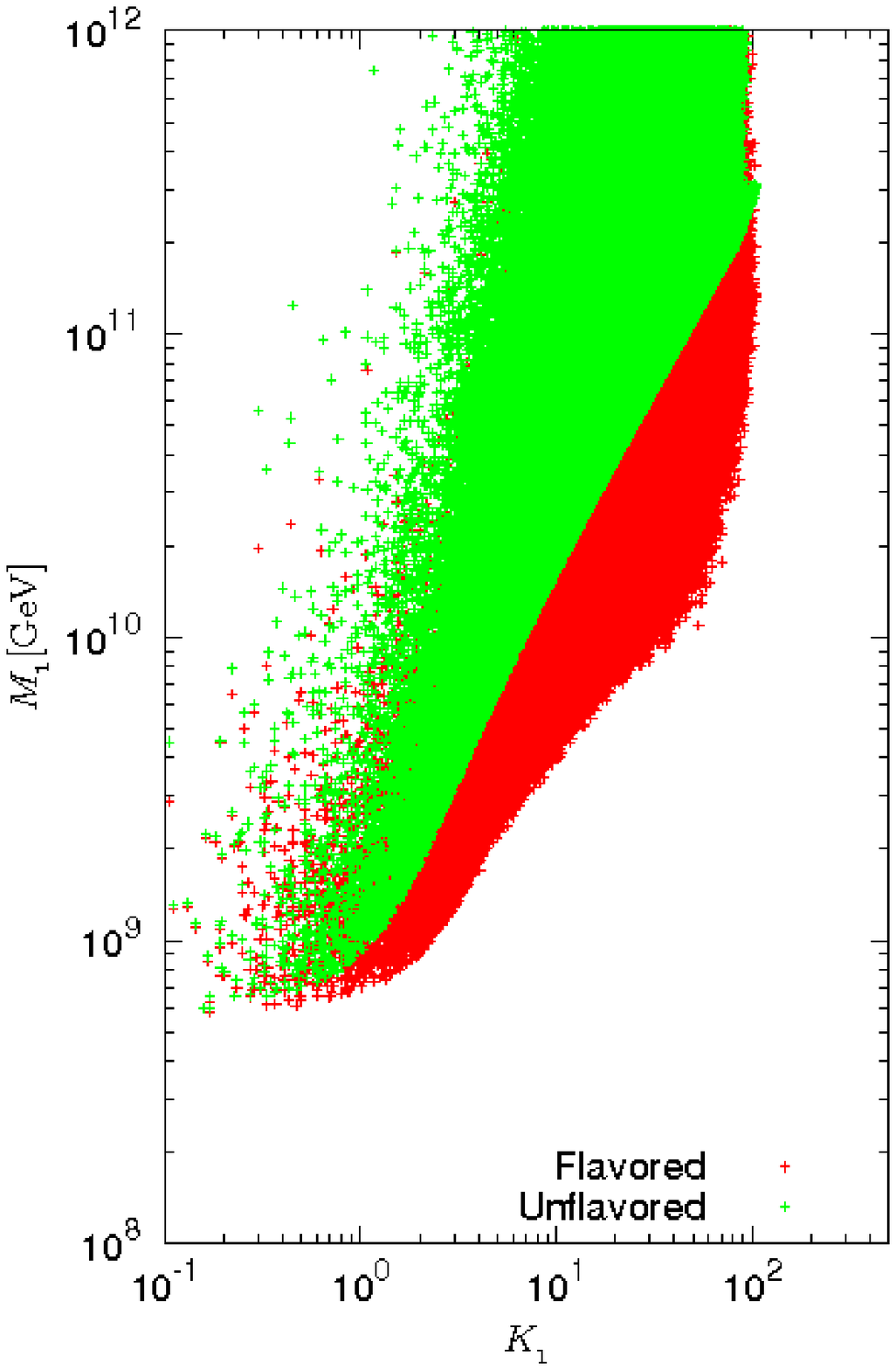,height=7cm,width=52mm}
\hspace{-1mm}
\psfig{file=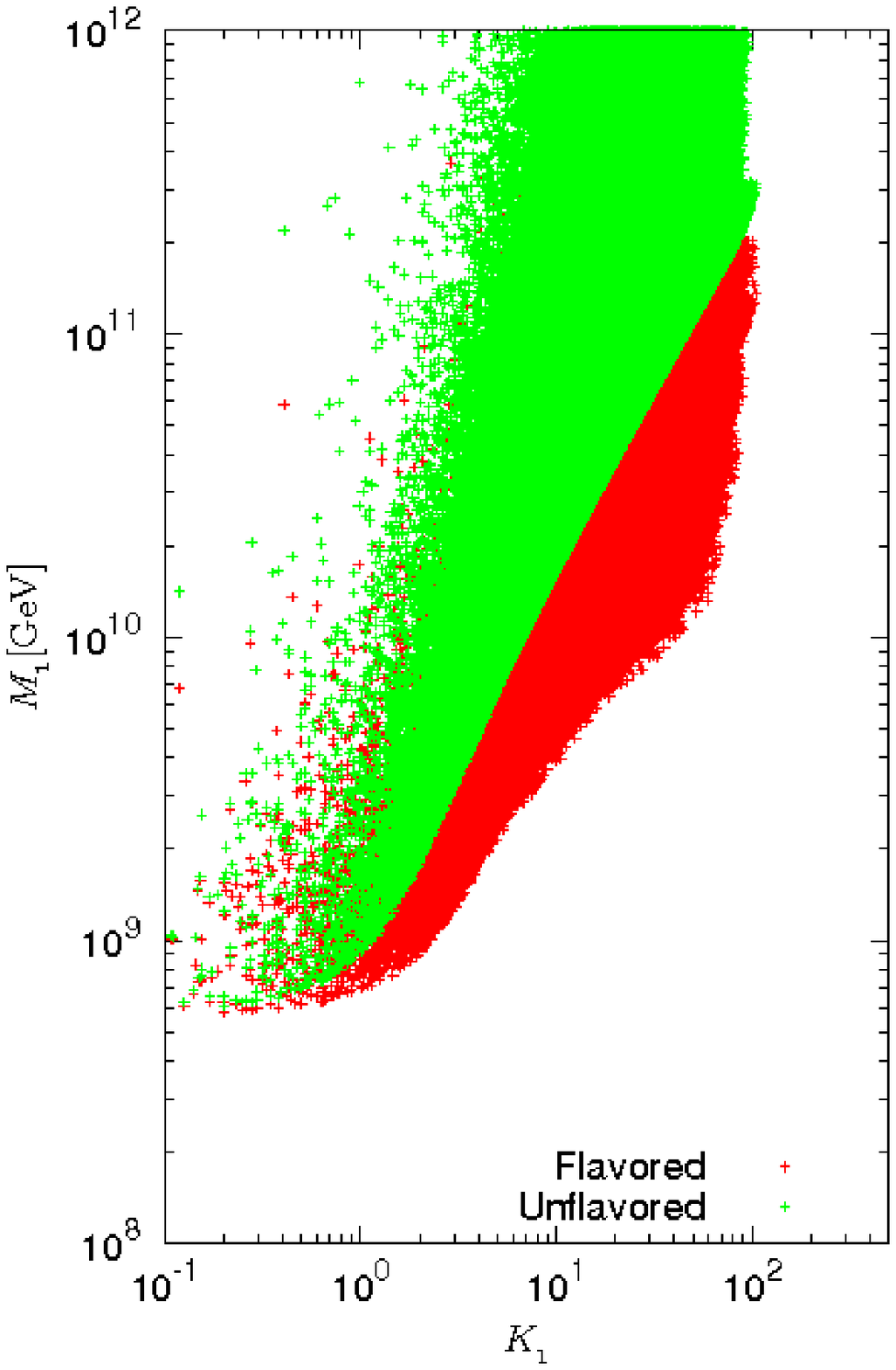,height=7cm,width=52mm}
\caption{Lower bound on $M_1$ versus $K_1$ for $m_1=0$ and imposing
$|\o_{ij}|<1$. We show the results  obtained in the fully flavored regime
(red points) comparing them with those obtained when they are neglected
(green points).  Top-left panel:
thermal initial $N_1$-abundance ($N_{N_1}^{\rm in}=1$). Top-center panel:
vanishing initial $N_1$-abundance ($N_{N_1}^{\rm in}=0$). Top-right panel:
points falling in the strong washout regime where the final asymmetry
depends on the initial $N_1$-abundance to a level less than $10\%$.
All bottom panels assume a thermal initial $N_1$-abundance. Bottom-left
panel: like top-left but removing all points that violate the condition~(\ref{condition}).
Bottom-center and right panel: inverted hierarchy keeping or removing
points violating the condition~(\ref{condition}) respectively.}
\label{lboundfl}
\end{figure}
The plots are obtained scanning  the seven free parameters,
$M_1$ and the 6 parameters in $\O$,
showing only the points where $\eta_B\geq\eta_B^{\rm CMB}$ at $3\,\s$.
This is equivalent  to search for the points where $M_1$ is larger
than the right-hand side of the Eq.~(\ref{M1}) with
$\overline{M}\simeq 5.7\times 10^8\,{\rm GeV}$.
In this subsection we are imposing $|\o_{ij}|\leq 1$,
implying an upper bound on $K_1$. We will study in Section~\ref{sec:largeom}
the effects of turning on large values of $|\o_{ij}|$.

In the plots the red region is the additional part of the allowed region
due to flavor effects within the fully flavored regime while the green
region is what one obtains within vanilla leptogenesis neglecting flavor
effects . In the top-left
panel the final asymmetry has been calculated for an initial
thermal $N_1$-abundance, while in the top-middle panel for an initial
vanishing $N_1$-abundance.
One can see again that flavor effects can relax the lower bound only in
the presence of washout, that means when $K_1\gtrsim 1$ and the amount of
the relaxation increases with $K_1$.
Essentially the lower bound we find coincides
with the lower bound found analytically in \cite{flavorlep}
that corresponds to neglect $\Delta r_{1\a}$ and
maximizing $\overline{r}_{1\a}$ and $\k_{1\a}^{\rm f}$ in the
Eq.~(\ref{M1}) in the case of one flavor dominance. In \cite{flavorlep} the lower bound
was numerically calculated only for a particular case, $\O=R_{13}$, and just a
small relaxation was found for vanishing $m_1$. Here, allowing for
$\o_{21}\neq 0$, we find a large relaxation also for vanishing $m_1$.
In the top-right panel, we selected only the points for which
there is independence of the initial conditions,
more exactly those for which there is a difference to a level
less than $10\%$ between thermal and vanishing initial $N_1$-abundance.
This generalizes the definition of strong washout regime
when flavor effects are taken into account.
The critical value of $K_1$ increases from $\sim 3$ in the
unflavored case to $\sim 7$ in the flavored case.

In the bottom left panel we finally show only the points that
further satisfy the condition of validity of the classic kinetic
equations in the fully flavored regime (cf. Eq.~(\ref{condition})), and one can
see how there are no  differences
between the unflavored and the fully flavored regime
since these arise for larger $K_1$ values.

In the bottom center and right panels we show the results for inverted hierarchy
without imposing the condition of validity and imposing it, respectively. One can
see that the situation is not very different from the normal hierarchy case. Slightly larger
values of $K_1$ are allowed, up to about $2\times K_{\rm atm}\simeq 93$. This implies
that the relaxation of the lower bound on $M_1$ can be as large as one order of magnitude for the maximal
allowed value of $K_1$. Notice that these points still satisfy the condition of
validity~(\ref{condition}).

In Fig.~\ref{fig:speciallbM1} we consider two special cases
for the orthogonal matrix when $m_1=0$. In the left panel $\O=R_{13}$
(i.e. $\o_{21}=\o_{32}=0$). In this case it is easy to show \cite{flavorlep} that
$\Delta P_{1\a}=0$ and therefore in the Eq. (\ref{Nf}) only the
first term survives and all flavor effects reduce to an enhancement
of the final asymmetry compared to vanilla leptogenesis given
just by $N_{\rm f}\sim 2$. In the right panel we consider the
case of very large $M_3\gg 10^{14}\,{\rm GeV}$ to be compared with
the right panel of Fig.~2 in the unflavored case. One can see that
this time inverted hierarchy (green and purple areas)
is not so suppressed compared to normal hierarchy (red and blue areas)
as in the unflavored case.
This is due to the presence of the PMNS phases
that give a further contribution to the asymmetry.
Indeed when phases are switched off (purple and blue areas)
again the allowed region for inverted hierarchy is strongly reduced
compared to normal hierarchy, similarly, though to a minor extent, to what
happened in the unflavored case.
In this specific case  $M_3\gg 10^{14}\,{\rm GeV}$,
the effect of phases has been recently studied in \cite{molinaro}.
The fact that PMNS phases can give a dominant contribution
to the final asymmetry was first noticed in the case
$\O=R_{13}$  and $m_1\neq 0$ in \cite{flavorlep}.
\begin{figure}
\begin{center}
\psfig{file=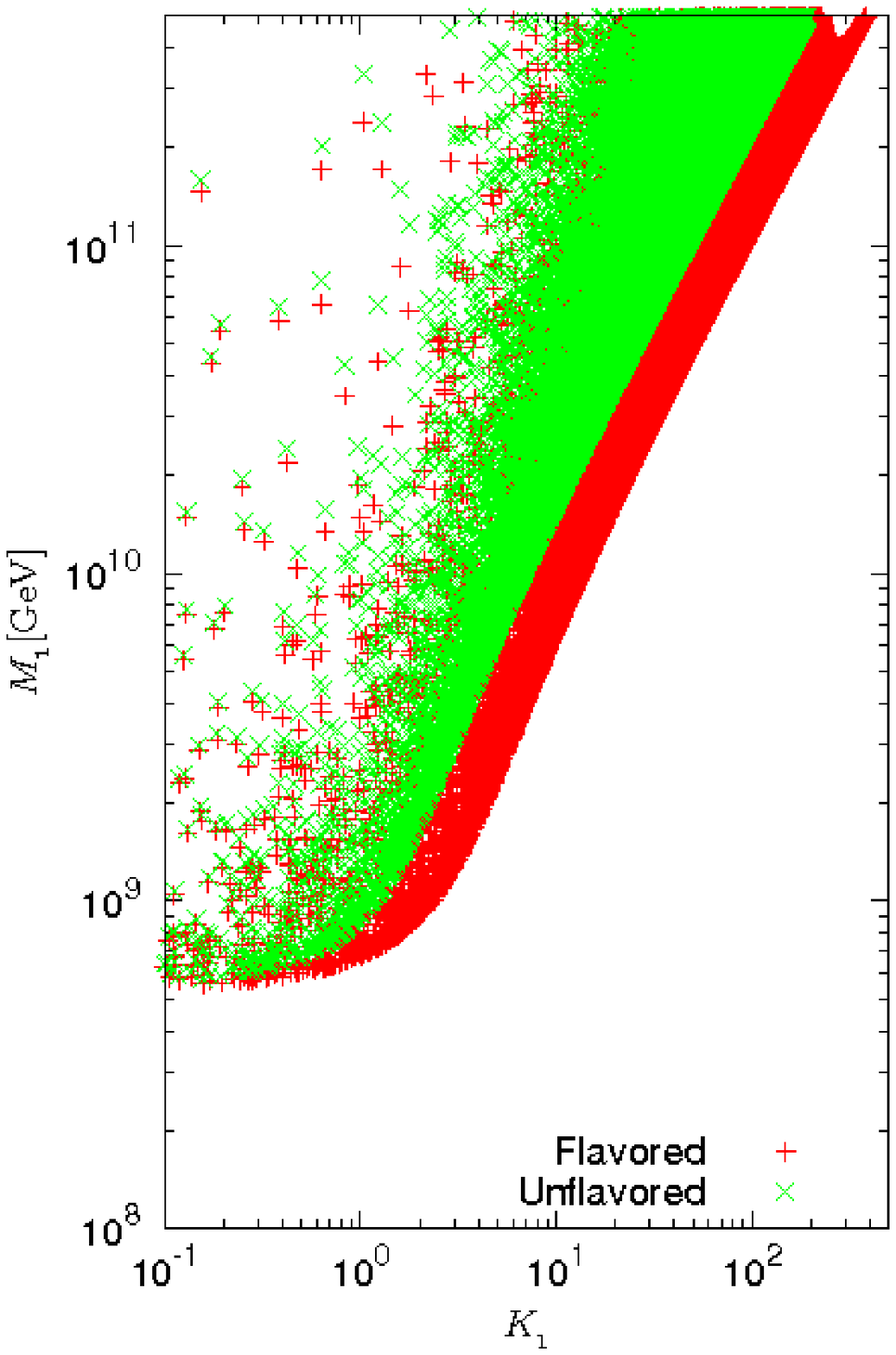,height=7cm,width=7cm}
\hspace{-1mm}
\psfig{file=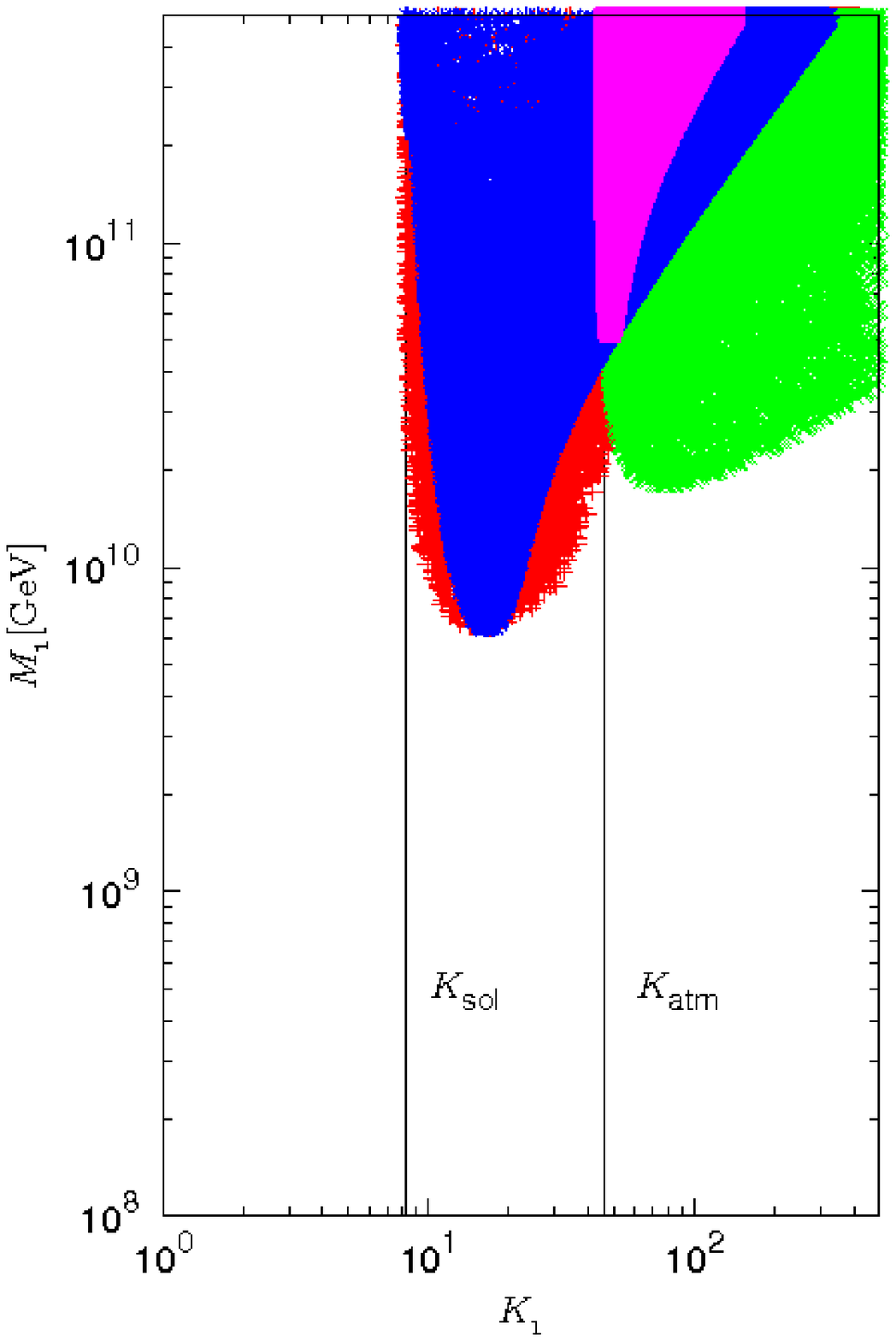,height=7cm,width=7cm}
\caption{Lower bound on $M_1$ versus $K_1$ for $\O=R_{13}$
(left) and for $M_3 \gg 10^{14}\,{\rm GeV}$ (right). In the right panel
the red and blue regions correspond to normal hierarchy and the purple
and the green to inverted hierarchy. The blue and purple regions are obtained
switching off the PMNS phases.}
\label{fig:speciallbM1}
\end{center}
\end{figure}

\subsection{Upper bound on $m_1$}

In this subsection we allow $m_1\neq 0$, investigating
how the upper bound on $m_1$ and its dependence on $M_1$
changes when flavor effects are included.
For this purpose, we show plots in the ($M_1,m_1$) plane.
As in the last subsection, we impose the condition $|\o_{ij}|\leq 1$.

In all figures we distinguish three different kinds of points:
those characterized by a strong one-flavor dominance
for which $P^0_{1\t}<0.1$ or $P^0_{1e}+P^0_{1\m}<0.1$ (red crosses),
those characterized by a mild one-flavor dominance for which
$0.1<P^0_{1\t}<0.45$ or $0.1<P^0_{1e}+P^0_{1\m}<0.45$ (green x) and
finally those for which a democratic scenario is realized
such that $0.45<P^0_{1\t},P^0_{1e}+P^0_{1\m}<0.55$ (blue stars).
This will make possible to understand under which circumstances the
upper bound on $m_1$ can be evaded when flavor effects
are taken into account.

In Fig.~\ref{fig:general} the results are shown both for
normal (left panel) and inverted (right panel) hierarchy
and have been obtained using the approximation $C=I$.
One can see that there is no upper bound on $m_1$,
as first pointed out in \cite{abada1}. The evasion of the bounds
occurs in a one-flavor dominance, as expected. Note that the results have
been obtained without imposing the condition of validity of the
fully flavored regime Eq.~(\ref{condition}).
In the case of inverted hierarchy (right panel) the bounds do not
change significantly.

\begin{figure}
\begin{center}
\includegraphics[width=0.33\textwidth,angle=-90]{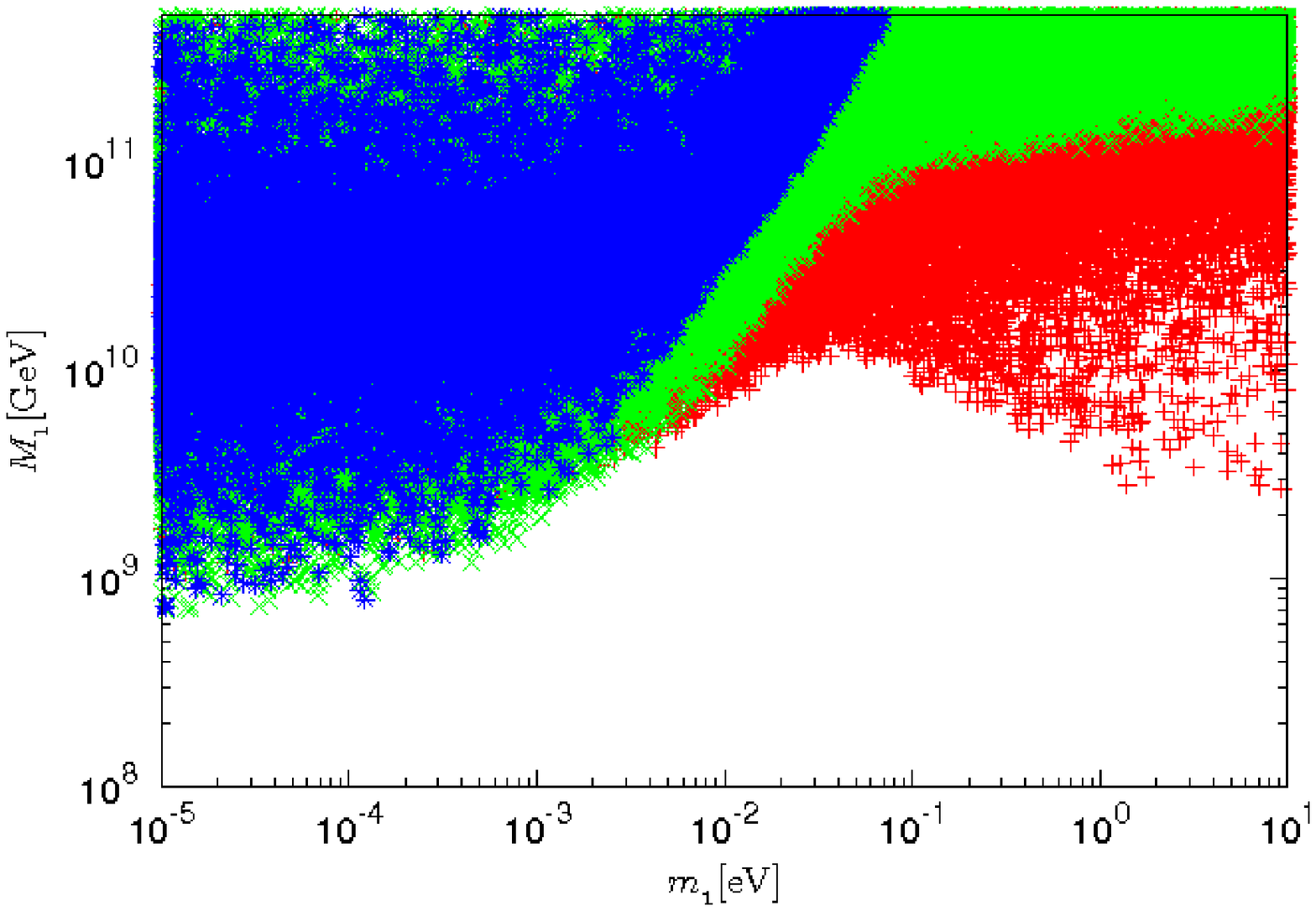}
\hspace{5mm}
\includegraphics[width=0.33\textwidth,angle=-90]{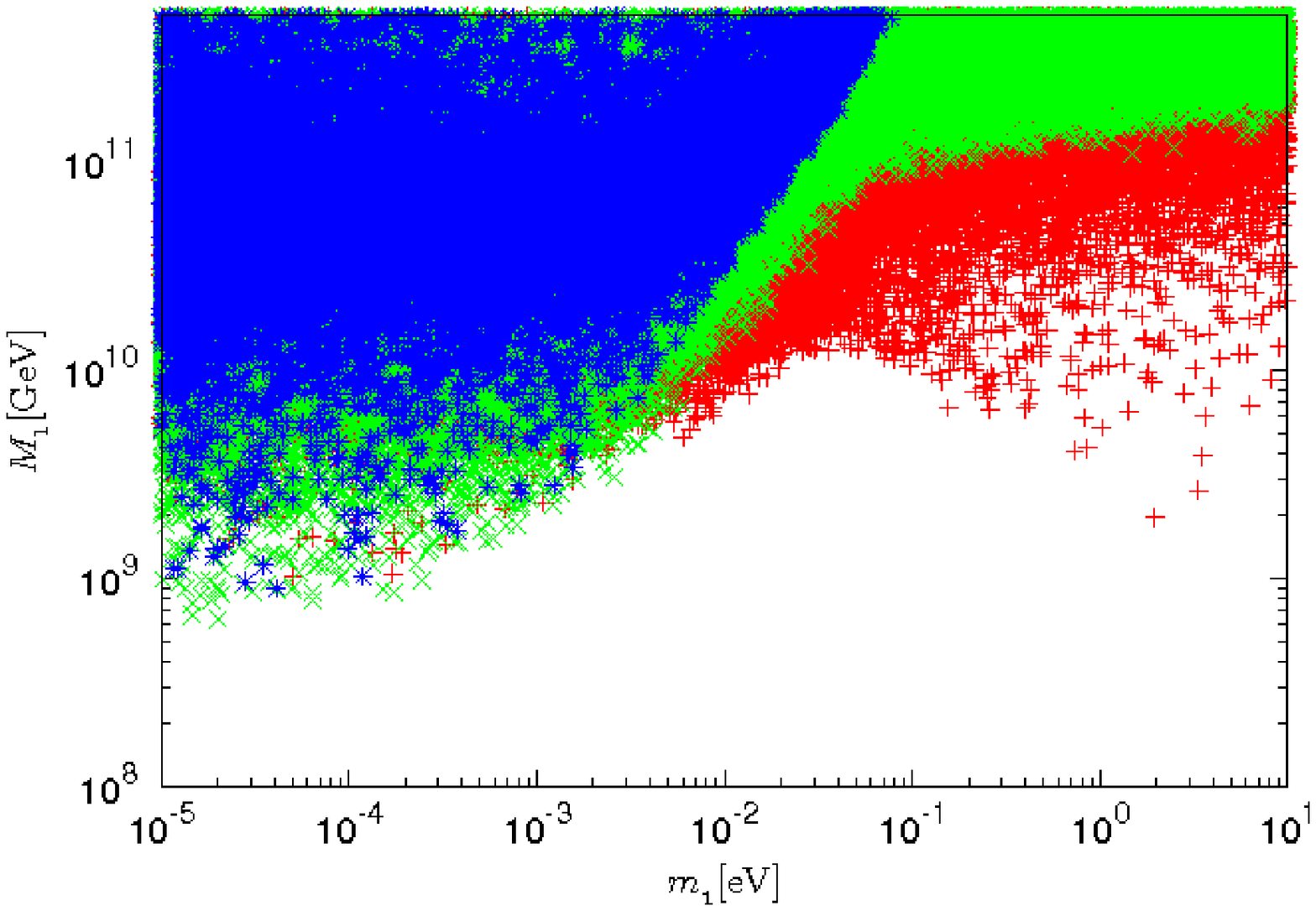}
\caption{$M_1$ vs. $m_1$ for normal (left panel) and inverted (right panel)
hierarchy. The red straight crosses denote
a projector $P^0_{1\t}<0.1$ or $P^0_{1e}+P^0_{1\m}<0.1$ , the green x's
$0.1<P^0_{1\t}<0.45$ or $0.1<P^0_{1e}+P^0_{1\m}<0.45$  and the blue stars
 $0.45<P^0_{1\t}<0.5$ or $0.45<P^0_{1e}+P^0_{1\m}<0.5$.}
\label{fig:general}
\end{center}
\end{figure}
In Fig.~\ref{fig:R13} we consider the special case $\O=R_{13}$,
corresponding to $\o_{21}=\o_{32}=0$ in the Eq.~(\ref{R}).
The number of free parameters gets
therefore reduced to 6 (one of the PMNS phases, $\F_2$, is irrelevant in this model).
One can see that the bounds are very similar to the general case, just slightly
more stringent at large $m_1$. Therefore, the special case  $\O=R_{13}$
gives an approximate condition for the saturation of the bounds in the $(m_1,M_1)$ plane.
\begin{figure}
\begin{center}
\includegraphics[width=0.33\textwidth,angle=-90]{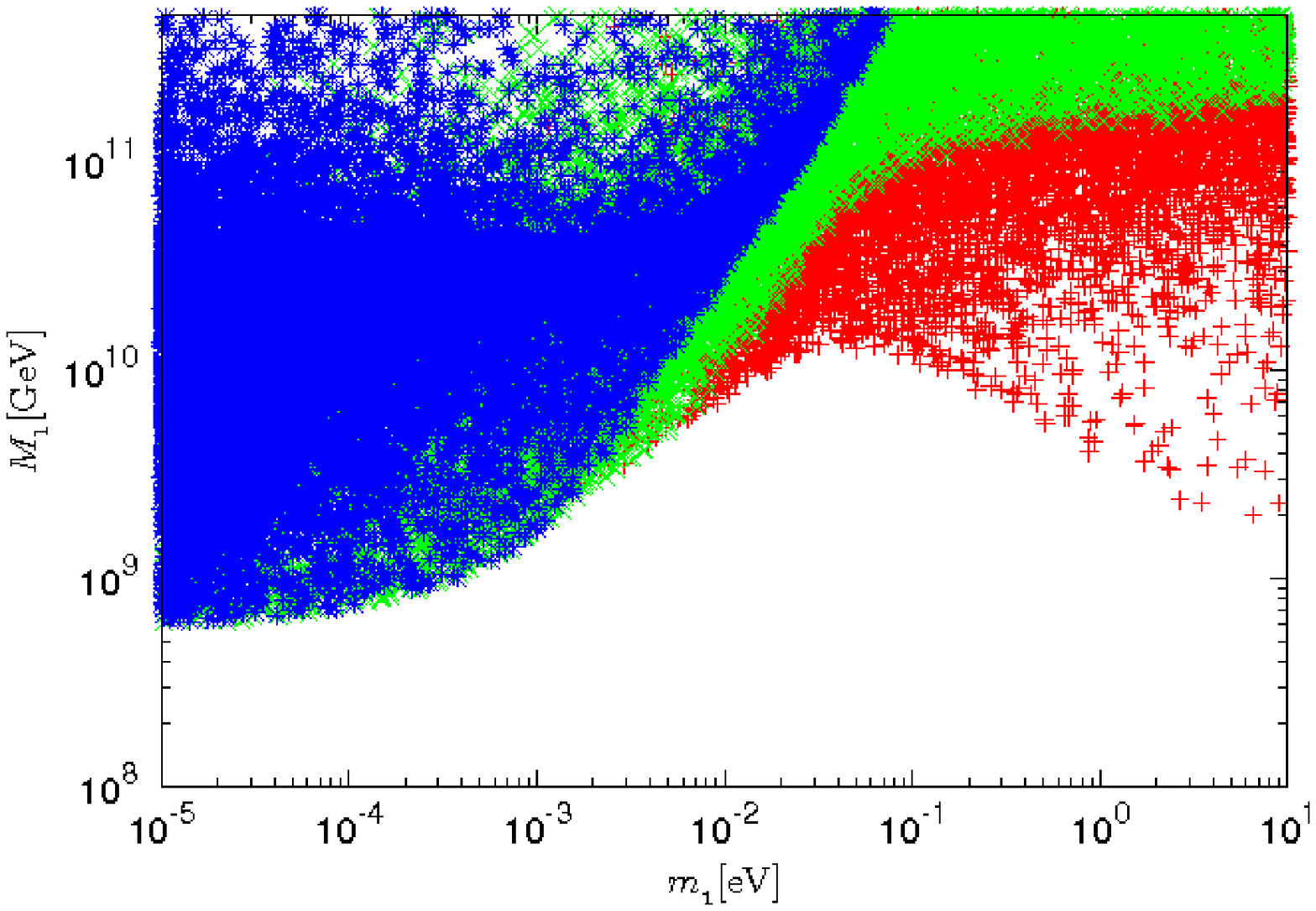}
\hspace{5mm}
\includegraphics[width=0.33\textwidth,angle=-90]{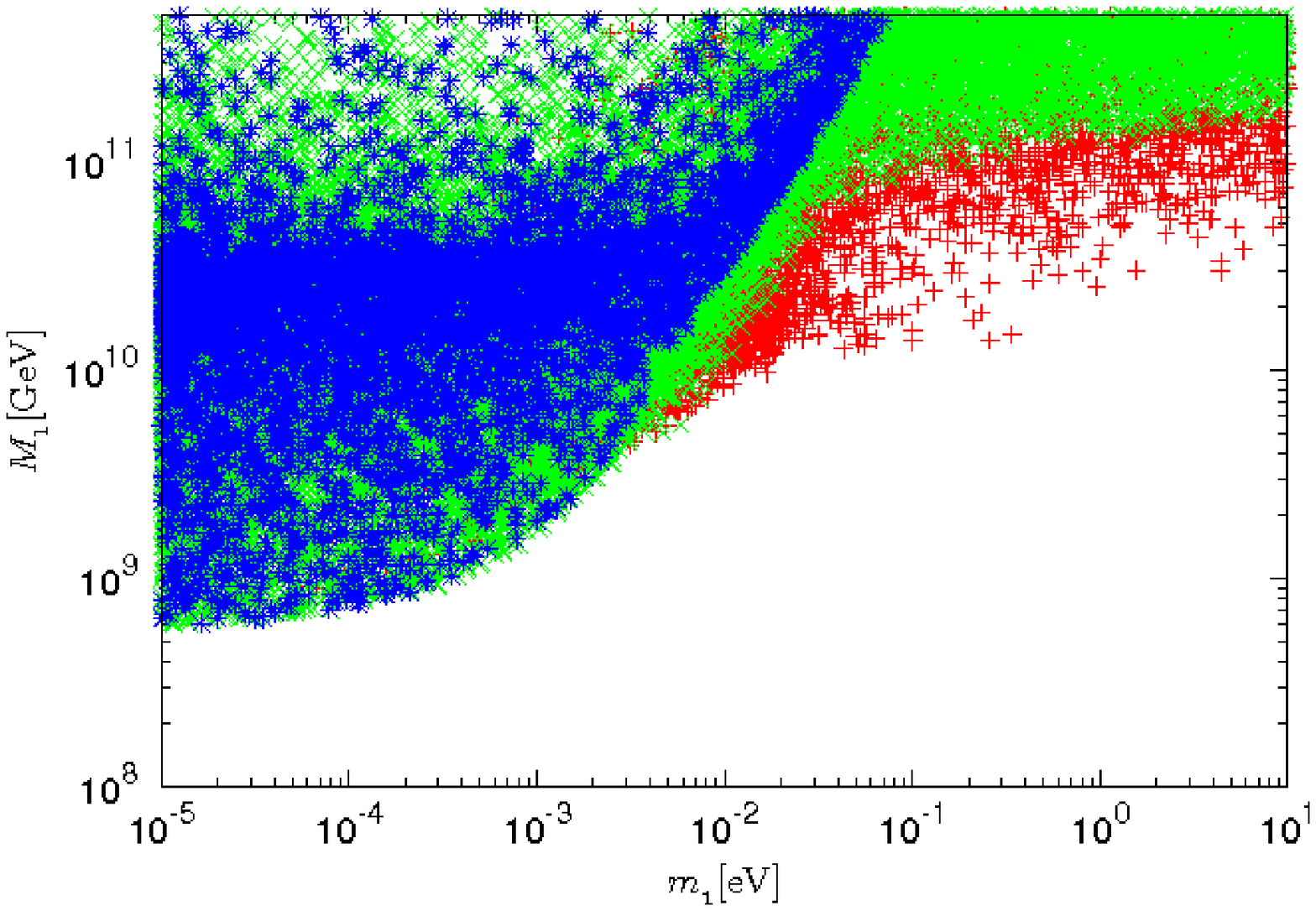}
\caption{Same as in Fig.~\ref{fig:general} but for the special case $\O=R_{13}$.}
\label{fig:R13}
\end{center}
\end{figure}
The phases in the PMNS matrix can play a crucial role in the fully flavored regime.
This was first emphasized in \cite{nardi1} and then analyzed in more detail in
\cite{abada2, flavorlep}. Here we want to show the effects of the PMNS phases
on the lower bound found in the previous figures comparing the previous
results with those obtained when the PMNS phases are turned off.
The result for a general $\O$ matrix is not shown because it is given precisely
by Fig.~\ref{fig:general}. There are indeed enough phases present in the
$\O$ matrix in this case to realize a strong one-flavor dominance and to
saturate the general lower bound even though
the PMNS phases are set to zero. The situation is different when one considers
special cases like $\O=R_{13}$ as first found in \cite{flavorlep}.
The result is shown in Fig.~\ref{fig:R13NoPhase}
and comparing with Fig.~\ref{fig:R13} one can see clearly that the PMNS phases are
responsible for the saturation of the bounds at large $m_1$.
The reason is that there is in this case only one phase in the $\O$ matrix,
and this is not enough to fulfill the conditions for the saturation of the lower bound.
Therefore, the Majorana phase $\F_1$ and the Dirac phase $\d$ play a
crucial role in this case analogously to the case $M_3 \gg 10^{14}\,{\rm GeV}$
for inverted hierarchy, as seen in the previous subsection and
recently pointed out in \cite{molinaro}.
\begin{figure}
\begin{center}
\includegraphics[width=0.33\textwidth,angle=-90]{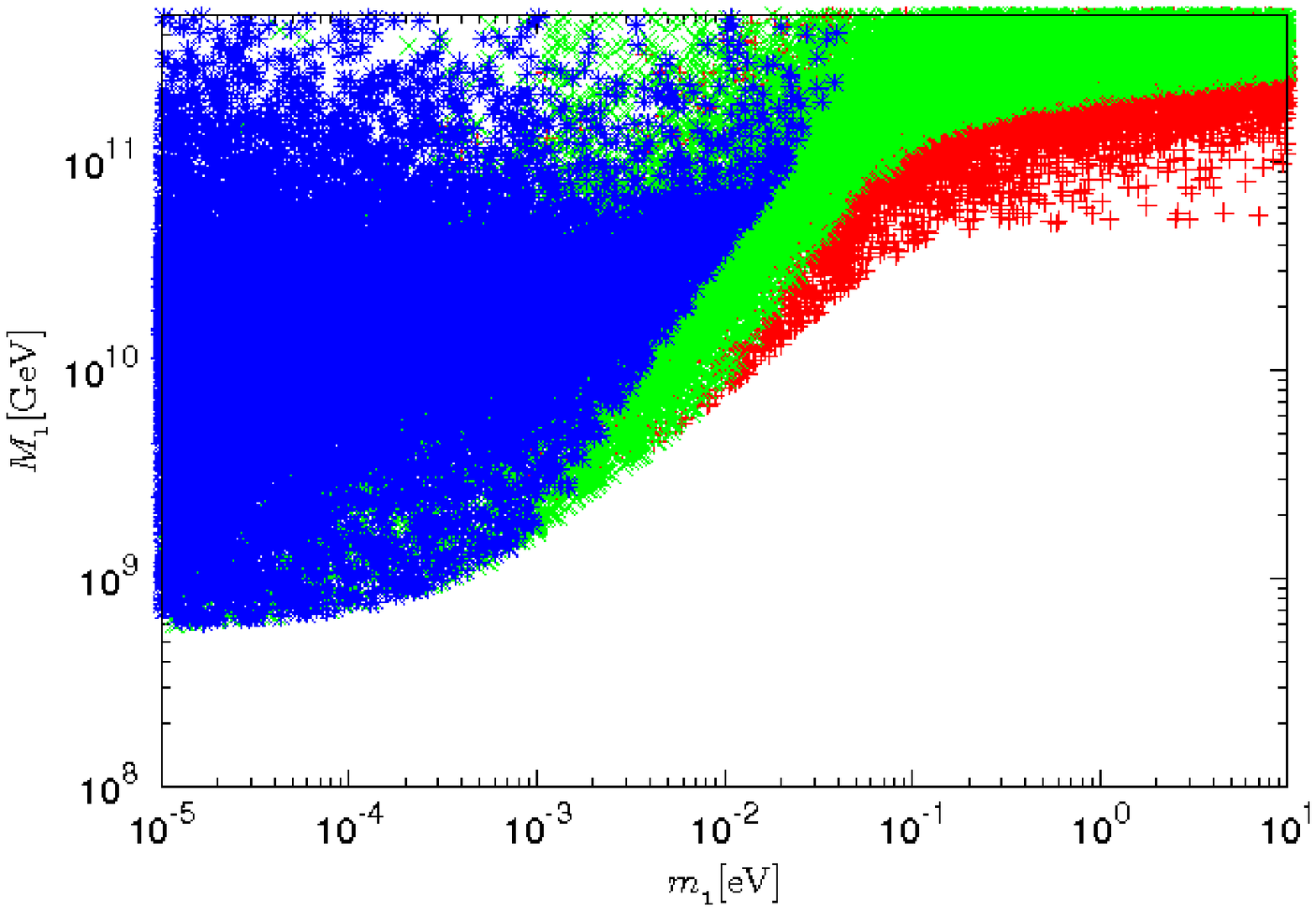}
\hspace{5mm}
\includegraphics[width=0.33\textwidth,angle=-90]{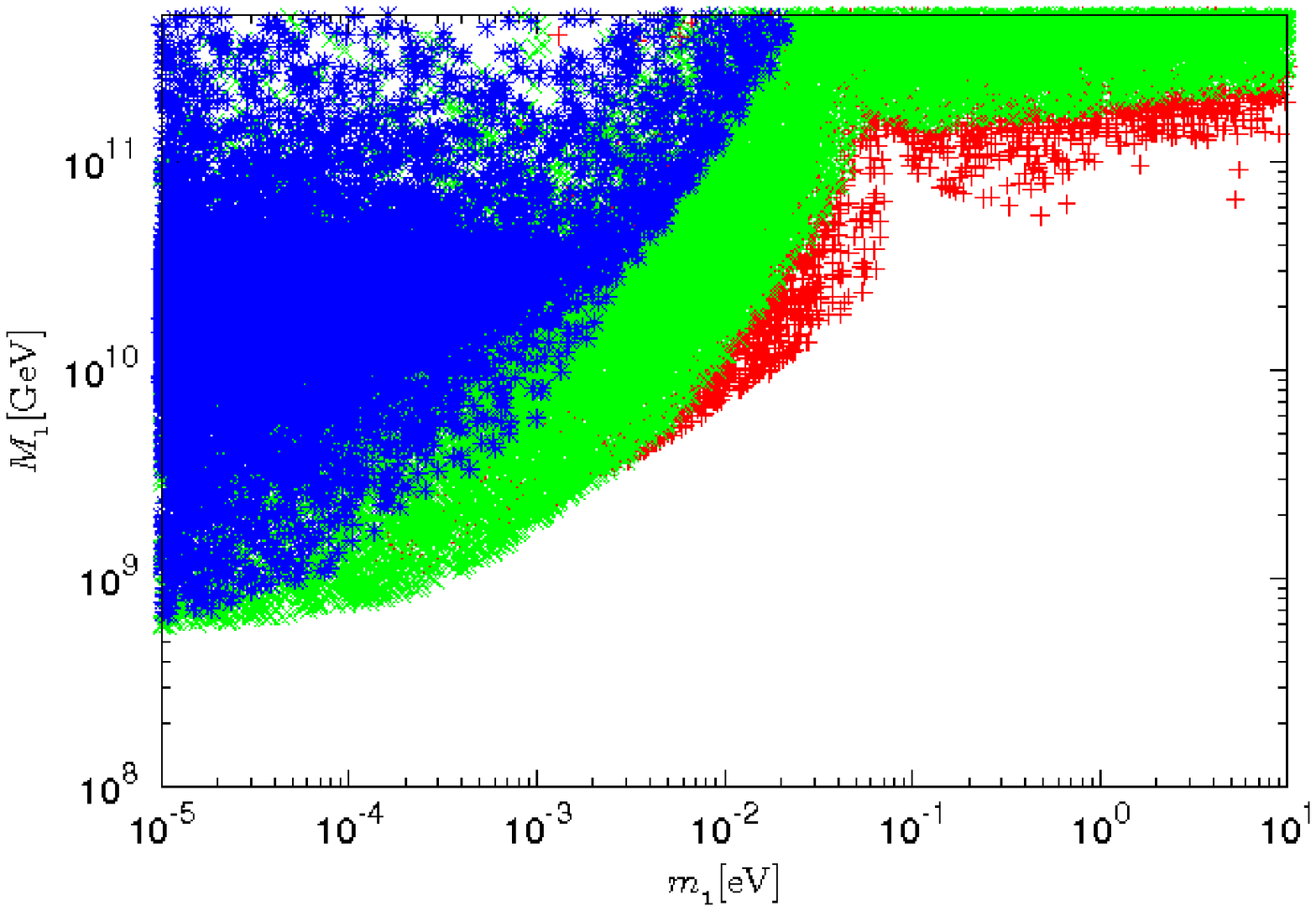}
\caption{Same as Fig.~\ref{fig:general}, for $\O=R_{13}$ with all PMNS phases turned off.}
\label{fig:R13NoPhase}
\end{center}
\end{figure}

Let us now discuss the consequence of imposing the condition
of validity of the fully flavored regime, Eq.~(\ref{condition}).
The result is shown in the left panel of Fig.~\ref{fig:condition}, to be compared with
Fig.~\ref{fig:general} where the same parameters were varied but all
points were kept. One notices clearly that many points at large $m_1$
disappear and an upper bound on $m_1$, given by $m_1\lesssim 0.15~{\rm eV}$,
appears again and is comparable to the upper bound holding
in the unflavored regime, $m_i\lesssim 0.12\,{\rm eV}$.
In the case of inverted hierarchy  similar results apply.
However, it should be clarified that $m_1\lesssim 0.15~{\rm eV}$ is not an upper bound on
the neutrino masses but a limit of validity of the fully flavored regime.
One should therefore solve more general kinetic density matrix equations
in order to describe the intermediate regime and to see whether
there is or not an upper bound on neutrino masses.
\begin{figure}
\begin{center}
\includegraphics[width=0.33\textwidth,angle=-90]{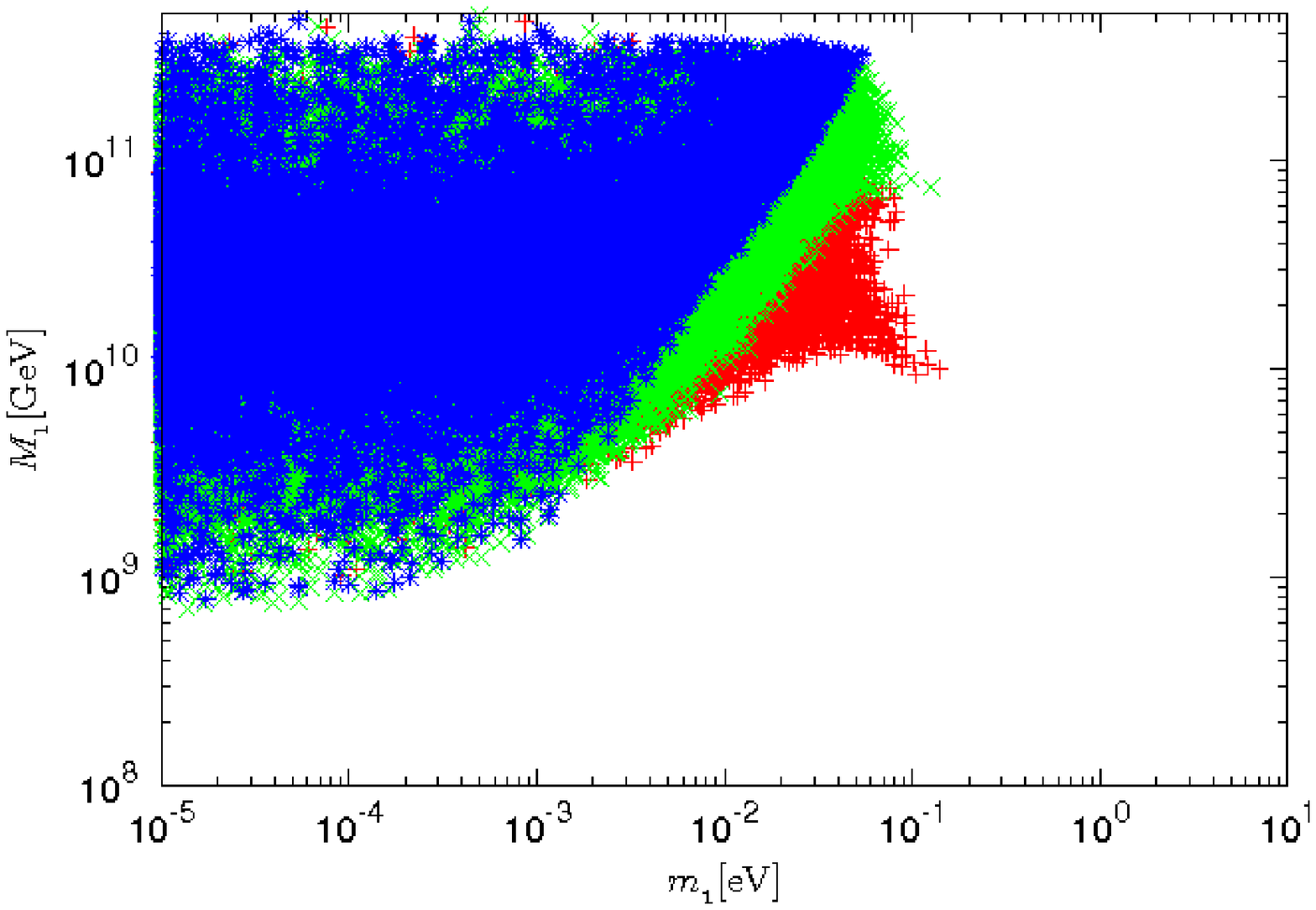}
\hspace{5mm}
\includegraphics[width=0.33\textwidth,angle=-90]{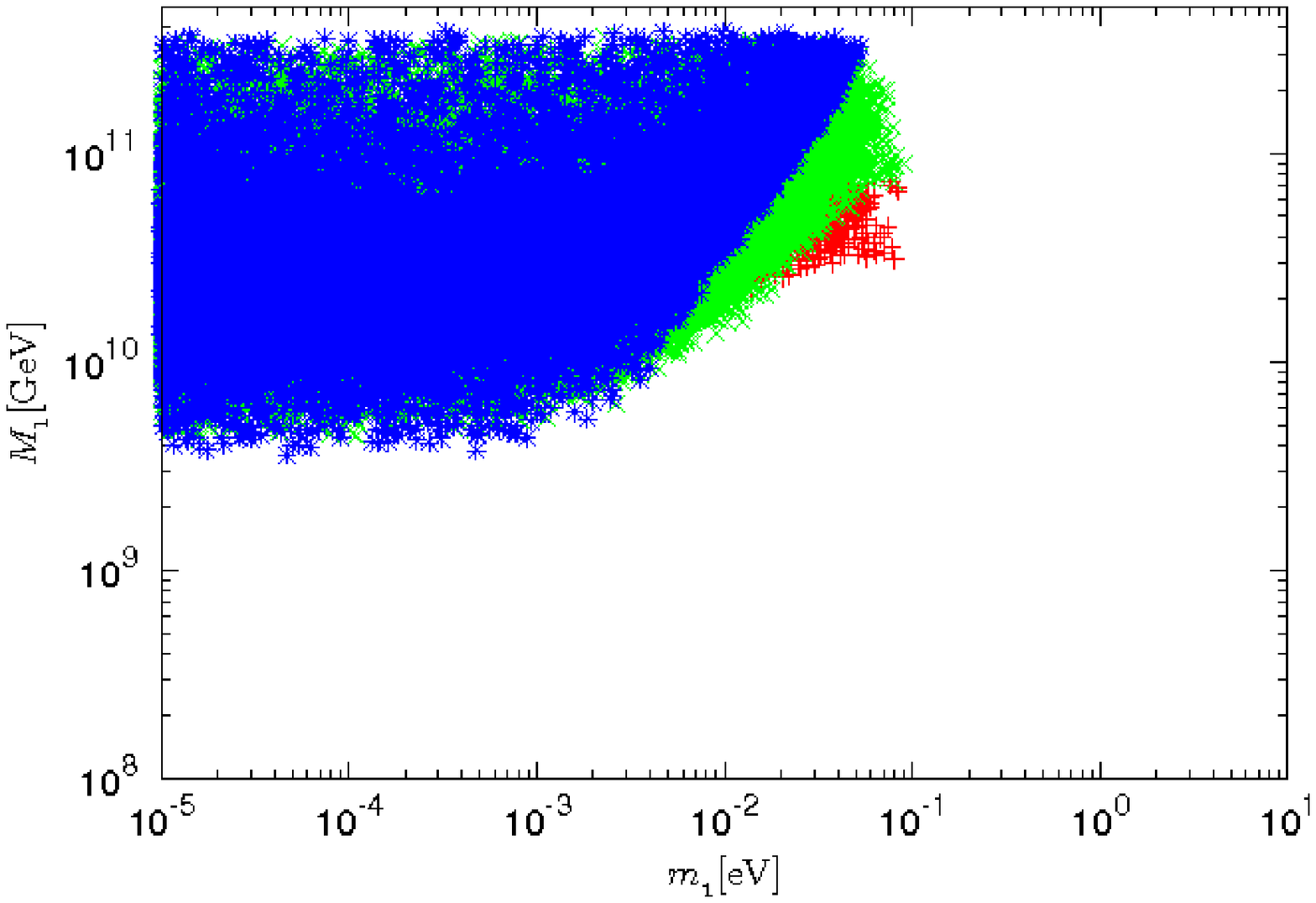}
\caption{Same as Fig.~\ref{fig:general}, imposing
the condition of validity of the fully flavored regime (left panel) and
the strong washout condition (right panel).}
\label{fig:condition}
\end{center}
\end{figure}
In the right panel of the same figure, we also imposed the strong
washout regime condition and one can see that the allowed region
gets further reduced and the upper bound on $m_1$ even more
stringent, $m_1\lesssim 0.10\,{\rm eV}$.

So far all results have been obtained making use  of the approximation $C=I$.
In the upper panels of Fig.~\ref{fig:higgs},
left for normal hierarchy and right for inverted hierarchy,
we show the results without taking into account the Higgs asymmetry
solving the Eq.'s (\ref{flke}) neglecting $C^H$ in $C$ that reduces to
$C^l$  given by the expression (\ref{Cl}). One can see that
the upper bound on $m_1$ gets relaxed when $C^H$ is neglected.
In the bottom panels we used the Eq.~(\ref{C}) for $C$ and
one can see that the results agree very well with those obtained
when we used the approximation $C=I$. Therefore, we will
continue to use this approximation in the following analysis.
\begin{figure}
\begin{center}
\psfig{file=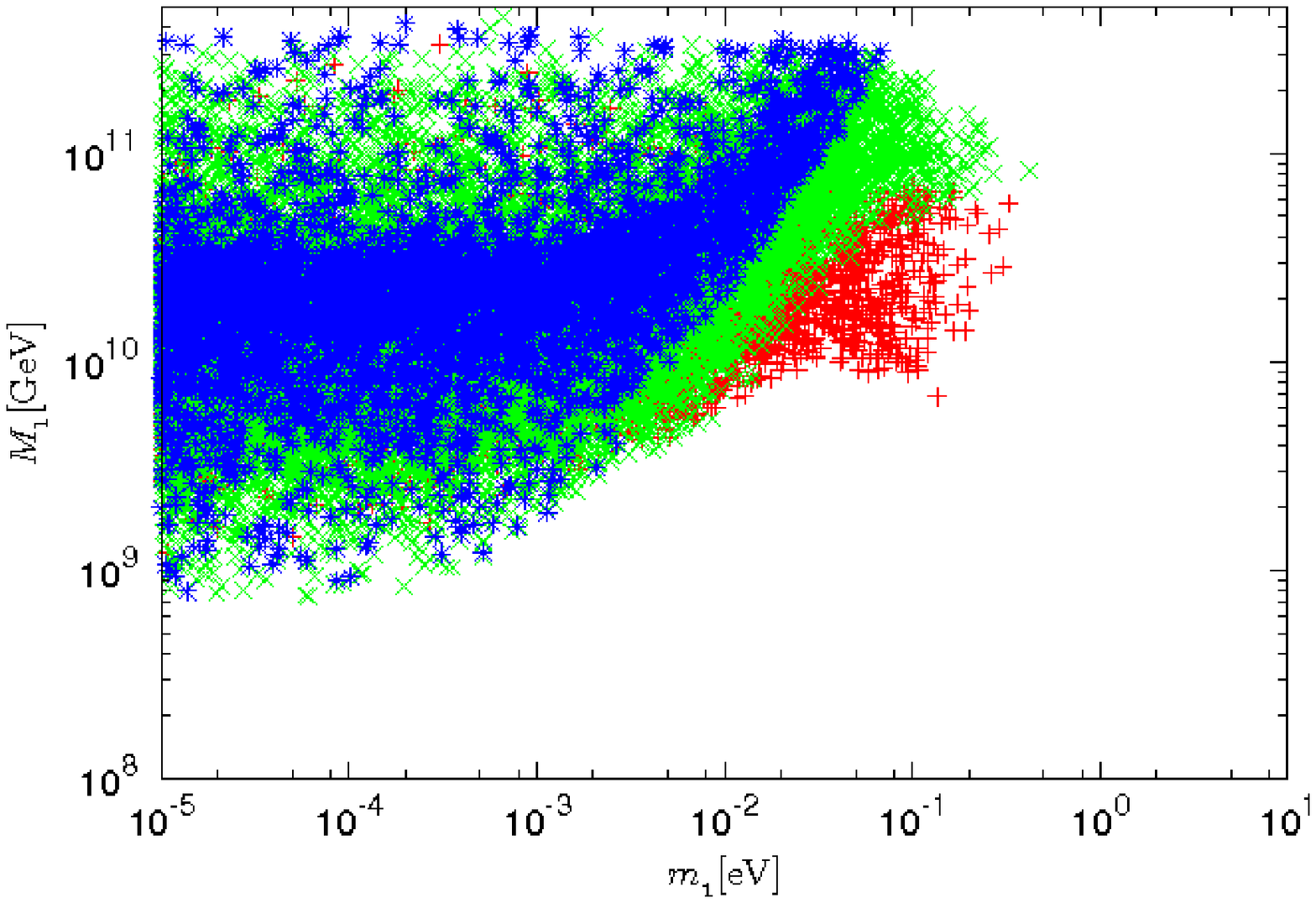,height=7cm,width=7cm,angle=-90}
\hspace{-1mm}
\psfig{file=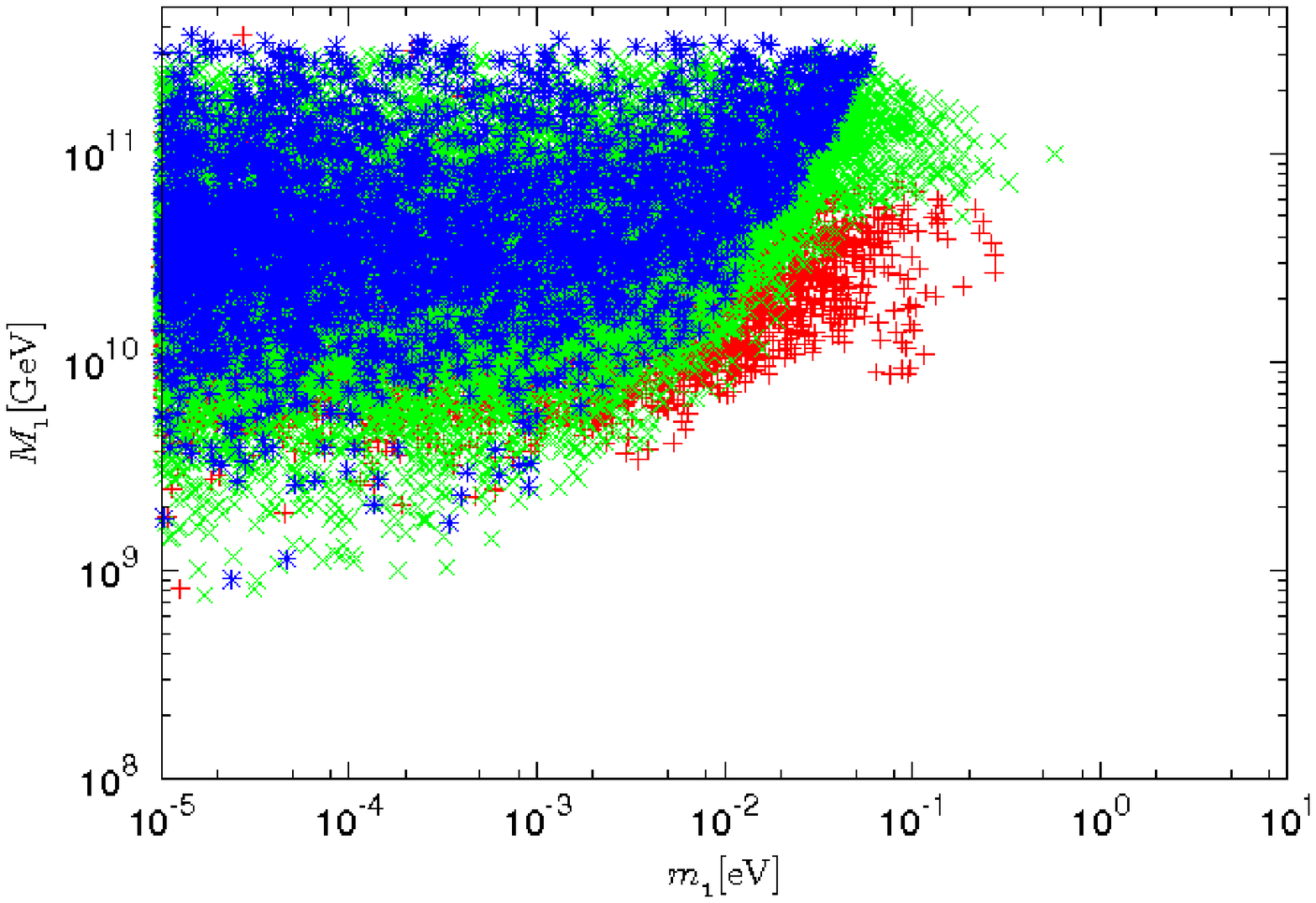,height=7cm,width=7cm,angle=-90}
\\
\psfig{file=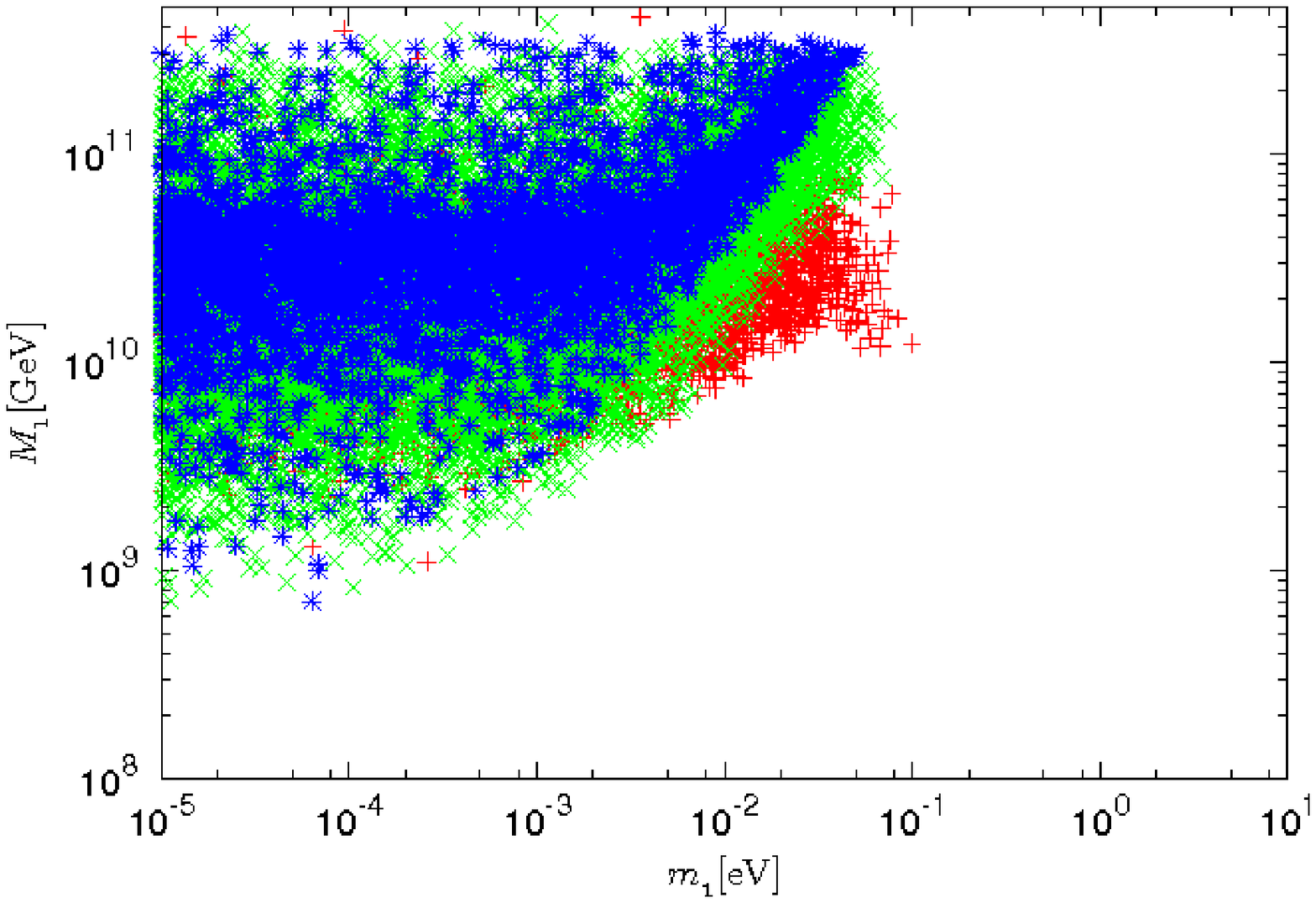,height=7cm,width=7cm,angle=-90}
\hspace{-1mm}
\psfig{file=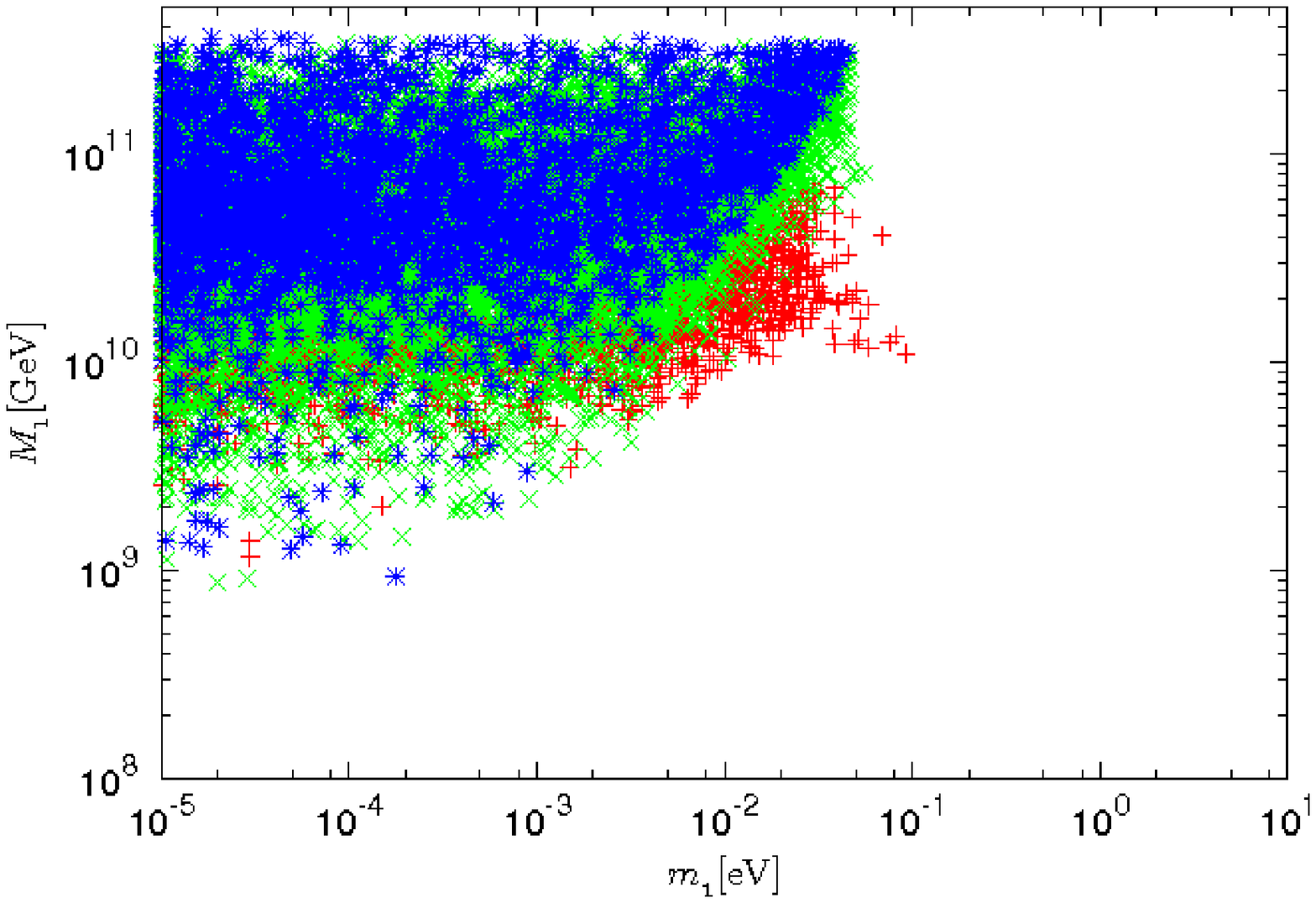,height=7cm,width=7cm,angle=-90}
\caption{Check of the approximation $C=I$. In the upper panels
we neglected the Higgs asymmetry solving the Eq.'s (\ref{flke})
with $C^H=0$. In the bottom panels we solved the Eq.'s (\ref{flke})
with $C$ given by the Eq. (\ref{C}).}
\label{fig:higgs}
\end{center}
\end{figure}

A particularly interesting case to be discussed is when all the high-energy phases
are switched off, i.e. imposing the $\O$ matrix to be real.
In this case the PMNS phases act as the only source of $C\!P$ violation responsible
for the explanation of the observed matter-antimatter asymmetry \cite{nardi1}.
It is therefore interesting to understand whether
there is an allowed region at all \cite{flavorlep}.
The results are shown in Fig.~\ref{fig:OmReal} for $M_3=M_2=3\,M_1$.
We show both the results obtained when both Majorana and the Dirac phase
are switched on (red points) and those obtained when only the
Dirac phase is non-vanishing (green points) and for $\sin\theta_{13}=0.2$,
the $3\,\s$ experimental upper bound. This even more special case is the Dirac phase
leptogenesis scenario \cite{diraclep}. The Dirac phase being the
only phase that we can realistically measure
in the near future in neutrino oscillation experiments, it is therefore
particularly relevant to understand whether
this testable source of $C\!P$ violation can be also responsible for
the observed matter-antimatter asymmetry.

The top-left panel is for normal hierarchy while the top-right is for
inverted hierarchy. One can notice that in this last case the lower bound on $M_1$
tends to infinite for vanishing $m_1$, meaning that the final asymmetry tends to vanish
\footnote{Similar results have been recently obtained in \cite{molinaro2}.}.
In both cases there is an upper bound on $m_1$, as first discussed in \cite{branco}.
In the bottom-left panel we impose the condition of validity of the
fully flavored regime given by the Eq.~(\ref{condition}) and one can see
that the allowed region gets reduced in a relevant way. Imposing even the
strong washout regime condition (i.e. independence of the initial conditions)
one can see that the allowed regions gets further reduced and it practically disappears
for the case of Dirac phase leptogenesis, confirming, more generally,
what was found in \cite{diraclep}.
\begin{figure}
\begin{center}
\psfig{file=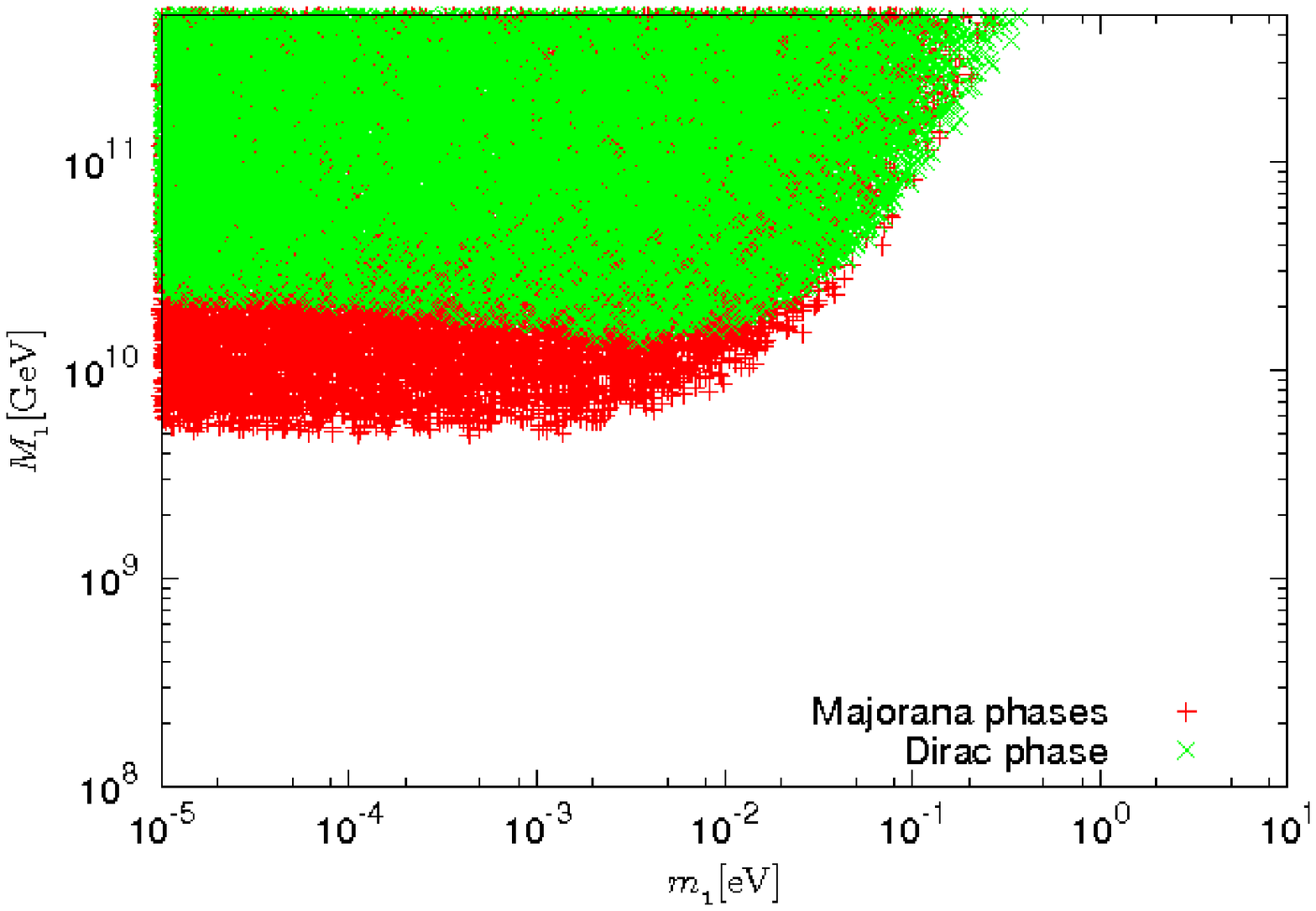,height=7cm,width=7cm,angle=-90}
\psfig{file=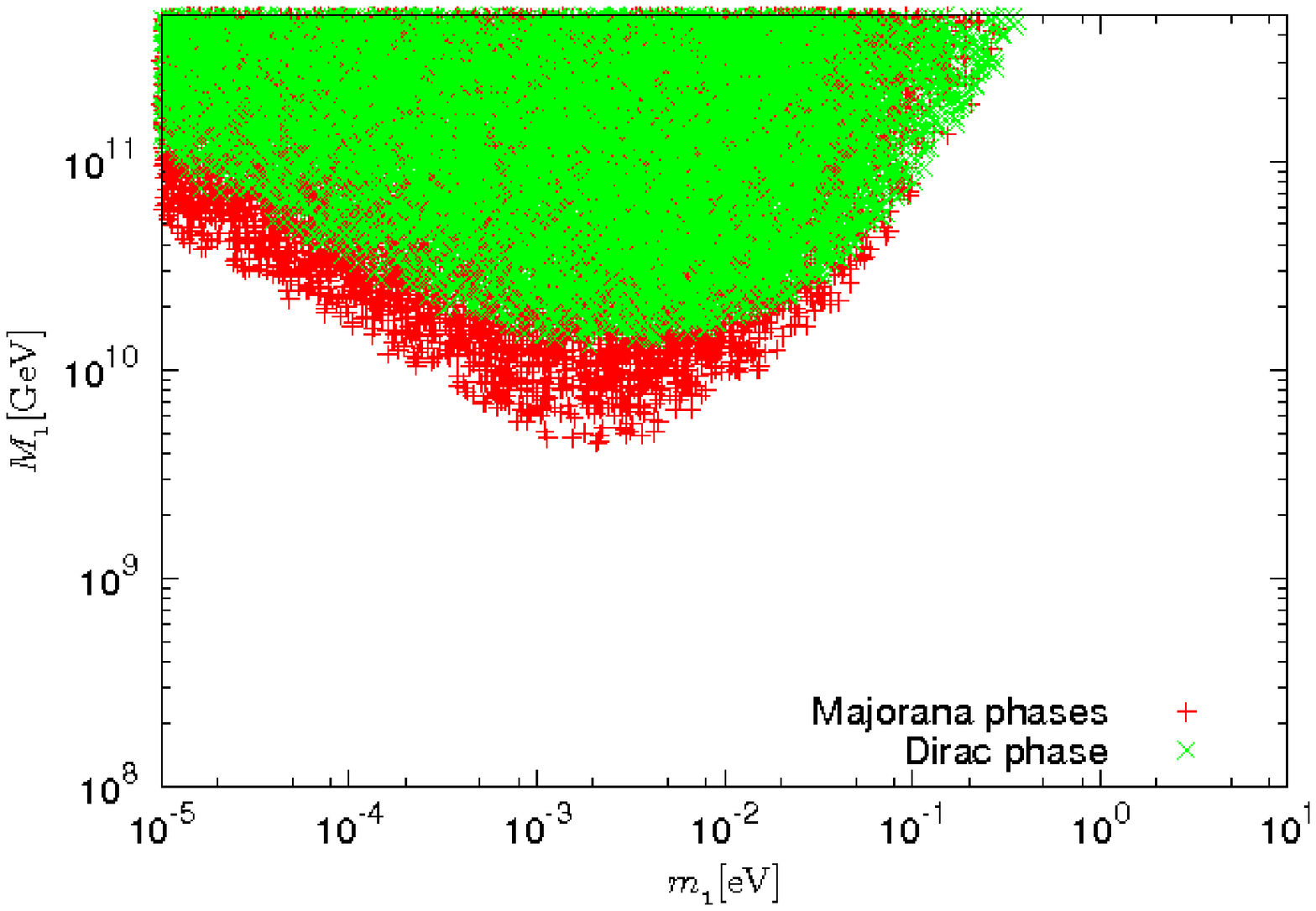,height=7cm,width=7cm,angle=-90}
\\
\psfig{file=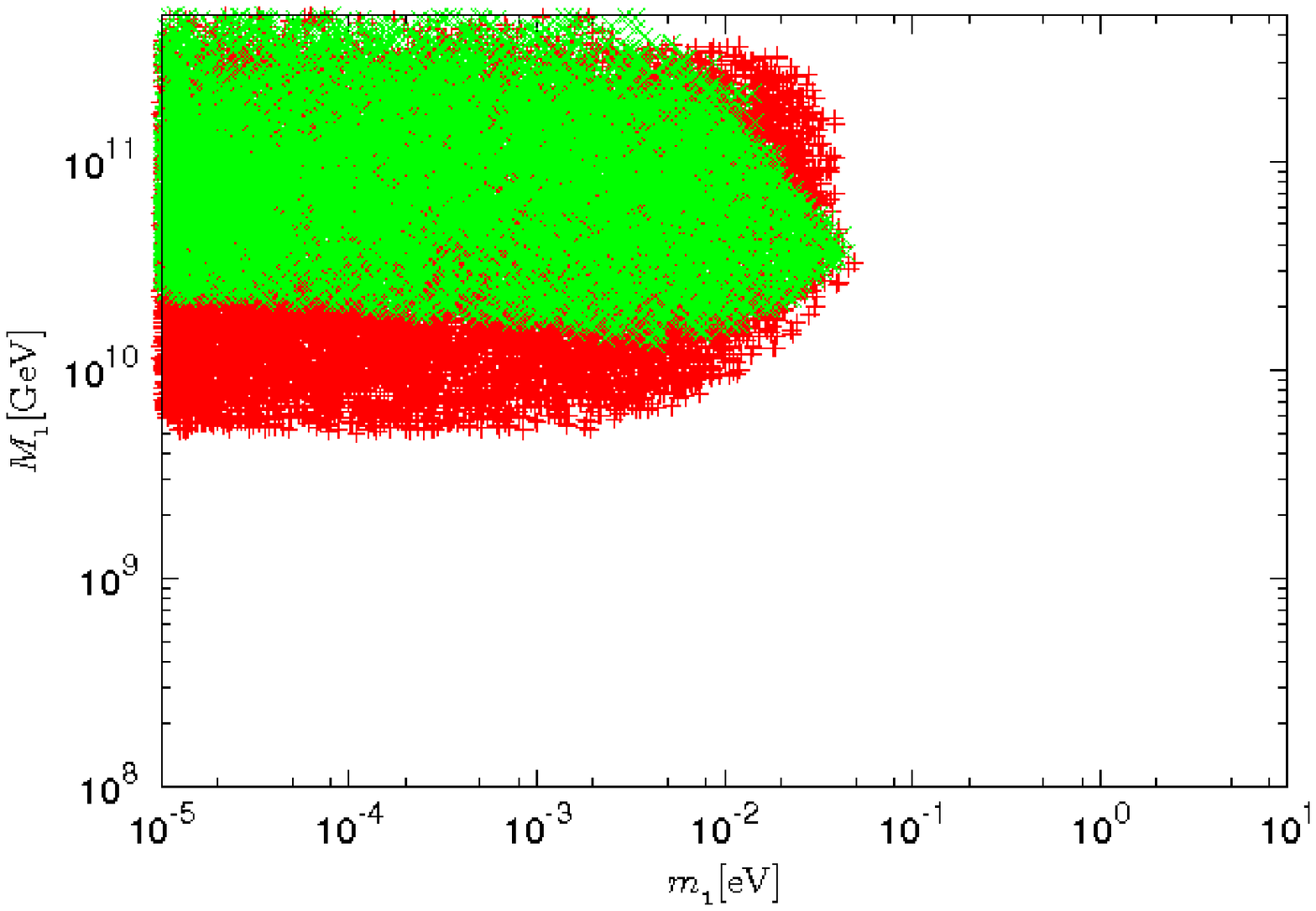,height=7cm,width=7cm,angle=-90}
\psfig{file=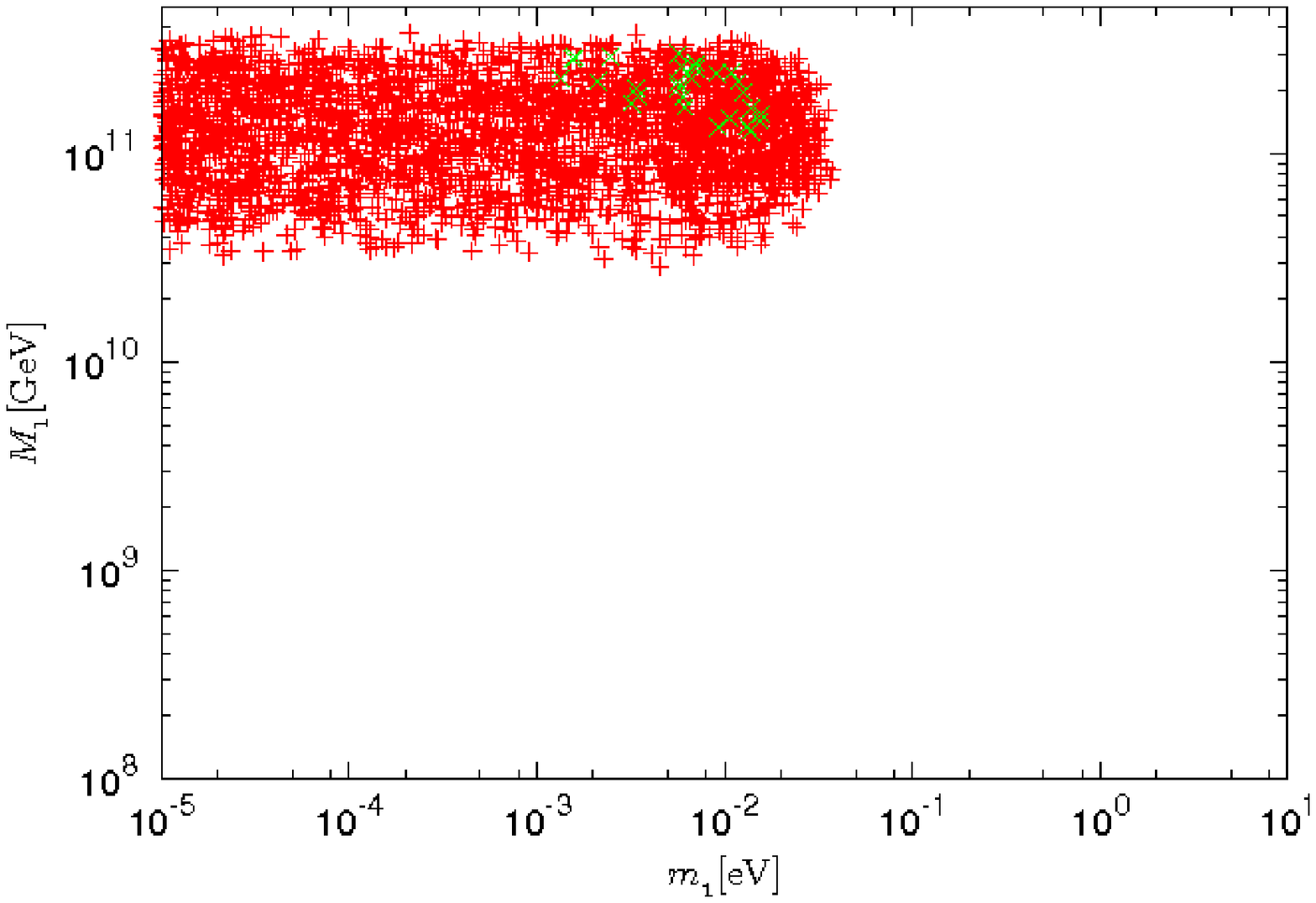,height=7cm,width=7cm,angle=-90}
\caption{Leptogenesis only with Majorana phases (red points)
and only with Dirac phase (green points). Top-left and top right panel: allowed region
in the plane $(m_1,M_1)$ for normal hierarchy and inverted hierarchy respectively.
Bottom panels: the same as top-left but imposing the condition Eq.~(\ref{condition})
(left) and the strong washout condition as well (right panel).}
\label{fig:OmReal}
\end{center}
\end{figure}

\subsection{Effect of large $|\o_{ij}|$ values}
\label{sec:largeom}

So far, we imposed the restriction $|\o_{ij}|\leq 1$.
In this subsection we relax this restriction
studying the effect of allowing $|\o_{ij}|$ as large as 10 on the bounds.
The main effect is that the extra-term $\Delta\ve_{1\alpha}$
in the flavored $C\!P$ asymmetries (cf. Eq.~(\ref{Dveps1a}))
can become dominant. The upper bound
Eq.~(\ref{CPbound}) does not apply to this term.
Since this term is suppressed like $M_1/M_2$  the dominance
is possible for a mild hierarchical spectrum,
such that $M_2/M_1 < {\cal O}(100)$. In any case notice that its
dominance applies under conditions that are much less restrictive
than those found for the dominance of $\D\ve_1$, that is suppressed
like $(M_1/M_2)^2$ and that cancels when $M_2=M_3$.

\begin{figure}
\hspace*{-5mm}
\psfig{file=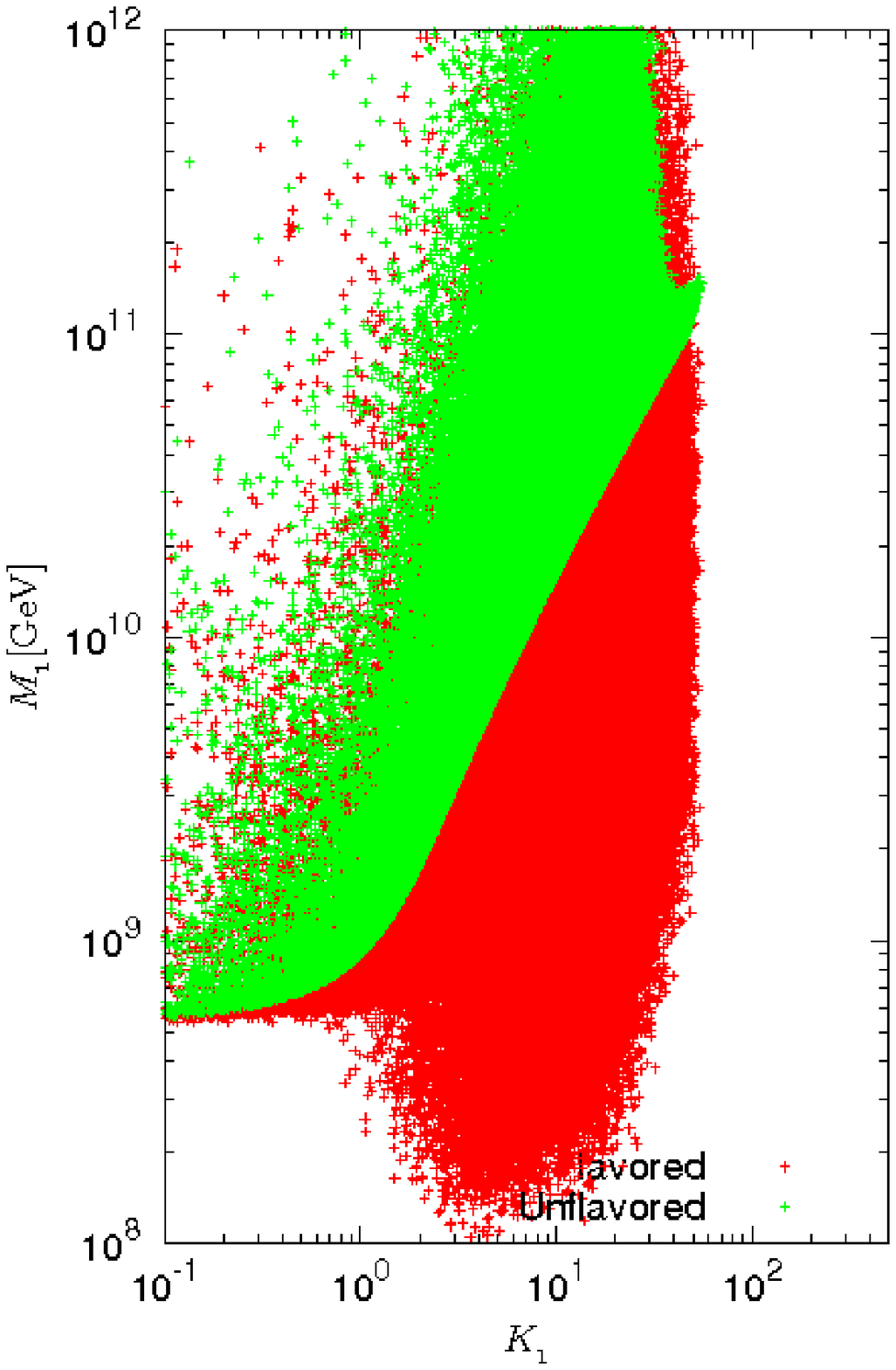,height=7cm,width=53mm}
\hspace{-1mm}
\psfig{file=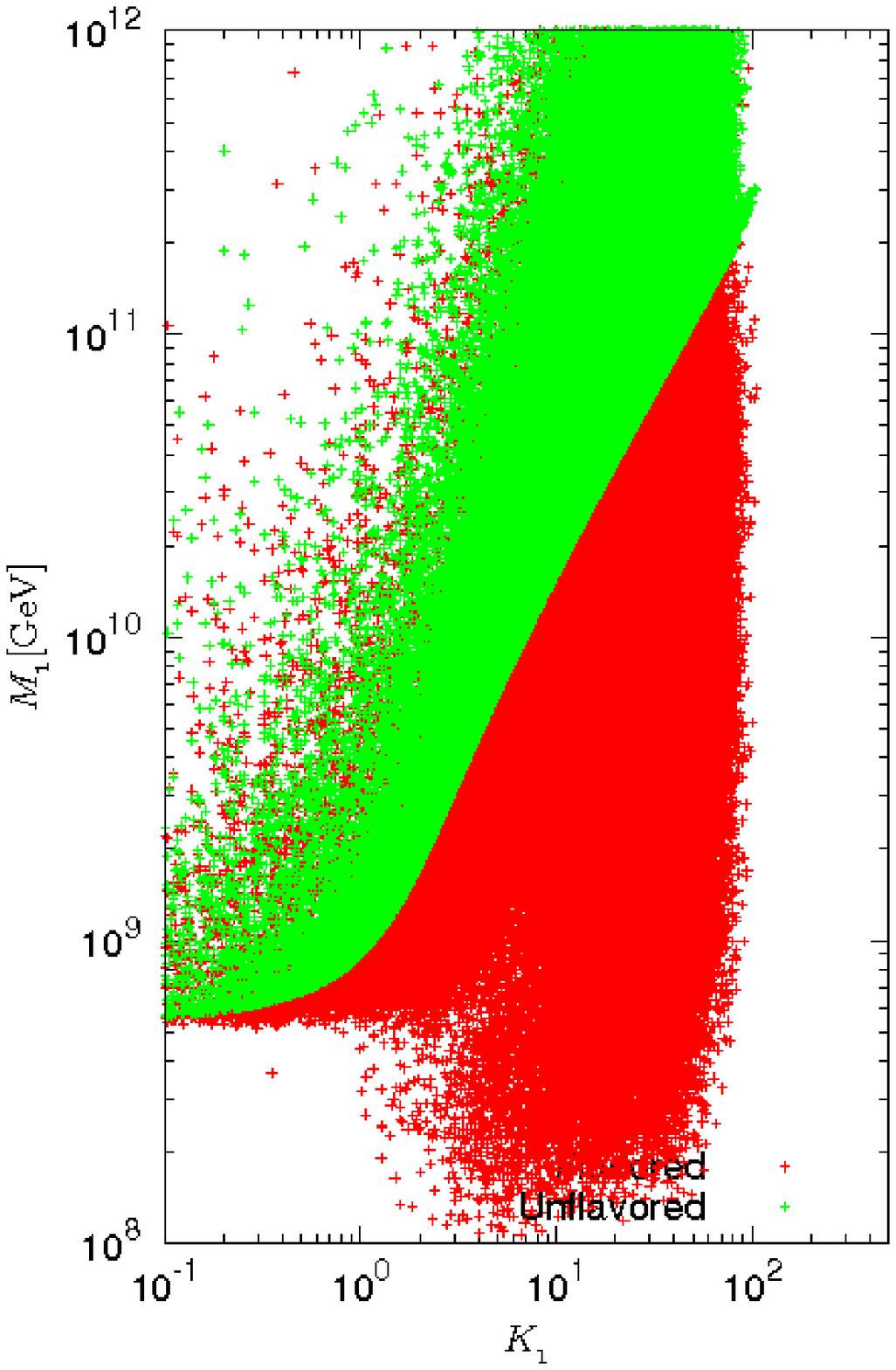,height=7cm,width=53mm}
\hspace{-1mm}
\psfig{file=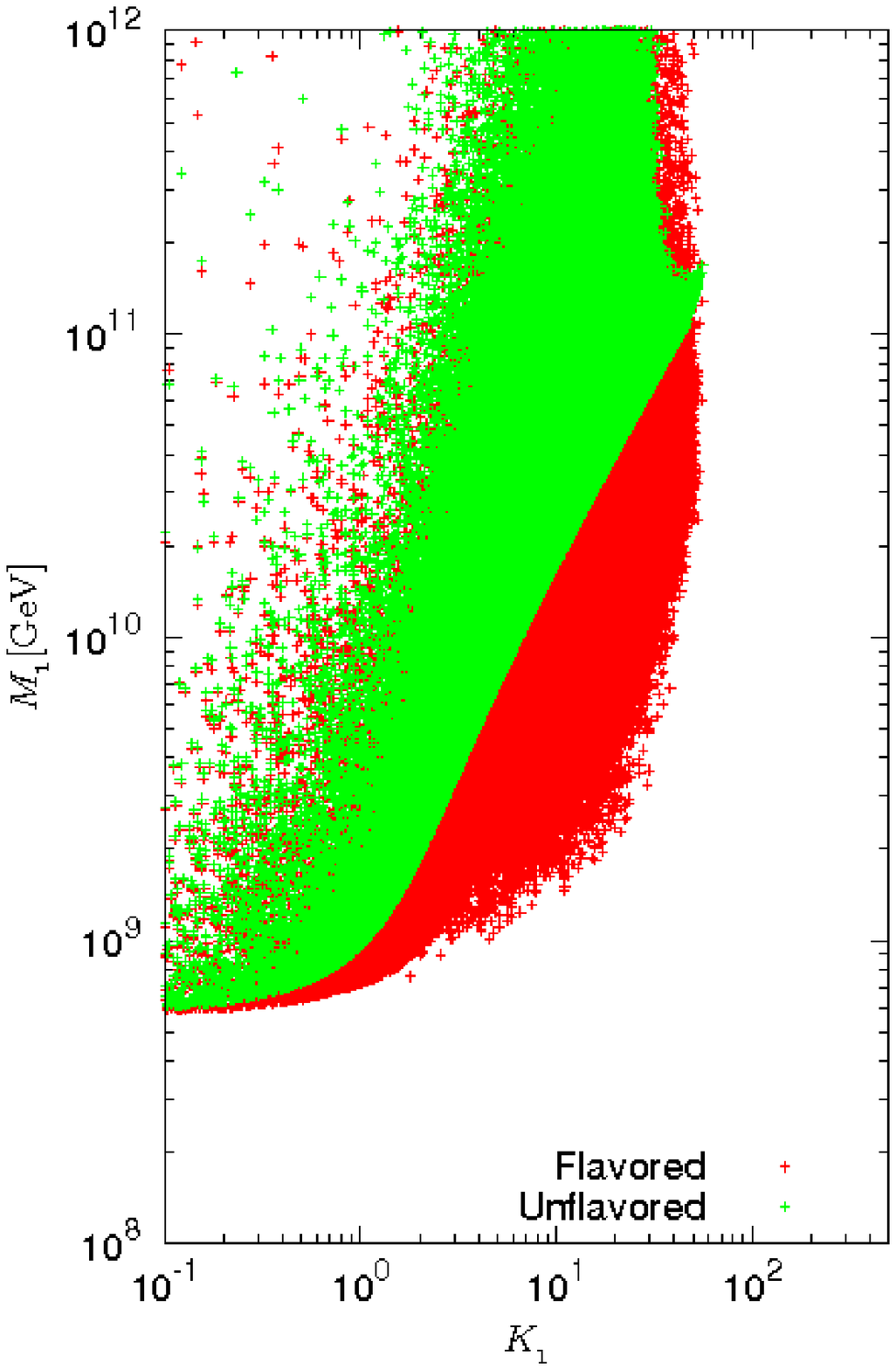,height=7cm,width=53mm}
\\
\psfig{file=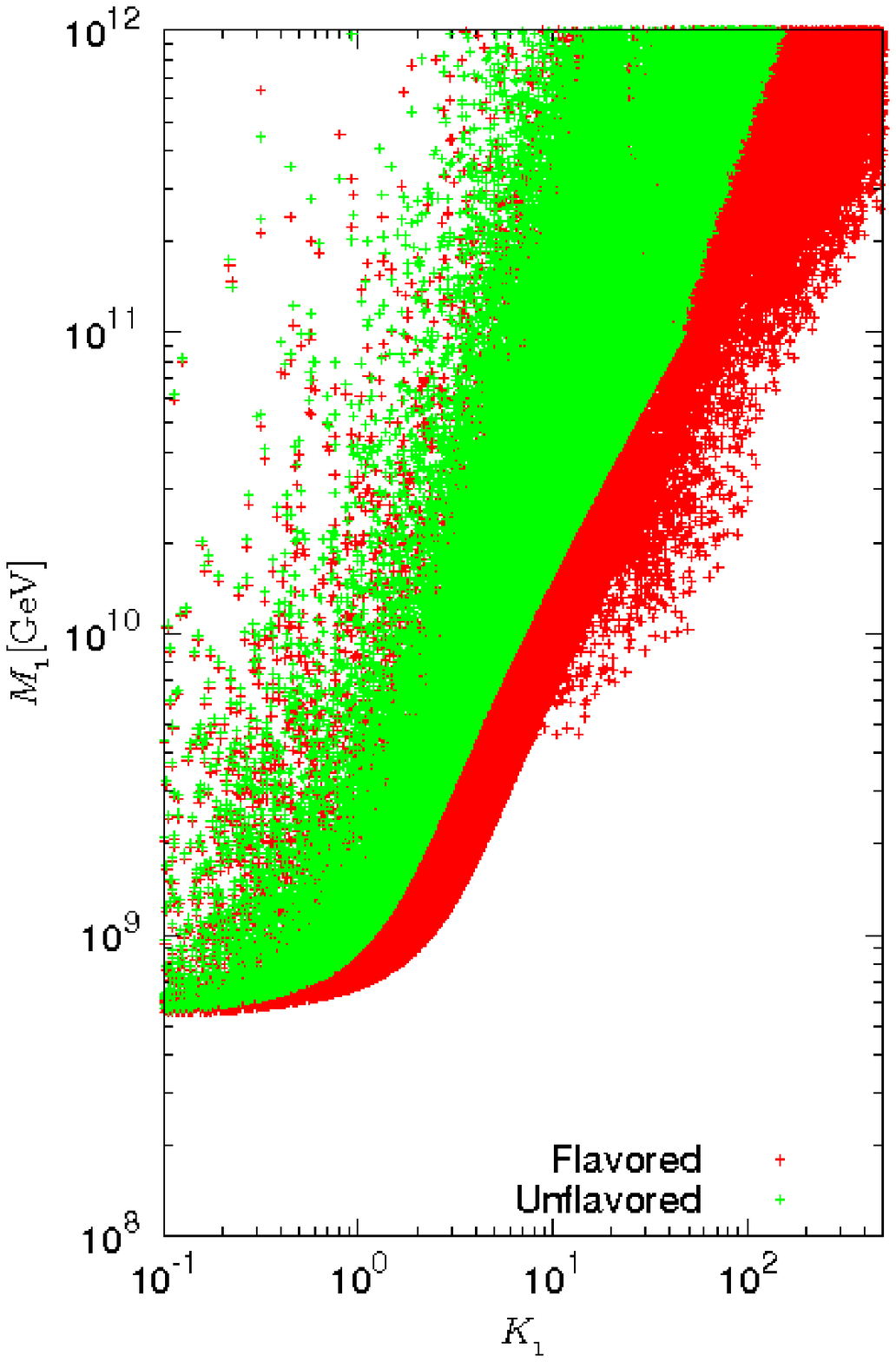,height=7cm,width=53mm}
\hspace{-1mm}
\psfig{file=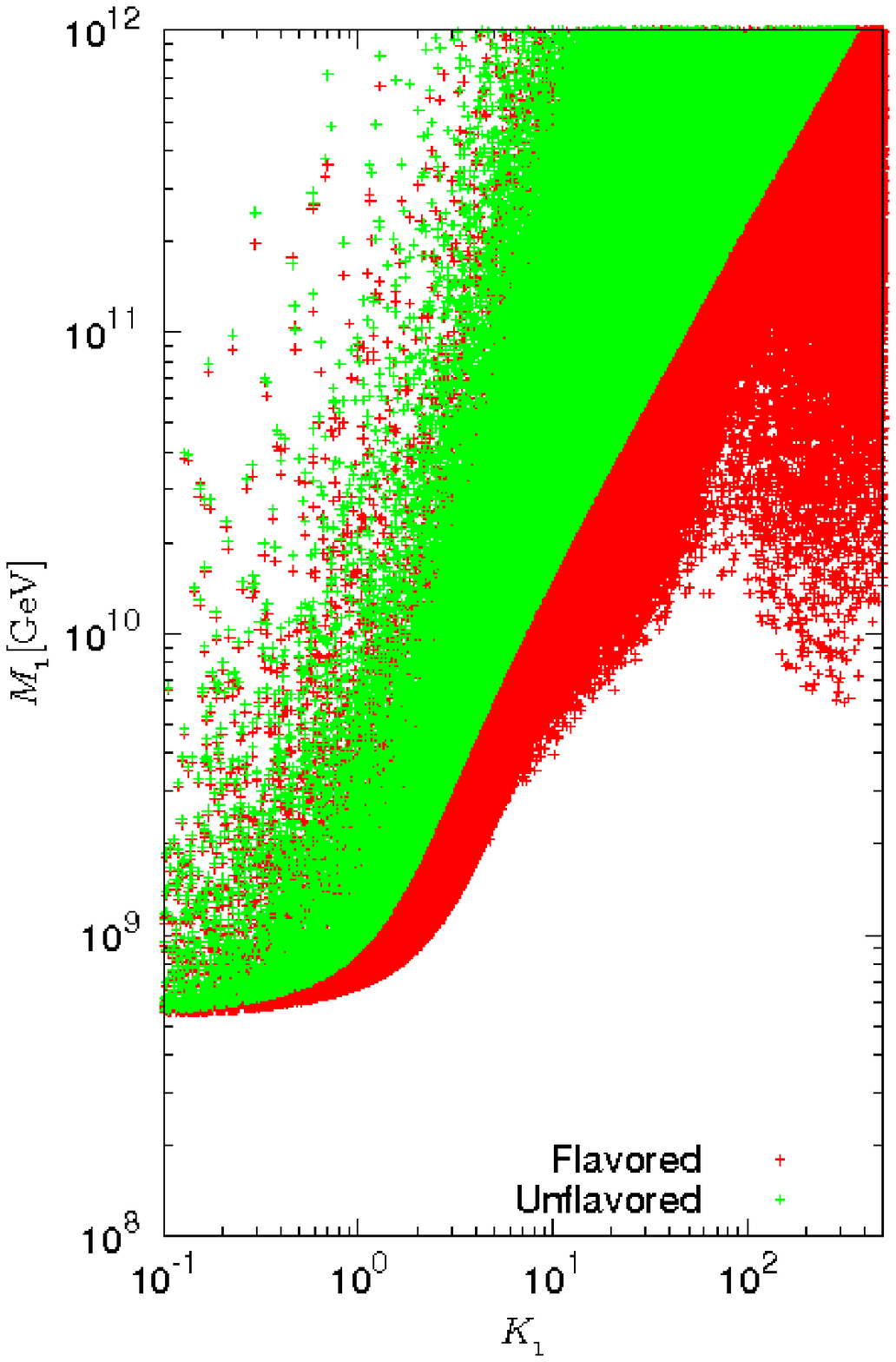,height=7cm,width=52mm}
\hspace{-1mm}
\psfig{file=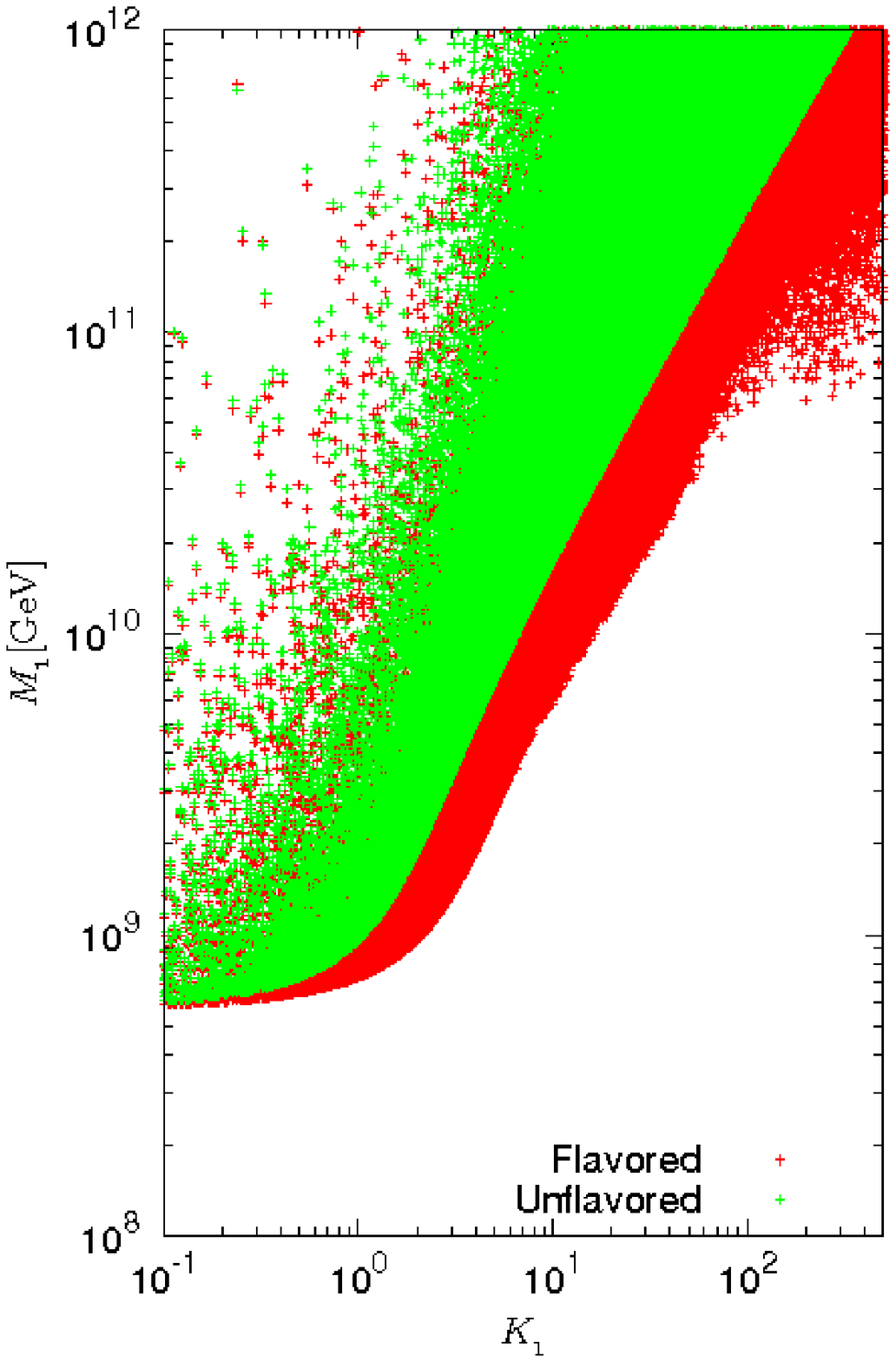,height=7cm,width=52mm}
\caption{Lower bound on $M_1$ versus $K_1$ for $m_1=0$. The results are
compared when flavor effects are neglected (green points)
and when they are taken into account (red points) while all
have been obtained for a thermal initial $N_1$-abundance ($N_{N_1}^{\rm in}=1$).
Top panels: $|\o_{21}|, |\o_{31}|\leq 1$, $|\o_{32}|\leq 10$; Top left and center
panels: normal and inverted hierarchy, respectively, and $M_2=M_3=3\,M_1$; Top
right panel: normal hierarchy and $M_2=M_3=30 M_1$. Bottom panels:  $|\o_{31}|,
|\o_{32}|\leq 1$, $|\o_{21}|\leq 10$; Bottom left and center panels:
normal and inverted hierarchy, respectively, and $M_2=M_3=3\,M_1$;
Bottom right panel: inverted hierarchy and $M_2=M_3=30\,M_1$.}
\label{lboundflLarge}
\end{figure}
In Fig.~\ref{lboundflLarge} we show the allowed region in the plane $(K_1,M_1)$
for $m_1=0$ and for $M_3=M_2$ so that $\D\ve_1$ does not give any contribution.
The results have to be compared with those in Fig.~\ref{lboundfl}.
The maximum effect is obtained again, like for $\D\ve_1$,
when $\o_{32}$ is large. In the top panels only $|\o_{32}|$ is allowed
to be as large as 10, while $|\o_{21}|, |\o_{31}| \lesssim 1$. In the top-left
panel and in the top-center panel, for normal and inverted
hierarchy respectively and for $M_2=3\,M_1$. One can see how the lower bound on $M_1$
gets relaxed of one order of magnitude so that it can be as low as $10^8\,{\rm GeV}$.
Analogous relaxation applies to the lower bound on the reheat temperature $T_{\rm reh}$.
Notice that  in the limit of no washout, for $K_1\rightarrow 0$,
the usual unflavored lower bound is recovered. In the top-right panel
$M_2=30\,M_1$, such that the extra-term $\Delta\ve_{1\a}$ is suppressed
enough not to be able to relax the lowest bound on $M_1$.  In the bottom panels
we allowed $|\o_{21}|$ to be as large as 10 while $|\o_{31}|, |\o_{32}| \lesssim 1$.
One can see that the effect is much more reduced since now $\D\ve_{1\a}$ increases
but at the same time $K_1$ and the washout increase as well.

\section{On the validity of the $N_1$-dominated scenario}

In general, the final asymmetry receives a contribution from
the decays of all 3 RH neutrinos: $N_{B-L}^{\rm f}=\sum_i\,N_{B-L}^{\rm f}(N_i)$.
In this Section we want to find a condition that guarantees the validity of the
bounds within the  $N_1$-dominated scenario, where
$N_{B-L}^{\rm f}\simeq N_{B-L}^{\rm f}(N_1)$.

If $M_3\gtrsim 10^{12}\,{\rm GeV}$,  then the asymmetry produced from
the heaviest RH neutrino decays occurs always in the unflavored regime.
If moreover  we impose $M_3\gg M_2$, then
it can be always safely neglected since the total $C\!P$ asymmetry
 $\ve_3\propto (M_{1,2}/M_3)^2$ is strongly suppressed
and at the same time a strong washout from $N_1$ and $N_2$ is unavoidable \cite{geometry}.

On the other hand, $\ve_2$ is not necessarily suppressed.
A condition that guarantees the possibility to neglect
$N_{B-L}^{\rm f}(N_2)$ in the determination
of the bounds is equivalent to impose
\be\label{condition2}
\eta_B(N_2)\equiv a_{\rm sph}\,{N_{B-L}^{\rm f}(N_2) \over N_{\gamma}} \ll \eta_B^{\rm CMB} \, .
\ee
From the Eq.~(\ref{CPas}) and using the orthogonal parametrization,
the $C\!P$  asymmetry $\ve_2$  can be recast as \cite{geometry}
\be\label{ve2beta}
\ve_2 = \overline{\ve}(M_2)\,\left[\b_{23}(m_1,\O)+
{4\over 3}\,{M_1^2\over M_2^2}\,\ln\left({M_2\over M_1}-1\right)
\b_{21}(m_1,\O)\right] \,,
\ee
where we defined
\be
\b_{ij}(m_1,\O)\equiv {{\rm Im}[\sum_h \, m_h\,\O^{\star}_{hi}\,\O_{hj}]^2
\over \mti \,m_{\rm atm}} \, .
\ee
There are two special cases for which the $N_1$-dominated scenario is certainly valid.
The first is to have $\O=R_{13}$ (i.e. $\o_{21}=\o_{32}=0$), since in this case
it is simple to see that $\ve_2=0$. The second case is the limit
$M_3\gg 10^{14}\,{\rm GeV}$. In this case $\O$ is
given by the Eq.~(\ref{2effRH}) and it is easy to see that $\ve_2=0$.
This result can be understood  considering that in this limit
the heaviest RH neutrino decouples and so necessarily the interference term
$(m_D^{\dagger}\,m_D)_{23}\rightarrow 0$.

The opposite case is for $\O=R_{23}$, since now one has $\ve_1=0$ while
\be
\ve_2\leq \bar{\ve}(M_2)\,{m_3-m_2\over m_{\rm atm}}\,f(m_2,\mtt) \, ,
\ee
the same maximum value holding for $\bar{\ve}_1$ (cf. Eq.~(\ref{beta}))
but where $(m_1,M_1,\mt)$ are replaced by $(m_2,M_2,\mtt)$.
Notice moreover that for $\O=R_{23}$ one has
$m_1\ll m_{\star}$, so that the asymmetry produced from $N_2$-decays is
certainly not washed out by $N_1$-inverse decays.
In this situation a $N_2$-dominated scenario is realized, with
$N_{B-L}^{\rm f}\simeq N_{B-L}^{\rm f}(N_2)$ and with no
lower bound on $M_1$ \cite{geometry}.

Between these two well-defined special cases,
one has to take into account both a contribution $N_{B-L}^{\rm f}(N_1)$
from $N_1$-decays and a contribution $N_{B-L}^{\rm f}(N_2)$
from $N_2$-decays. For example, choosing $\O=R_{12}$ one has
\be
\ve_2={4\over 3}\,{M_1\over M_2}\,\ve_1 \, .
\ee
Assuming that $M_2\gtrsim 5\times 10^{11}\,{\rm GeV}$, such that the
asymmetry from $N_2$-decays is produced in the unflavored regime
and that  $M_2\gg M_1$, one can see that
even neglecting the washout from $N_1$-inverse processes the contribution
$N_{B-L}^{\rm f}(N_2)$ from $N_2$-decays can be safely neglected in the
determination of the bounds.

This example shows that if the first term in the Eq.~(\ref{ve2beta})
 $\propto {\rm Im}[(m^{\dagger}_D\,m_D)_{23}]^2$
vanishes, then $\eta_B(N_2)$ can be neglected in the determination of
the bounds in the hierarchical limit where $M_2\gg M_1$.
Therefore, a condition ${\rm Im}[(m_D^{\dagger}\,m_D)_{23}]^2=0$ certainly
guarantees the validity of the $N_1$-dominated scenario
but is quite a restrictive one. However, this condition enlightens that
an asymmetry generated from $N_2$-decays requires an interference of the heaviest
RH neutrino $N_3$ with $N_2$. Indeed, as we said, this condition is
certainly verified in the limit $M_3\gg 10^{14}\,{\rm GeV}$,
when the heaviest RH neutrino decouples and $(m_D^{\dagger}\,m_D)_{23}=0$.

Now we want to see whether, starting from $\O=R_{13}$,
one can turn on a rotation $R_{12}$ ($\o_{21}\neq 0$) still having
a negligible $\eta_B(N_2)$. Since both for $\O=R_{13}$
and $\O=R_{12}$ one has $(m_D^{\dagger}\,m_D)_{23}=0$, one could
naively think that this is still true for $\O=R_{12}\,R_{13}$ and therefore
that $\o_{32}= 0$ is a sufficient condition for the validity of the
$N_1$-dominated scenario. However it is easy to check  it is not true that
$(m_D^{\dagger}\,m_D)_{23}=0$ and therefore it is not
guaranteed that the asymmetry from $N_2$-decays can be neglected.
This has to be done by inspection. Since
we are assuming the hierarchical limit, $M_2\gtrsim 3\,M_1$,
the calculation of $N^{\rm f}_{B-L}(N_2)$ factorizes in two terms
\be\label{NmLN2}
\eta_B(N_2) = \left. \eta_B(N_2)
\right|_{T\sim T_B(K_2)}\, \times w_1(T\sim M_1) \, .
\ee
The first term is the asymmetry produced at $T_B\simeq M_2/z_B(K_2)$ from
$N_2$-decays, while the second term is the washout from $N_1$-inverse processes.

A precise calculation of $w_1(T\sim M_1)$ has to take into account two types of flavor effects.
If $M_1\lesssim 10^9\,{\rm GeV}$, then the  asymmetry produced
from $N_2$-decays is fully projected on the three-flavor basis at the time
when the asymmetry is washed out by $N_1$-inverse processes  \cite{vives}.
Assuming moreover that $M_2\gtrsim 5\times 10^{11}\,{\rm GeV}$,
so that the asymmetry from $N_2$ decays is produced in the unflavored regime,
the contribution to the final asymmetry from $N_2$-decays can be calculated as
\be\label{w1}
N^f_{B-L}(N_2)= \ve_{2}\,\k(K_2) \sum_\a\,P^0_{2\a}\,\,
e^{-{{3\pi}\over 8}\,P^0_{1\a}\,K_1} \,.
\ee
On the other hand, if $M_2\lesssim 5\times 10^{11}\,{\rm GeV}$,
then the asymmetry is produced in the fully flavored regime and
\be
N^f_{B-L}(N_2)= \sum_\a \ve_{2\a}\,\k(K_{2\a}) \,
e^{-{{3\pi}\over 8}\,P^0_{1\a}\,K_1}\,.
\ee
Notice, however, that since we are interested in finding the
condition for the $N_1$-dominated scenario to hold, then
$M_1\gtrsim 10^9\,{\rm GeV}$ and at the time
of the washout from $N_1$-inverse processes the asymmetry is projected
on a two-flavor basis:
the $\tau$ flavor and an orthogonal combination of electron and muon flavors.
In this situation $N_1$-inverse processes have the effect to further project
the asymmetry stored in the $e+\mu$ flavor  on a two-flavor basis where
one direction is determined by $l_1$ and the other is the orthogonal component
\cite{bcst,nardinir}.
The washout from $N_1$-inverse decays does not touch this orthogonal component
and therefore, accounting for this effect, the value of the final asymmetry has to lie
somewhere between the value of the asymmetry produced at $T\sim T_B(K_2)$  and the
value of the asymmetry calculated neglecting the effect of projection along ${\ell}_1$,
\be
\left. N^{\rm f}_{B-L}(N_2)\right|_{T\sim z_B(K_2)}\, \times w_1(T\sim M_1)
\lesssim
N^{\rm f}_{B-L}(N_2)
\lesssim
\left. N^{\rm f}_{B-L}(N_2)\right|_{T\sim z_B(K_2)}  \, .
\ee
As already mentioned,  two special cases where the validity of the
$N_1$-dominated scenario is guaranteed are $M_3\gg M_2\gtrsim 10^{14}\,{\rm GeV}$
and $\O=R_{13}$. In Fig.~\ref{fig:N2DS} we show the results obtained in a more general case.
We allow $\O=R_{12}(\o_{21})\,R_{13}(\o_{31})$ where $\o_{21}\neq 0$ but still $\o_{32}=0$.
In the left panels we show the results in the plane $(M_2,\eta_B(N_2))$.
The final asymmetry is calculated both neglecting the washout (red points) and
with the washout calculated in the two-flavor regime (green points)
but neglecting the effect of projection of the asymmetry due to
$N_1$-inverse processes: as we said an account of this effect should
give a result that has to be somehow in between.
\begin{figure}
\psfig{file=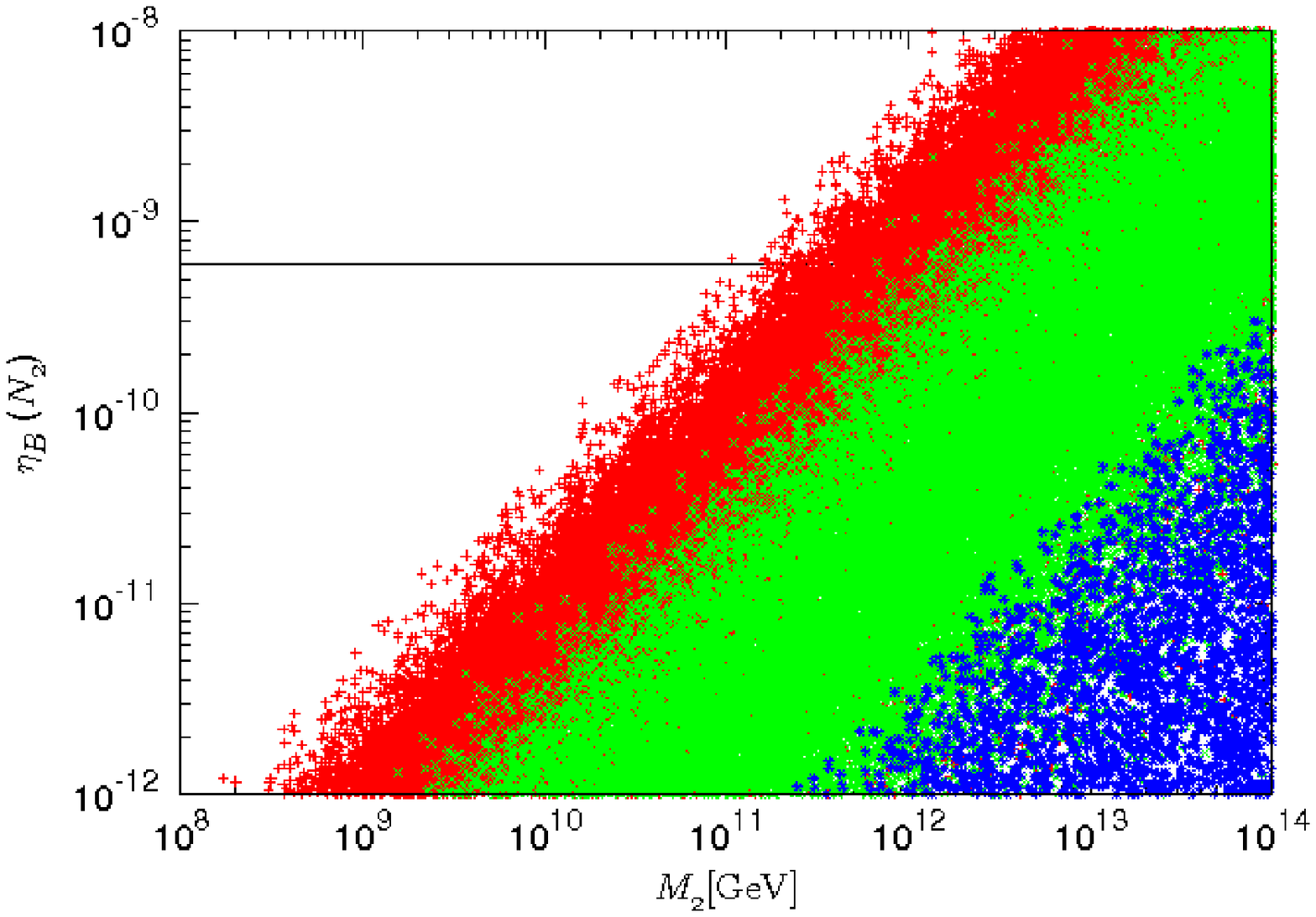,height=75mm,width=75mm,angle=-90}
\hspace{5mm}
\psfig{file=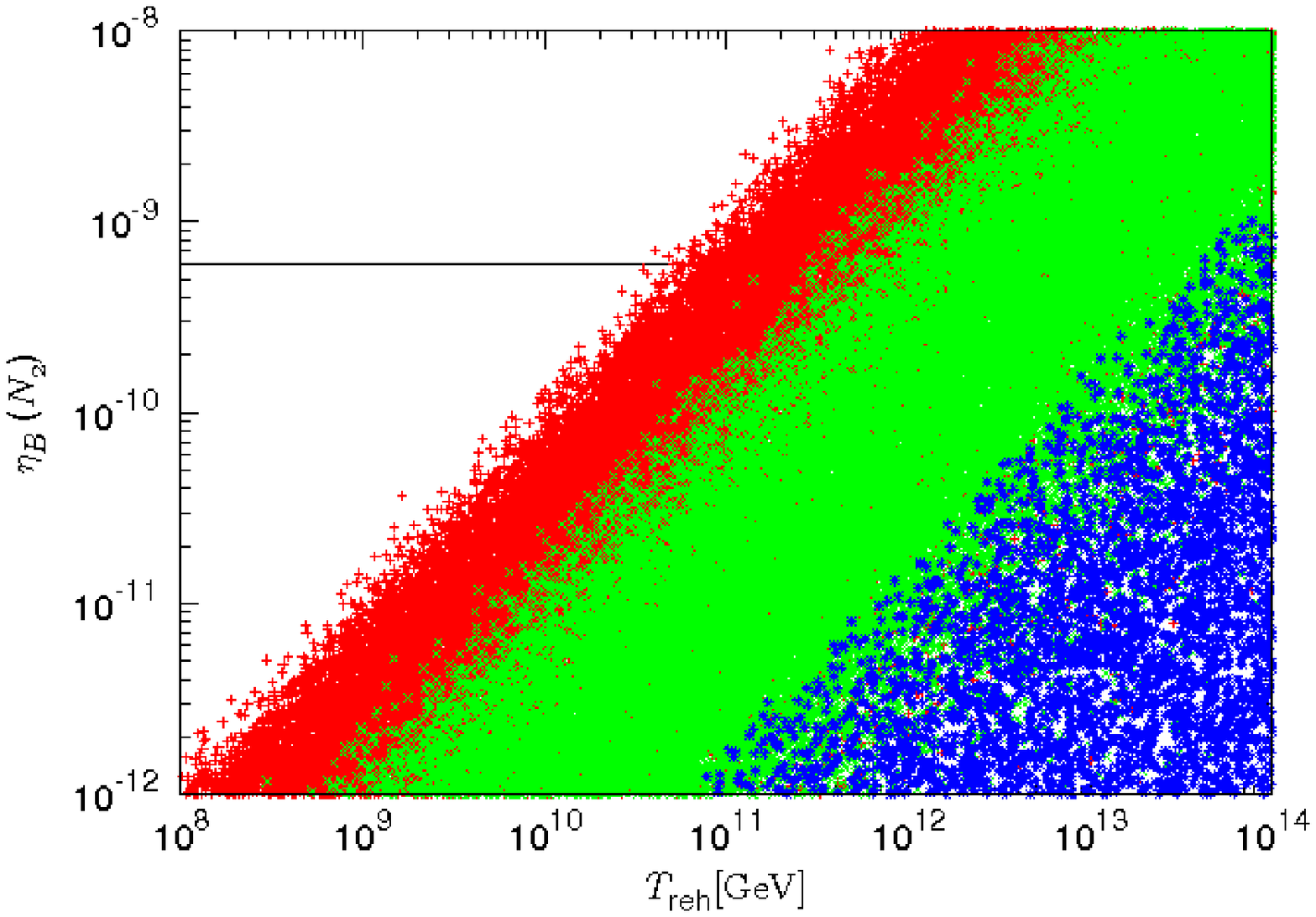,height=75mm,width=75mm,angle=-90} \\
\psfig{file=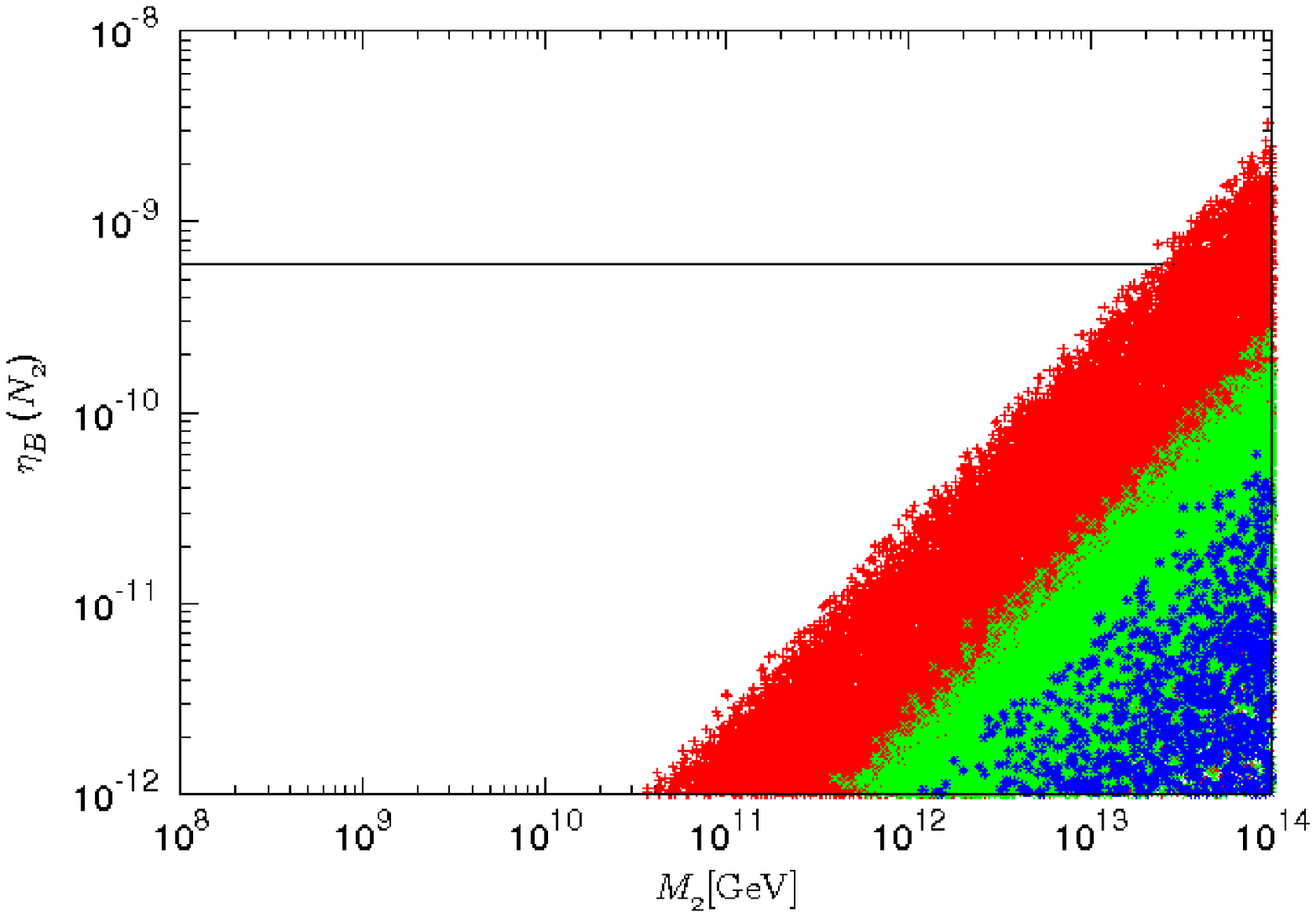,height=75mm,width=75mm,angle=-90}
\hspace{5mm}
\psfig{file=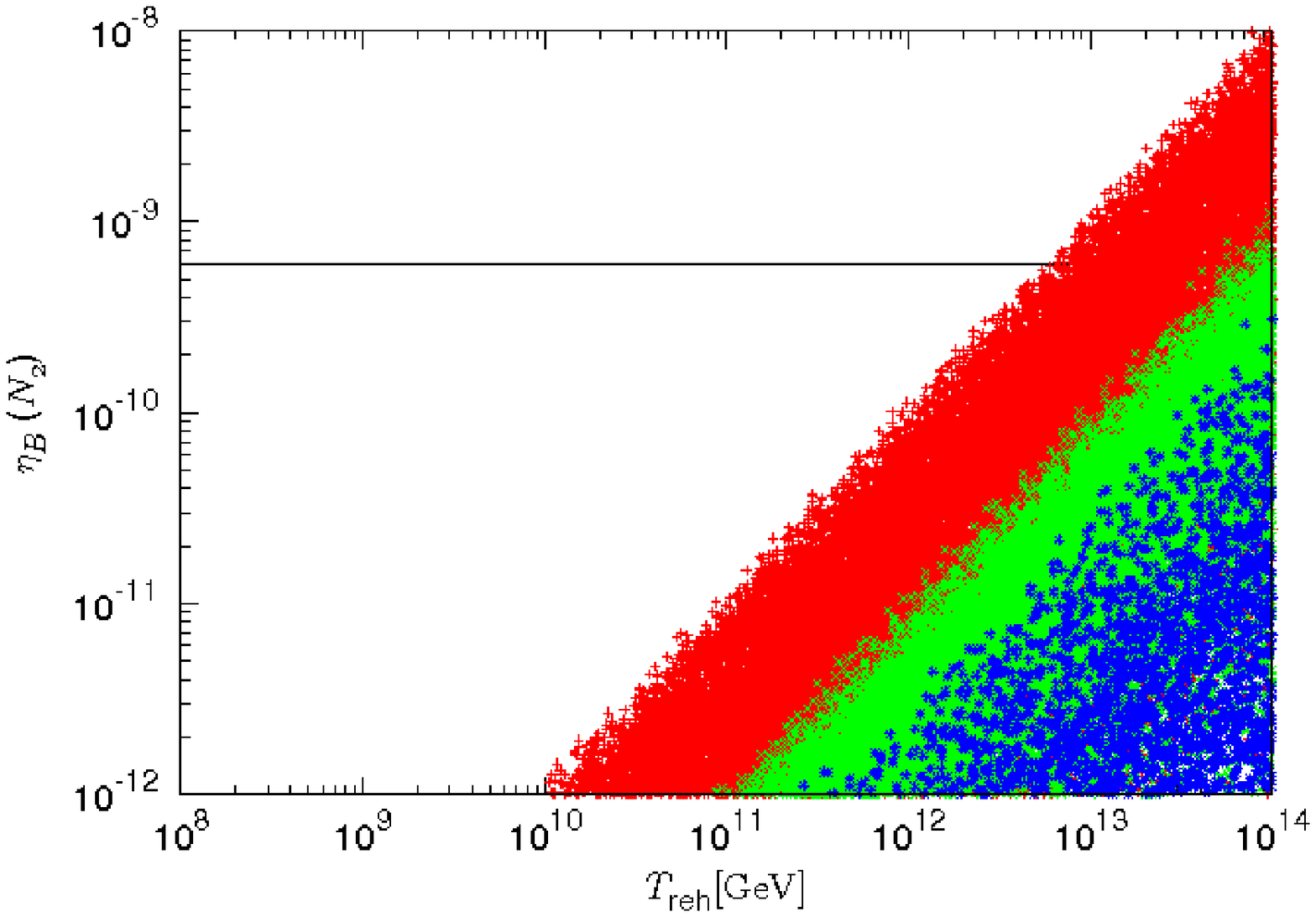,height=75mm,width=75mm,angle=-90}
\caption{final asymmetry from $N_2$-decays versus $M_2$ (left)
and versus the reheat temperature $T_{\rm reh}$ (right) for
$|\o_{32}|=0$ and $\o_{31}\leq 1$. In the top panels $\o_{21}\leq 1$, while
in the bottom panels $\o_{21}\leq 0.1$. The final asymmetry is obtained
neglecting the washout term $w_1(T\sim M_1)$ in the Eq.~(\ref{NmLN2}) (red points),
calculating it taking into account flavor effects (cf. Eq.~(\ref{w1})) (green points)
and neglecting them (blue points).}
\label{fig:N2DS}
\end{figure}
One can see in the top-left panel that for $M_2\lesssim 10^{11-12}\,{\rm GeV}$
the asymmetry produced by $N_2$-decays falls below the observed one indicated by
the horizontal solid line. Here we are imposing $\o_{21}\leq 1$
and we are assuming the reheat temperature to be
higher than $\sim T_B(N_2)\sim M_2/z_B(K_2)$. This result can be actually also
translated into a condition on the reheat temperature given by
 $T_{\rm reh}\lesssim T_B(N_2)\sim M_2/z_B(K_2)$,
while $M_2$ this time is free.
The result is shown in the right panels of the same figure and one can
see that the lower bound on the reheat temperature is approximately
five time more relaxed compared to the lower bound on $M_2$.

In the bottom panels we imposed $|\o_{21}|\lesssim 0.1$. One can see that even
for such small angles there is still a marginal allowed region. Therefore, only for
very small $|\o_{21}|$ values one recovers the special case $\O=R_{13}$.

This result shows that while for $\O=R_{13}$ or for $\O=R_{12}$ the asymmetry
from $N_2$-decays is always suppressed and the $N_1$-dominated scenario holds,
as soon as one allows either $\o_{31}\neq 0$ or $\o_{21}\neq 0$ or both,
the asymmetry produced from $N_2$-decays can explain the observed asymmetry for
acceptable values of $M_2$. Even though we did not perform a systematic calculation
of the final asymmetry from $N_2$-decays in the whole  parameter space, in the light of this result
it emerges that, if $10^{11}\,{\rm GeV}\lesssim M_2\lesssim 10^{14}\,{\rm GeV}$, then
the $N_2$-dominated scenario seems to be the most natural choice and
the validity of the $N_1$-dominated scenario relies on unnaturally small
complex angles in order to suppress ${\rm Im}[(m_D^{\dagger}\,m_D)_{23}]^2$.
We can therefore confirm even in a stronger way what we said in the introduction:
it is misleading to talk of
a lower bound on $M_1$ in leptogenesis, while it is more correct to talk
of a lower bound on $T_{\rm reh}$ holding in the hierarchical limit for
$M_2\gtrsim 3\,M_1$.

We also notice that this result is not relying crucially on
an exact calculation of the washout from $N_1$-inverse processes
but more on the calculation
of the asymmetry produced at $T\sim M_2$. In the same Fig.~\ref{fig:N2DS}
we show (red points) the asymmetry
$\left. \eta_B(N_2)\right|_{T\sim T_B(K_2)}$, without the washout, while
the green points take into account the washout term $w_1(T\sim M_1)$ calculated
with the Eq.~(\ref{w1}). We also show (blue points) the results that would have
been obtained calculating the washout neglecting flavor effects.

One can see that the main result relies on the observation that
$\left. \eta_B(N_2)\right|_{T\sim T_B(K_2)}$ is large already when
small $|\o_{21}|$ are turned on. A proper calculation of the washout
is an important step but somehow a secondary one. Certainly a proper account
of flavor effects greatly enhances the asymmetry from $N_2$-decays
but primarily it is important that there is an asymmetry to be washed out.
Notice that an account of the effect envisaged in \cite{nardinir}
would produce results somewhere in between the green and the red points.
It is certainly important for a precise evaluation but
it does not seem to play a crucial role.

Our results show that there is a continuous increase of the asymmetry
going from $\O=R_{13}$, where it vanishes,
toward $\O=R_{23}$. What is important is that
as soon as one switches on small $|\o_{21}|$ or $|\o_{31}|$ the contribution
from $N_2$-decays is sufficient to explain the observed asymmetry. Going
toward the case $\O=R_{23}$ ($\o_{21}=\o_{31}=0$), the allowed region for the
$N_2$-dominated scenario increases and the lower bound on $M_2$ relaxes
down to a few's $\times 10^{10}\,{\rm GeV}$ \cite{geometry}. We do not see any
discontinuity or qualitatively different regime as envisaged in \cite{review}
where the authors distinguish a decoupled regime from a strong washout
regime. A proper evaluation of the washout term  $w_1(T\sim M_1)$, taking into account
all kinds of flavor effects \cite{vives,nardinir} is in any case an important
ingredient for a correct determination of the border between
the domain of validity of the $N_1$-dominated
scenario and that one of the  $N_2$-dominated scenario.

\section{Beyond the hierarchical limit}

Finally, in this section we discuss how the
bounds change going beyond the hierarchical limit,
when $\delta_2\equiv (M_2-M_1)/M_1\lesssim 2$.
We first neglect flavor effects induced by charged lepton interactions.
Notice that there is a second type of flavor effects in the heavy neutrino
sector itself \cite{bcst,hambye,nardinir} due to the fact that for example
the lepton quantum state $|l_2\rangle $ produced by a RH neutrino $N_2$
does not coincide with $|l_1\rangle$ produced by a RH neutrino $N_1$
and in particular $|l_2\rangle$
does not in general inverse decays with a Higgs
to produce a $N_1$ with the same rate as $|l_1\rangle$.
We will neglect this effect that would imply
that washout terms do not simply add up in the Eq.~(\ref{unflke}).

Under these assumptions, there are three different effects that change
the bounds compared to the hierarchical limit \cite{beyond}.
First of all, in general, one cannot neglect the contribution from the heavier RH
neutrinos. Therefore, one cannot just relax the assumption of hierarchical spectrum
without also considering the heavier RH neutrino decays. Second, now the addition of washout
in the Eq. (\ref{unflke}) cannot be neglected.
Third, the total $C\!P$ asymmetries get typically enhanced compared to their value
in the hierarchical limit. The first and third effect tend to increase the
final asymmetry relaxing the bounds but the second effect tends to reduce the final
asymmetry, making the bounds more stringent.
The first two effects saturate for $\d_2\lesssim 0.01$, the so called degenerate limit,
and therefore when $\d_2$ decreases below $0.01$ only the third effect is left and it
changes the bounds in quite a simple way \cite{beyond},
except for some effects studied in \cite{quantumres} in the extreme case of resonant leptogenesis.
However, notice that in the case of resonant leptogenesis the bounds simply disappear.
Therefore, here we focus especially on the transition between the hierarchical
limit and the degenerate limit, for  $0.01\lesssim\d_2\lesssim 2$.
The final asymmetry is given by  $N^{\rm f}_{B-L}= \sum_i\,\ve_i\,\k_i^{\rm f}$,
where $\ve_1$ is given by the Eq.'s (\ref{ve1xi})-(\ref{extra}) while, from the Eq. (\ref{CPas}),
the total $C\!P$ asymmetries $\ve_2$ and $\ve_3$ are given by
\be\label{ve2}
\ve_2={3\over 16\pi}\, \sum_{j\neq 2}\,{{\rm Im}\,
\left[(h^{\dagger}\,h)^2_{2j}\right] \over
(h^{\dagger}\,h)_{22}} \,{\xi(x_j/x_2)\over \sqrt{x_j/x_2}}
\hspace{8mm}
{\rm and}
\hspace{8mm}
\ve_3=
{3\over 16\pi}\, \sum_{j\neq 3}\,{{\rm
Im}\,\left[(h^{\dagger}\,h)^2_{3j}\right] \over
(h^{\dagger}\,h)_{33}} \,{\xi(x_j/x_3)\over \sqrt{x_j/x_3}} \, .
\ee
The unflavored efficiency factors are given by the Eq.~(\ref{efial}) with $P^0_{i\a}=1$,
\be
\k_i^{\rm f}(K_j,\d_2,\d_3)=
-\int_{z_{\rm in}}^\infty\,dz'\,{dN_{N_i}\over dz'}\,
{\rm e}^{-\int_{z'}^\infty\,dz''\,[\D W(z'')+\sum_j\,W_j^{\rm ID}(z'';K_j)]} \, ,
\ee
where we defined $\d_3\equiv (M_3-M_1)/M_1$.
Let us assume that the $N_i$-abundances track closely the equilibrium value,
so that $dN_{N_i}/dz \simeq dN_{N_i}^{\rm eq}/dz$. This approximation works
well for $K_i\gtrsim 1$, as we will assume in the following.
An approximated expression for $\k_i^{\rm f}$ is  then obtained
\be \label{kisss}
\k_i^{\rm f}(K_j,\d_2,\d_3)  \simeq
 - \int_{0}^{\infty}\, dz'\
{dN_{N_i}^{\rm eq}\over dz'}\,
{\rm exp}\left\{-\,\int_{z'}^{\infty}\ dz''\,[\D\,W(z'')+\sum_j\,W_j^{\rm ID}(z'')]\right\}\, .
\ee
In \cite{geometry,beyond} it was shown that conservatively for
\be
\d_i \gtrsim \d_{\rm HL} \equiv  {z_B(K_i)+2\over z_B(K_1)-2}-1 \, ,
\ee
one can assume the hierarchical limit for the calculation of $\k_i^{\rm f}$.
In this case the washout from the RH neutrinos
$l\neq i$ lighter than $N_i$ is factorized and (cf. Eq.~(\ref{k}))
\be
\k_i^{\rm f} \simeq \,
\k(K_i)\,e^{-\,\int_{0}^{\infty}\ dz'\,\sum_l\,W_{l}^{\rm ID}(z')} \,  .
\ee
On the other hand the washout from the $N_i$-inverse processes on
the asymmetry produced by $N_l$-decays is negligible and
\be
\k_{l_{\star}}^{\rm f}\simeq
-\int_{z_{\rm in}}^\infty\,dz'\,{dN_{N_{l_{\star}}}\over dz'}\,
{\rm e}^{-\int_{z'}^\infty\,dz''\,
[\D W(z'')+\sum_{l}\,W_{l}^{\rm ID}(z''; K_{l})]} \, ,
\ee
where with $N_{l_{\star}}$ we indicated one particular $N_l$.
Typically one has $\d_{\rm HL}\simeq 2$ and this is the reason why we always
used $M_2\gtrsim 3\,M_1$ as a condition for the hierarchical limit to hold.
On the contrary, if $(M_i-M_l)/M_i \lesssim 0.01$, then one recovers the
degenerate limit where $\k_i^{\rm f}\simeq \k_l^{\rm f}(K_i+K_l,K_{j\neq i,l})$.
There are three different cases: a partial degenerate limit with
$\d_2\lesssim 0.01$, a partial degenerate limit with $(M_3-M_2)/M_2 \lesssim 0.01$
and a full degenerate limit where $\d_3\lesssim 0.01$.
In this last case one has simply
\be
\k_1^{\rm f}=\k_2^{\rm f}=\k_3^{\rm f} \simeq \k(K_1+K_2+K_3) \, .
\ee
We will now focus on two particular choices for $M_3$
but without imposing any restriction on $\d_2$.
The {\em first case} we consider is $M_3\gg M_2\simeq M_1$. This was also studied in \cite{beyond}
but for two particular choices of the orthogonal matrix, while here we do not make any
assumption on $\O$.
The contribution from the heaviest RH neutrino is negligible
both because $\k_3^{\rm f}\ll \k_1^{\rm f}, \k_2^{\rm f}$
and because $\ve_3\ll \ve_1, \ve_2$. The washout of
$N_3$-inverse processes on the asymmetry produced from the
two lightest RH neutrino decays is negligible as well.

Two convenient fits for $\k_1^{\rm f}$ and $\k_2^{\rm f}$ can be used \cite{beyond},
\be\label{fit1}
\k_1^{\rm fit}(K_1,K_2,\d_2)={2\over z_B(K_1+K_2^{(1-\d_2)^3})\,
(K_1+K_2^{1-\d_2})}
\ee
and
\be\label{fit2}
\k_2^{\rm fit}(K_1,K_2,\d_2)={2\,\left[1-\d_2 \over \left(1-\d_2\right)^2\right]
\over z_{\rm B}(K_2+K_1^{(1-\d_2)^3})(K_2+K_1^{1-\d_2})}
\times {\rm e}^{-{3\pi\over 8}\,K_1 \left({\d_2\over 1+\d_2}\right)^{2.1}} \, .
\ee
For $M_3\gg M_2$ the Eq. (\ref{ve2}) for $\ve_2$ can be further specialized into
\be
\ve_2\,\simeq \,
\overline{\ve}(M_2)\,{{\rm Im}\left[\sum_h\,m_h\,\O^{\star}_{h2}\O_{h3}\right]^2\over \mtt\,m_{\rm atm}}
+\overline{\ve}(M_1)\,{{\rm Im}\left[\sum_h\,m_h\,\O^{\star}_{h2}\O_{h1}\right]^2\over \mtt\,m_{\rm atm}}\,
\xi\left({1\over x_2}\right) \, .
\ee
If $\d_2\ll 1$ and if one maximizes over the $\O$ parameters,
the first term can be neglected and since $\xi(1/x_2)\simeq -\xi(x_2)$,
the expression simplifies into
\be
{\rm max}_{\O}[\ve_2] \simeq -\overline{\ve}(M_1)\,\xi(x_2)\,\b_2(m_1) \, , \hspace{7mm}
\b_2(m_1) \equiv
{\rm max}_{\O}
\left[{{\rm Im}\left[\sum_h\,m_h\,\O^{\star}_{h2}\O_{h1}\right]^2\over \mtt\,m_{\rm atm}}\right]\, .
\ee
While $\ve_1$ gets suppressed when $m_1$ increases (cf. Eq.~(\ref{beta})),
the function $\b_2(m_1)$ in general does not and therefore, at large $m_1$,
it gives the dominant contribution to the maximum asymmetry determining
the bounds in the plane $(m_1,M_1)$ that are shown in Fig.~\ref{fig:lepdeglargeM3}
for $\d_2=1,0.1,0.05,0.01$ and where we imposed a strong washout condition.
This automatically guarantees the validity of the fits (\ref{fit1}) and (\ref{fit2})
for the efficiency factors. The results in the unflavored case correspond to the green points.
\begin{figure}
\begin{center}
\psfig{file=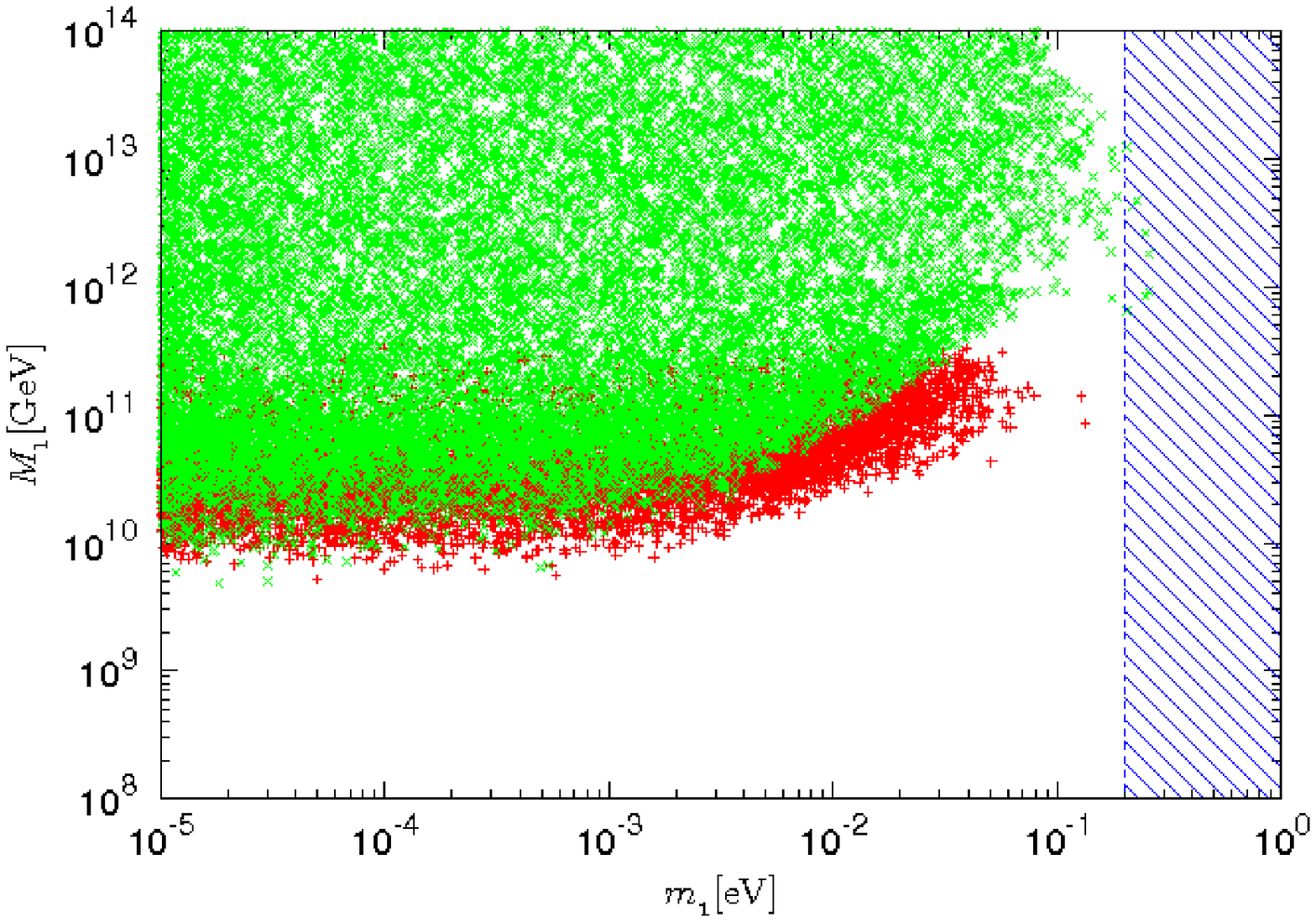,height=7cm,width=7cm,angle=-90}
\hspace{-1mm}
\psfig{file=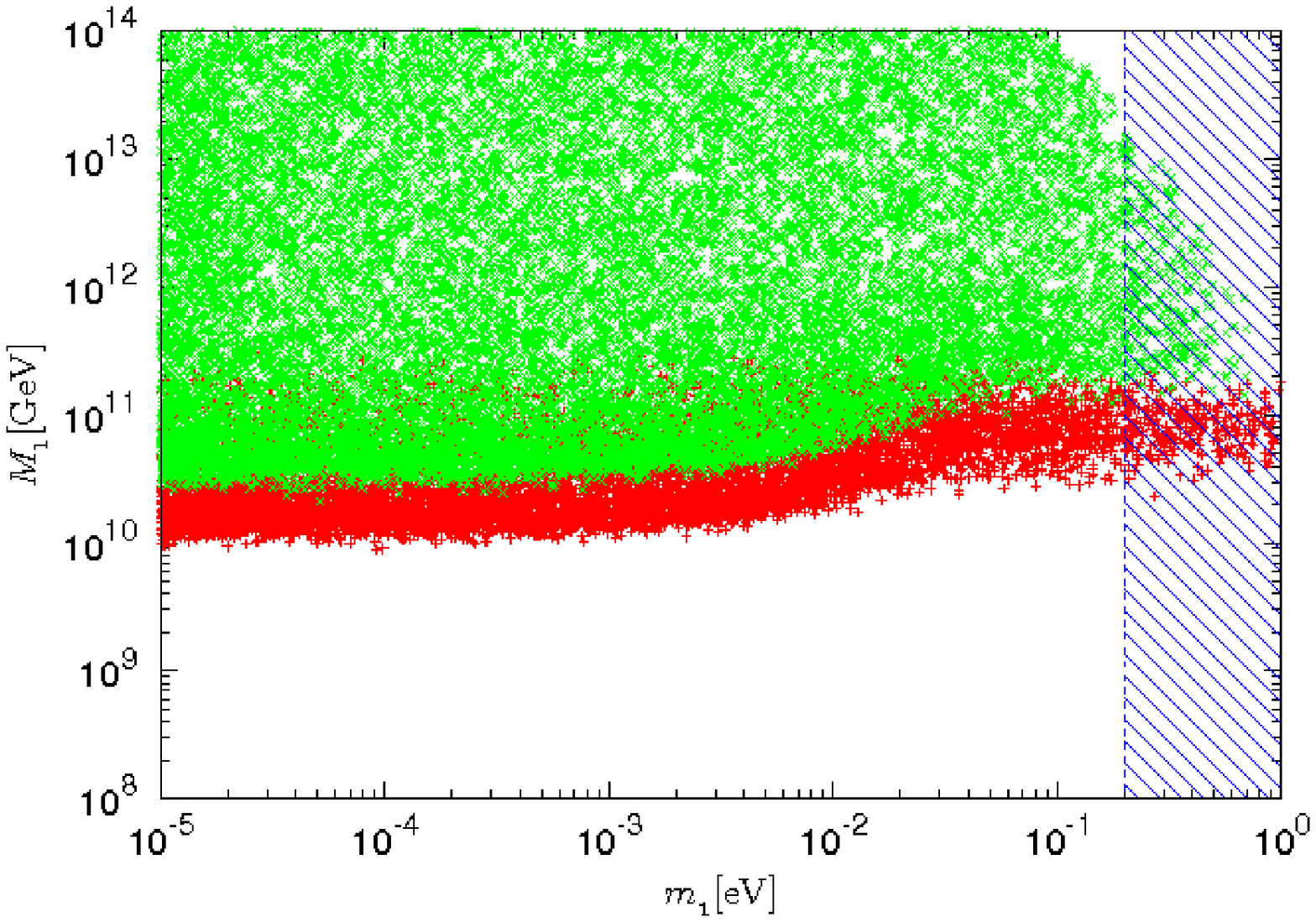,height=7cm,width=7cm,angle=-90}
\\
\psfig{file=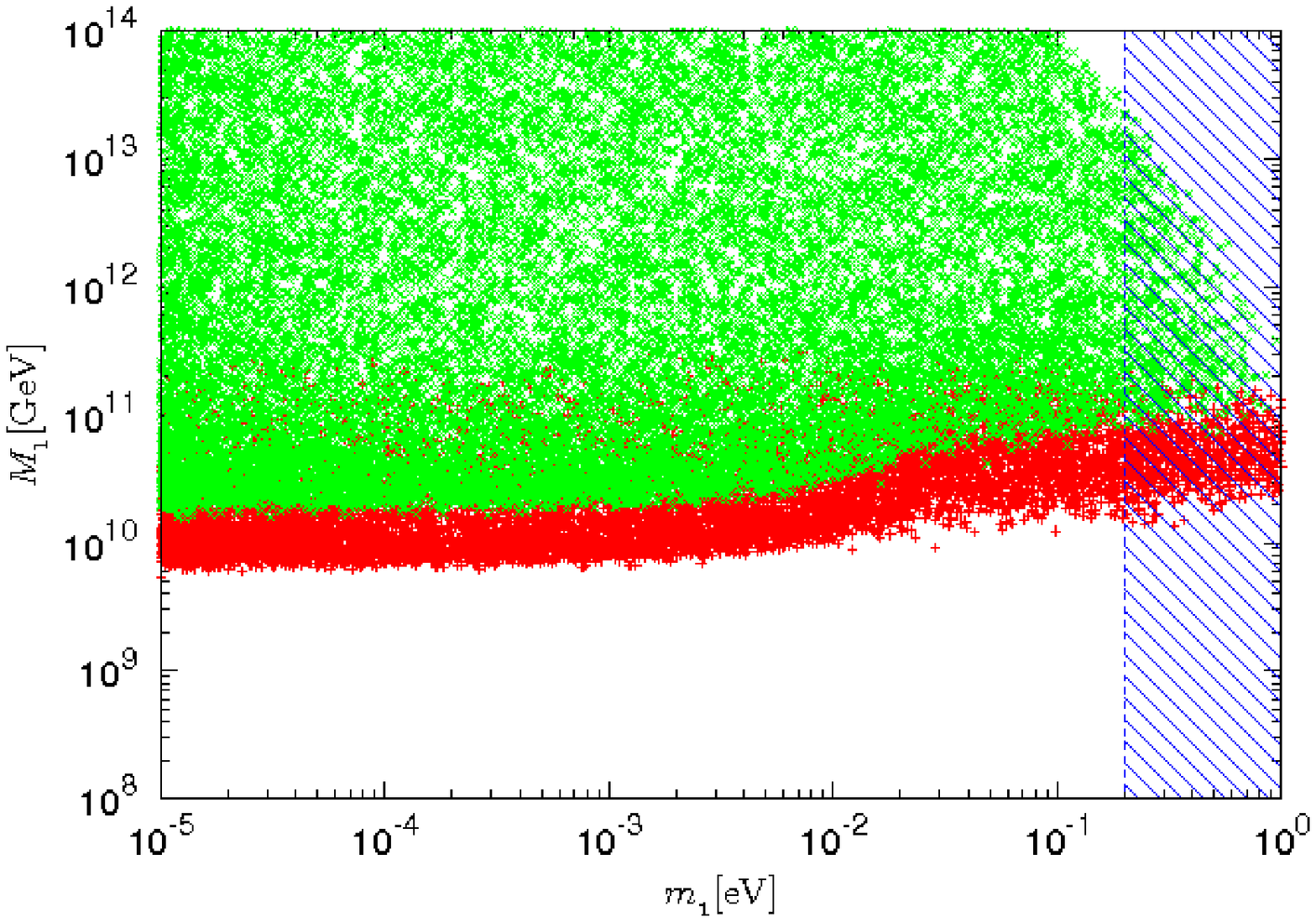,height=7cm,width=7cm,angle=-90}
\hspace{-1mm}
\psfig{file=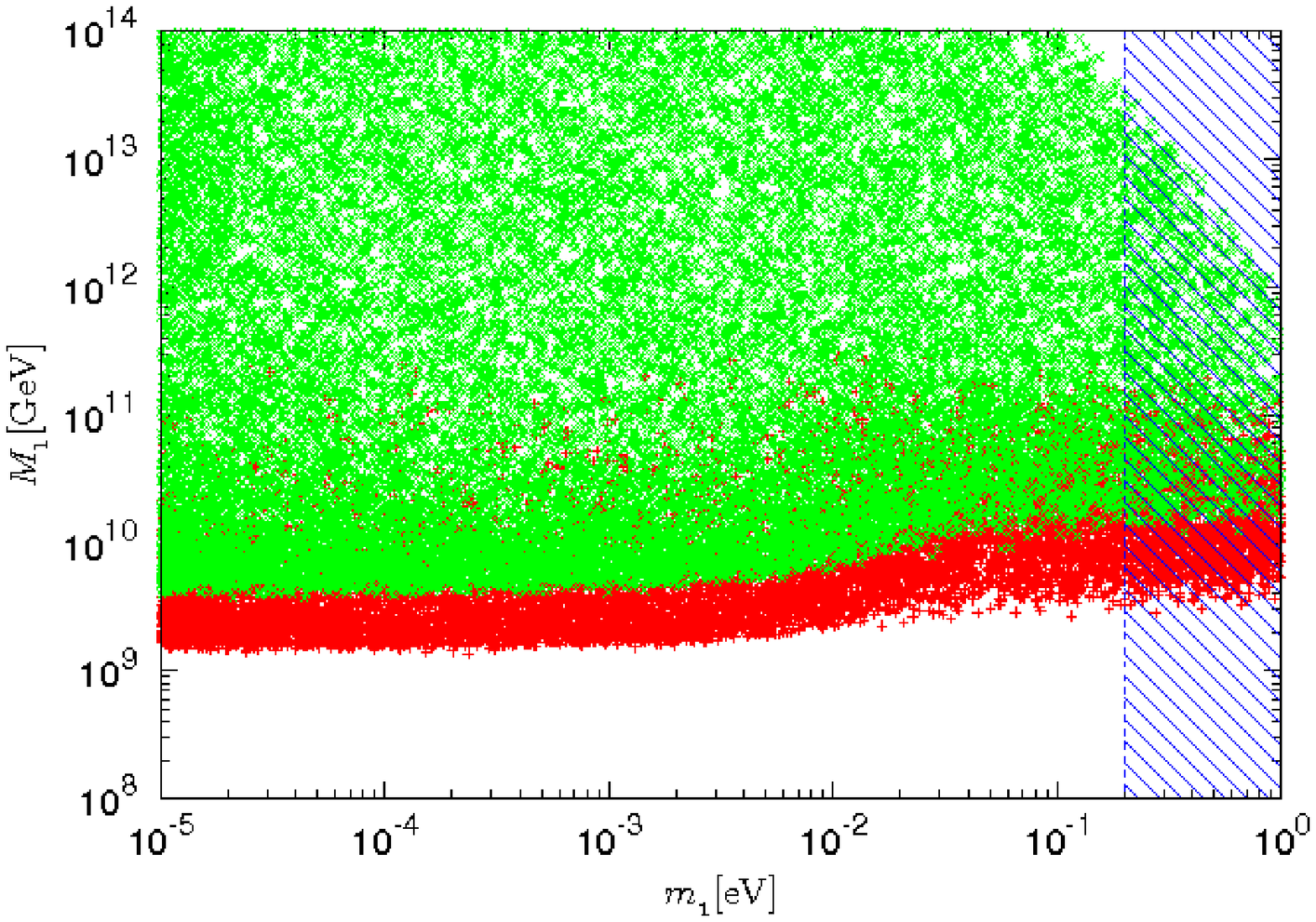,height=7cm,width=7cm,angle=-90}
\caption{Allowed region in the $(m_1,M_1)$ plane for the case $M_3\gg M_2$ in the
unflavored case (green points) and in the fully flavored regime
imposing the condition Eq.~(\ref{condition}) (red points) for
$\d_2=1$ (top-left), $0.1$ (top-right), $0.05$ (bottom-left)
and $0.01$ (bottom-right).}
\label{fig:lepdeglargeM3}
\end{center}
\end{figure}
One can notice a result already found in \cite{beyond}
that we confirm here in a more general way. For $\d_2\sim 0.1$
the washout addition makes the lower bound on $M_1$ even
slightly more stringent while only for $\d_2\lesssim 0.01$ the lower
bound gets clearly more relaxed. On the other hand, concerning
the upper bound on $m_1$, the fact that $\ve_2$ increases with $m_1$
implies that the upper bound on $m_1$ gets relaxed already at $\d_2\simeq 0.1$.

Let us now turn to study the {\em second case} when $M_3=M_2$.
This time the heaviest RH neutrinos contribute both to the
asymmetry production and to the washout.
Having the expressions for the case $M_3\gg M_2$,
it is easy to derive the efficiency factors in the
partial degenerate limit $(M_3-M_2)/M_2 \lesssim 0.01$.
Indeed if $i=2,3$, one has simply
\bea
\k_3^{\rm f}\simeq \k_2^{\rm f} & \simeq &
- \int_{0}^{\infty}\, dz'\
{dN_{N_{i}}^{\rm eq}\over dz'}\,\times \\  \nonumber
& &  \times \,
{\rm exp}\left\{-\,\int_{z'}^{\infty}\ dz''\,[\D\,W(z'')+W_1^{\rm ID}(z'';K_1)
+W_i^{\rm ID}(z'';K_2+K_3)]\right\} \\
& \simeq & \nonumber
\k_2^{\rm fit}(K_1,K_2+K_3,\d_2) \, .
\eea
The calculation of the total $C\!P$ asymmetries $\ve_2$ and $\ve_3$
is slightly more involved. If we assume that $M_2=M_3$ exactly, then
the interference term between $N_2$ and $N_3$ vanish,
though notice that the expression Eq.~(\ref{CPas}) for $\ve_2$
and $\ve_3$ diverge for $M_2=M_3$ because they hold
only for mass differences less than the decay widths.
Therefore, one has
\be
\ve_{i=2,3} \simeq \xi(x_2)\,\overline{\ve}(M_1)\,
\left[
{\sum_h \,m_h^2\,{\rm Im}[\O^{\star 2}_{hi}\,\O_{h1}^2] \over \mti\,m_{\rm atm}}
+2\,\sum_{h<l}\, m_h\,m_l\,
{{\rm Im}[\O_{h1}\,\O_{l1}\,\O^{\star}_{hi}\,\O^{\star}_{li}]\over \mti\,m_{\rm atm}}
\right] \, .
\ee
In passing let us notice that, without the interference
term between $N_2$ and $N_3$, the asymmetries
vanish for $M_1\ra 0$.
The final asymmetry can then be written as
\bea
N_{B-L}^{\rm f} & = & \xi(x_2)\,\overline{\ve}(M_1)\,
 \left\{\k_1^{\rm fit}(K_1,K_2;\d_2)\,\b(m_1,\O) +\k_2^{\rm fit}(K_1,K_2+K_3,\d_2)\,
 \times \right. \\
& & \nonumber  \\
& & \times  \nonumber
\left.\left[{\sum_h \,m_h^2\,{\rm Im}[\O^{\star 2}_{h2}\,\O_{h1}^2] \over \mtt\,m_{\rm atm}}
+{\sum_h \,m_h^2\,{\rm Im}[\O^{\star 2}_{h3}\,\O_{h1}^2] \over \mttt\,m_{\rm atm}} \right]\right. \\
& &
\left. +2\,\sum_{h<l}\, m_h\,m_l\,
{{\rm Im}[\O_{h1}\,\O_{l1}\,\O^{\star}_{h3}\,\O^{\star}_{l3}]\over m_{\rm atm}}
\left({\mtt-\mttt \over \mtt\,\mttt} \right)
\right\}
\, .
\eea
Notice that if $\mtt=\mttt$, then $\ve_2+\ve_3= (\mtt/\mt)\,\ve_1 \propto (m_3-m_1)$
and therefore the contribution from the two heavier RH neutrinos is suppressed as well.
This makes the bounds more stringent compared to the case $M_3\gg M_2$,
especially for $\d_2\sim 0.1$.
In Fig.~\ref{fig:lepdegM2eqM3} we show the bounds in the  case
 $M_2=M_3$ again for $\d_2=1,0.1,0.05,0.01$.
\begin{figure}
\begin{center}
\psfig{file=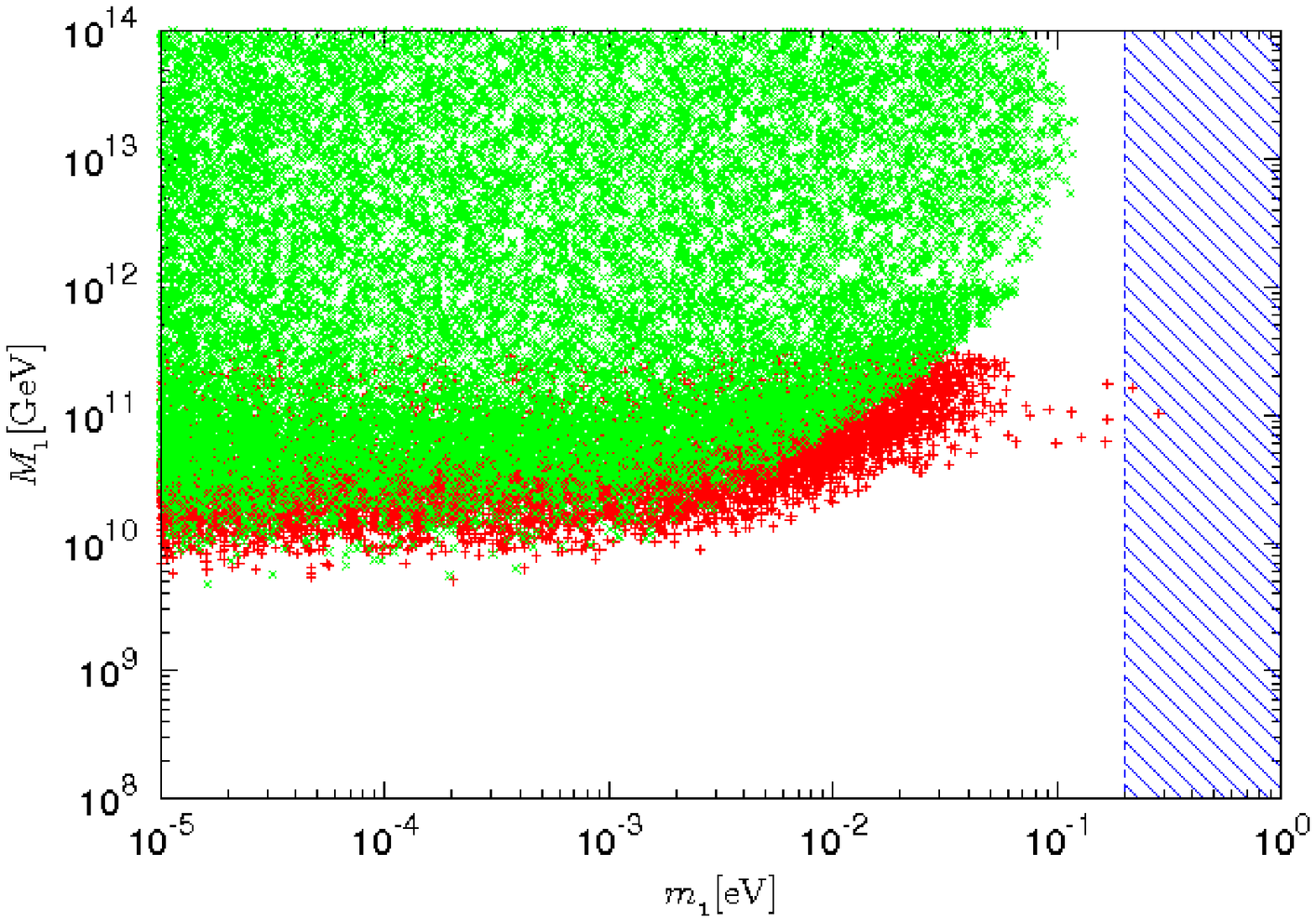,height=7cm,width=7cm,angle=-90}
\hspace{-1mm}
\psfig{file=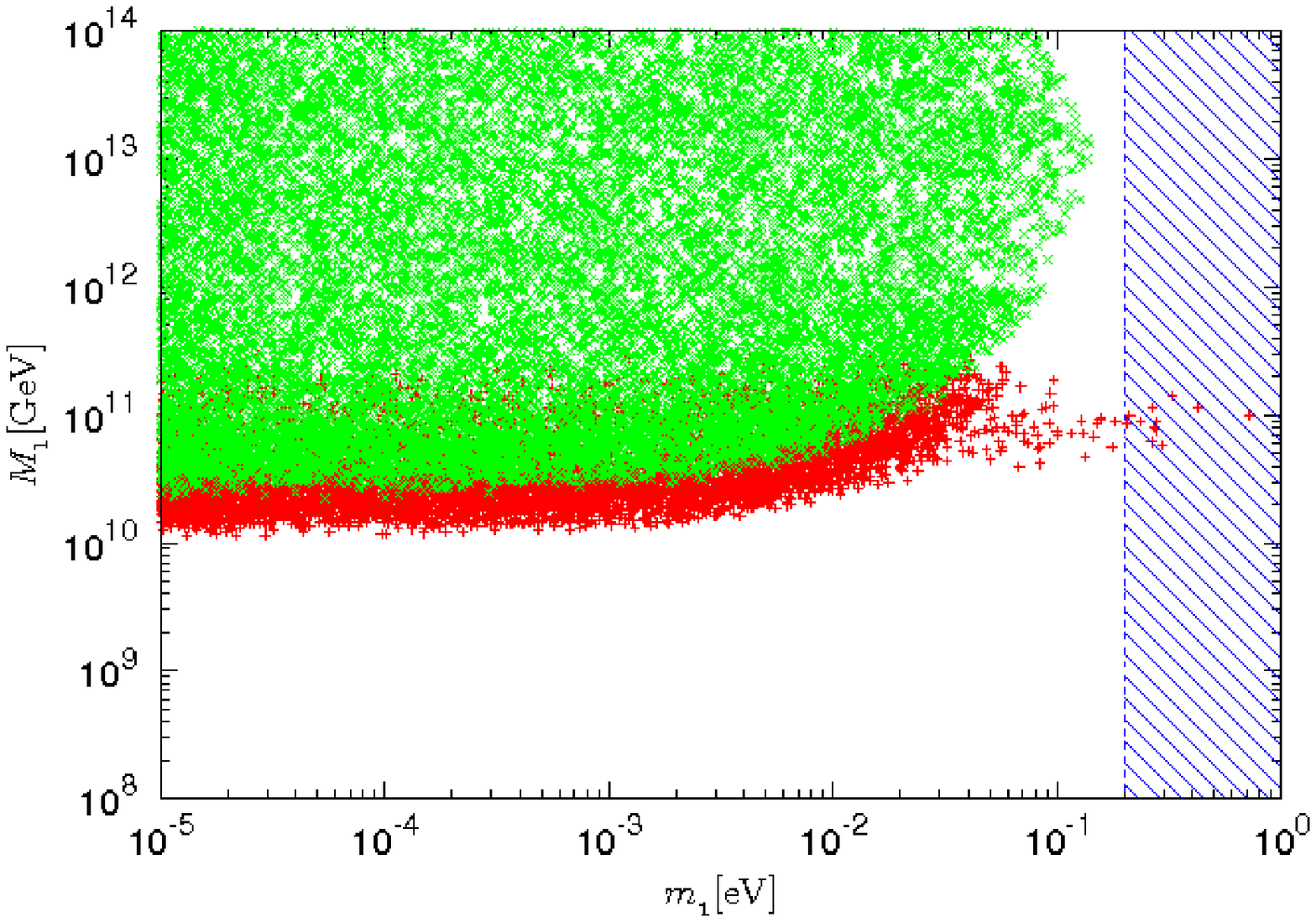,height=7cm,width=7cm,angle=-90}
\\
\psfig{file=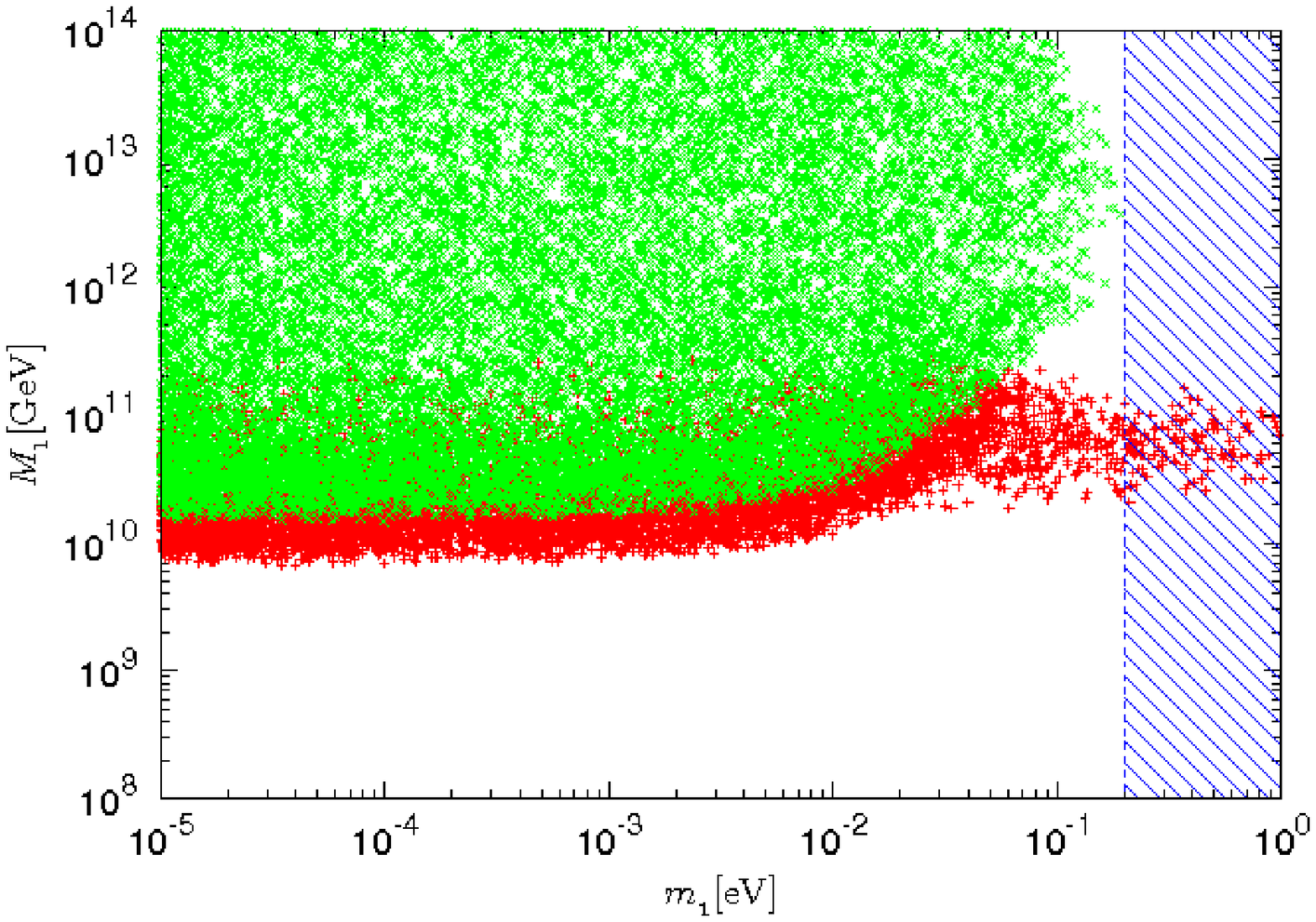,height=7cm,width=7cm,angle=-90}
\hspace{-1mm}
\psfig{file=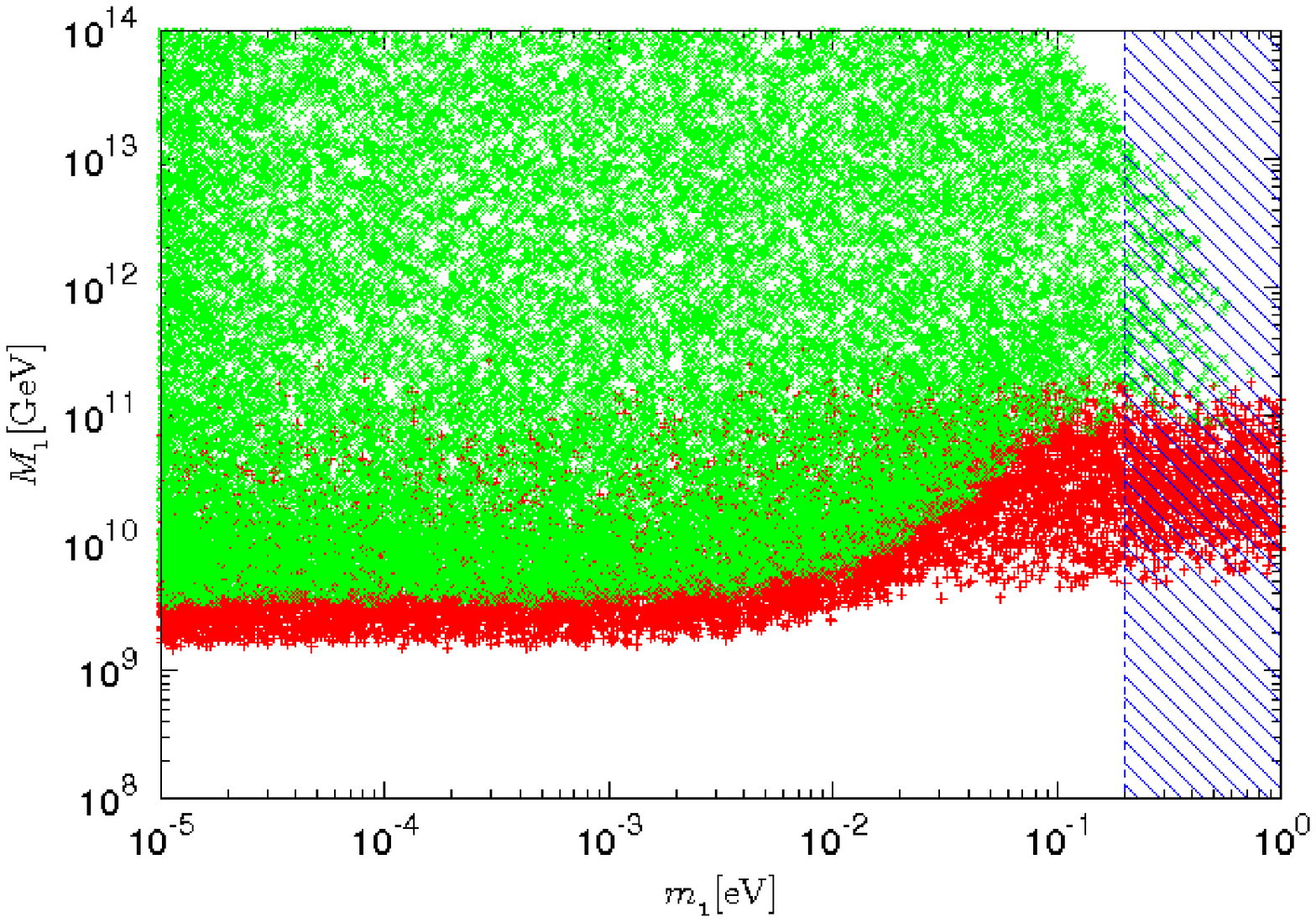,height=7cm,width=7cm,angle=-90}
\caption{Allowed region in the $(m_1,M_1)$ plane for $M_2=M_3$ in the
unflavored case (green points) and in the fully flavored regime
imposing the condition Eq.~(\ref{condition}) (red points) for
$\d_2=1$ (top-left), $0.1$ (top-right), $0.05$ (bottom-left)
and $0.01$ (bottom-right). We are also imposing $|\o_{ij}|\leq 1$
and $K_{\a i}\gtrsim 1$.}
\label{fig:lepdegM2eqM3}
\end{center}
\end{figure}
One can see that now for $\d_2\sim 0.1$ not only the
lower bound on $M_1$ does not get relaxed but also
the upper bound on $m_1$ (green points).
Again, for $\d_2\lesssim 0.01$,  both the $M_1$ lower
bound and the $m_1$ upper bound get clearly relaxed.

Both in the case $M_3\gg M_2$ and in the case $M_3=M_2$,
we repeated the calculations also in the fully flavored
regime. Now the final asymmetry has to be calculated using
\be \label{Nfi}
N_{B-L}^{\rm f}  =  \sum_{\a,i}\,\ve_{i\a}\,\kappa_{i\a}^{\rm f}
 \simeq
\sum_i N_{\rm fl}^i\,\bar{\ve}_i\,\k_i^{\rm f}
+{1\over 2}\,\sum_i\,\Delta P_{i\alpha}\,[\kappa_{i\a}^{\rm f}-\k_{i\b}^{\rm f}] \, ,
\ee
where we generalized the approximated form we already used in the
hierarchical limit (cf. Eq.~(\ref{Nf})).

In the case $M_3\gg M_2$, the contribution from $N_3$-decays
is this time less trivially negligible because in principle
the washout from the the two lighter RH neutrinos could
be weaker in one flavor and because moreover the flavored $C\!P$ asymmetries $\ve_{3\a}$,
contrarily to the total $\ve_3$, are not necessarily suppressed
in the hierarchical limit \cite{flavorlep}. However, it turns out
that the contribution from $N_3$-decays is too small to explain
the observed asymmetry and it can be therefore neglected in the
determination of the bounds.

Assuming $K_{\a i}\gtrsim 1$, expressions for the flavored efficiency factors can
be obtained from the unflavored ones just replacing $K_i\ra K_{i\a}\equiv P^0_{i\a}\,K_i$.
In this way we can  use the approximation (\ref{kisss}) and the fits
Eq.~(\ref{fit1}) and (\ref{fit2}) with $K_i$ replaced by $K_{i\a}$.

In Fig.~\ref{fig:lepdeglargeM3} one can see how flavor effects affect
the bounds (red points). For $\d_2=1$ there is no relaxation of the lower bound on $M_1$,
since the hierarchical limit  still holds and because at the onset of the strong washout
there is no relaxation  due to flavor effects for $|\o_{ij}|< 1$ \cite{flavorlep},
as we are imposing.
For $\d_2\ll 1$ and $m_1\lesssim 0.01\,{\rm eV}$ one can see that there is a factor two relaxation.
The reason is that the asymmetry is not maximized in a one flavor dominance case
but in a democratic case, that means for
values of the parameters where the only change to the final asymmetry from
flavor effects is described by the enhancement of a factor $N_{\rm fl}^i\simeq 2$,
while the additional terms  $\propto \D P_{i\a}\,(\k_{i\a}-\k_{i\b})$ vanish.
On the other hand, for $m_1\gtrsim 0.01\,{\rm eV}$,
one flavor dominance makes possible a large enhancement of
the asymmetry and this is why the upper bound on $m_1$ is
much more relaxed.
Similar results hold in the case $M_2=M_3$,
as one can see from Fig.~\ref{fig:lepdegM2eqM3}.

\section{Conclusions}

The simple vanilla leptogenesis scenario grasps
important features of leptogenesis bounds but misses
many important effects. Assuming the $N_1$-dominated scenario,
the lower bound on $M_1$ seems to be a solid feature and we have seen
that it resists even for heavy neutrino mass degeneracies as small as $\d_2 \sim 0.01$.
However, our analysis revealed that flavor effects introduce new $C\!P$
violating terms that relax the bound of  one order of magnitude
for acceptable choices of the parameters and still within the hierarchical
limit. Flavor effects modify the upper bound on $m_1$ as well but,
as we stressed, an ultimate answer requires solutions
of more general kinetic equations that should be able to describe the intermediate regime
where the coherence of the final quantum lepton state is lost but
a full decoherence is still not achieved. We have seen that an account of the
Higgs asymmetry supports such a prudent conclusion.

Still within the hierarchical limit but accounting for the asymmetry produced from
the $N_2$-decays, the lower bound on $M_1$ disappears and is replaced by
a lower bound on $M_2$ that still implies a lower bound on the
reheat temperature.
We showed some more general conditions for the validity of the $N_1$-dominated scenario
and of the lower bound on $M_1$. For example the $N_1$-dominated scenario
certainly applies in the popular
two effective RH neutrino limit, for $M_3\gg 10^{14}\,{\rm GeV}$,
that implies that the heaviest RH neutrino decouples.
However, apart from this case, our analysis indicates that a
$N_2$-dominated scenario is a more natural option and that neglecting
the asymmetry from $N_2$-decays can be a wrong assumption.

In conclusion, our analysis answered many different questions about the
bounds on neutrino masses that are obtained within the leptogenesis scenario based on the
simplest version of the see-saw mechanism. Despite many proposed extensions,
this still represent the  most attractive possibility since it realizes
a successful link between the neutrino masses and the observed asymmetry
where the measured values exhibit an interesting conspiracy.
The discovery of $C\!P$ violating effects in neutrino oscillations or in lepton decays,
the determination of the absolute neutrino mass scale and of the
neutrino mass spectrum ordering, normal or inverted, will likely provide further
interesting tests during next years, making current experimental `coincidences'
even stronger or forcing departures from the minimal picture.

\vspace{3mm}

\textbf{Acknowledgments}

We wish to thank A.~Riotto for many useful discussions. PDB is supported by the
Helmholtz Association of National Research Centres, under project VH-NG-006.


\begin{thebibliography}{99}

\bibitem{fy}
M.~Fukugita, T.~Yanagida, \pl{174}{1986}{45}.

 \bibitem{seesaw}
P.~Minkowski, Phys.\ Lett.\ B {\bf 67} (1977) 421;
T.~Yanagida, in {\it{Workshop on Unified Theories}}, KEK report
79-18 (1979) p.~95;
M.~Gell-Mann, P.~Ramond, R.~Slansky, in {\it{Supergravity}} (North Holland,
Amsterdam, 1979) eds. P.~van Nieuwenhuizen, D.~Freedman, p.~315;
S.L. Glashow, in {\it 1979 Cargese Summer Institute on Quarks and Leptons}
(Plenum Press, New York, 1980) 
p.~687;
R.~Barbieri, D.~V.~Nanopoulos, G.~Morchio and F.~Strocchi,
Phys.\ Lett.\ B {\bf 90} (1980) 91;
R.~N.~Mohapatra and G.~Senjanovic,
Phys.\ Rev.\ Lett.\  {\bf 44} (1980) 912.

\bibitem{di}
  S.~Davidson and A.~Ibarra,
  Phys.\ Lett.\ B {\bf 535} (2002) 25.


\bibitem{cmb}
  W.~Buchm\"{u}ller, P.~Di Bari and M.~Pl\"{u}macher,
  Nucl.\ Phys.\ B {\bf 643} (2002) 367.

\bibitem{geometry}
P.~Di Bari, Nucl.\ Phys.\ B {\bf 727} (2005) 318.

\bibitem{flavorlep}
  S.~Blanchet and P.~Di Bari, 
   JCAP {\bf 03} (2007) 018.


\bibitem{annals}
W.~Buchm\"uller, P.~Di Bari and M.~Pl\"{u}macher,
Annals Phys.\  {\bf 315} (2005) 305.

\bibitem{giudice}
G.~F.~Giudice, A.~Notari, M.~Raidal, A.~Riotto and A.~Strumia,
  Nucl.\ Phys.\  B {\bf 685} (2004) 89.

\bibitem{window}
  W.~Buchmuller, P.~Di Bari and M.~Plumacher,
  Nucl.\ Phys.\ B {\bf 665} (2003) 445.

\bibitem{nardi1}
 E.~Nardi, Y.~Nir, E.~Roulet and J.~Racker,
  JHEP {\bf 0601} (2006) 164.

\bibitem{abada1}
  A.~Abada, S.~Davidson, F.~X.~Josse-Michaux, M.~Losada and A.~Riotto,
  JCAP {\bf 0604} (2006) 004

\bibitem{bcst}
R.~Barbieri, P.~Creminelli, A.~Strumia and N.~Tetradis,
Nucl.\ Phys.\ B {\bf 575} (2000) 61.

\bibitem{seealso}
T.~Endoh, T.~Morozumi and Z.~h.~Xiong,
  Prog.\ Theor.\ Phys.\  {\bf 111} (2004) 123;
  A.~Pilaftsis and T.~E.~J.~Underwood,
  Phys.\ Rev.\ D {\bf 72} (2005) 113001
  O.~Vives,
  Phys.\ Rev.\ D {\bf 73} (2006) 073006.

\bibitem{abada2}
  A.~Abada, S.~Davidson, A.~Ibarra, F.~X.~Josse-Michaux, M.~Losada and A.~Riotto,
  JHEP {\bf 0609} (2006) 010.

\bibitem{pascoli1}
  S.~Pascoli, S.~T.~Petcov and A.~Riotto,
  Phys.\ Rev.\  D {\bf 75} (2007) 083511.

\bibitem{pascoli2}
  S.~Pascoli, S.~T.~Petcov and A.~Riotto,
  Nucl.\ Phys.\  B {\bf 774} (2007) 1.

\bibitem{branco}
  G.~C.~Branco, R.~Gonzalez Felipe and F.~R.~Joaquim,
  Phys.\ Lett.\  B {\bf 645} (2007) 432.

\bibitem{antusch}
S.~Antusch and A.~M.~Teixeira, JCAP {\bf 0702} (2007) 024.

 \bibitem{diraclep}
A.~Anisimov, S.~Blanchet and P.~Di Bari, JCAP {\bf 0804} (2008) 033.

\bibitem{zeno}
S.~Blanchet, P.~Di Bari and G.~G.~Raffelt,
 JCAP {\bf 03} (2007) 012.


\bibitem{desimone1}
  A.~De Simone and A.~Riotto,
  JCAP {\bf 0702} (2007) 005.

\bibitem{review}
 S.~Davidson, E.~Nardi and Y.~Nir,
  arXiv:0802.2962 [hep-ph].

\bibitem{CPbound}
K.~Hamaguchi, H.~Murayama and T.~Yanagida,
  Phys.\ Rev.\  D {\bf 65} (2002) 043512.

\bibitem{hambye}
  T.~Hambye, Y.~Lin, A.~Notari, M.~Papucci and A.~Strumia,
  Nucl.\ Phys.\  B {\bf 695} (2004) 169

\bibitem{kitano}
S.~Davidson and R.~Kitano, JHEP {\bf 0403} (2004) 020.

\bibitem{gonzalez}
 M.~C.~Gonzalez-Garcia and M.~Maltoni,
 Phys. \ Rept. {\bf 460} (2008) 1.

\bibitem{WMAP5}
E.~Komatsu {\it et al.}  [WMAP Collaboration], arXiv:0803.0547 [astro-ph].

\bibitem{sakharov}
  A.~D.~Sakharov,
  Pisma Zh.\ Eksp.\ Teor.\ Fiz.\  {\bf 5} (1967) 32

\bibitem{sphalerons}
V.~A.~Kuzmin, V.~A.~Rubakov,  M.~E.~Shaposhnikov, \pl{155}{1985}{36}.

\bibitem{crv}
L.~Covi, E.~Roulet, F.~Vissani, \pl{384}{1996}{169}.

\bibitem{Campbell:1992jd}
  B.~A.~Campbell, S.~Davidson, J.~R.~Ellis and K.~A.~Olive,
  Phys.\ Lett.\ B {\bf 297} (1992) 118;
 J.~M.~Cline, K.~Kainulainen and K.~A.~Olive,
  Phys.\ Rev.\  D {\bf 49} (1994) 6394.

\bibitem{dolgov}
A.~D.~Dolgov and Ya.~B.~Zeldovich, Rev. Mod. Phys. {\bf 53} (1981) 1;
E.~W.~Kolb and S.~Wolfram, Nucl. Phys. B{\bf 172} (1980) 224, ibid. {\bf B 195} (1982) 542 (E).

\bibitem{luty}
 M.~A.~Luty, Phys.\ Rev.\ D {\bf 45} (1992) 455.

\bibitem{buchplum}
  W.~Buchm\"{u}ller and M.~Pl\"{u}macher,
  Phys.\ Lett.\ B {\bf 511} (2001) 74

\bibitem{nardi2}
  E.~Nardi, Y.~Nir, J.~Racker and E.~Roulet,
JHEP {\bf 0601} (2006) 068

\bibitem{casas}
  J.~A.~Casas and A.~Ibarra,
  Nucl.\ Phys.\ B {\bf 618} (2001) 171.

\bibitem{PDG}
  S.~Eidelman et al., Phys.\ Lett. {\bf B592}, 1 (2004)
  (URL:http://pdg.lbl.gov/).

\bibitem{fhy}
 M.~Fujii, K.~Hamaguchi and T.~Yanagida,
  Phys.\ Rev.\  D {\bf 65} (2002) 115012.

\bibitem{colliders}
F.~del Aguila, J.~A.~Aguilar-Saavedra, A.~Martinez de la Ossa and D.~Meloni,
  Phys.\ Lett.\  B {\bf 613} (2005) 170;
J.~Kersten and A.~Y.~Smirnov,
  Phys.\ Rev.\  D {\bf 76} (2007) 073005.

\bibitem{2effRH}
P.~H.~Frampton, S.~L.~Glashow and T.~Yanagida,
Phys.\ Lett.\ B {\bf 548} (2002) 119;
 M.~Raidal and A.~Strumia,
  Phys.\ Lett.\  B {\bf 553} (2003) 72;
A.~Ibarra and G.~G.~Ross, Phys.\ Lett.\ B {\bf 575} (2003) 279.


\bibitem{turzynski}
  P.~H.~Chankowski and K.~Turzynski,
  Phys.\ Lett.\  B {\bf 570} (2003) 198.

\bibitem{seealso2}
 P.~H.~Chankowski, J.~R.~Ellis, S.~Pokorski, M.~Raidal and K.~Turzynski,
  Nucl.\ Phys.\  B {\bf 690} (2004) 279;
 S.~T.~Petcov, W.~Rodejohann, T.~Shindou and Y.~Takanishi,
  Nucl.\ Phys.\  B {\bf 739} (2006) 208.

\bibitem{beyond}
S.~Blanchet and P.~Di Bari, JCAP {\bf 0606} (2006) 023.

\bibitem{proc}
P.~Di Bari, AIP Conf.\ Proc.\  {\bf 655} (2003) 208
  [arXiv:hep-ph/0211175]. P.~Di Bari, [arXiv:hep-ph/0406115].

\bibitem{aspects}
W.~Buchmuller, P.~Di Bari and M.~Plumacher,
  New J.\ Phys.\  {\bf 6} (2004) 105.


\bibitem{molinaro}
E.~Molinaro and S.~T.~Petcov, arXiv:0803.4120 [hep-ph].

\bibitem{molinaro2}
E.~Molinaro, S.~T.~Petcov, T.~Shindou and Y.~Takanishi,
  Nucl.\ Phys.\  B {\bf 797} (2008) 93.

\bibitem{nardinir}
G.~Engelhard, Y.~Grossman, E.~Nardi and Y.~Nir,
  Phys.\ Rev.\ Lett.\  {\bf 99} (2007) 081802.

\bibitem{vives}
O.~Vives, Phys.\ Rev.\  D {\bf 73} (2006) 073006.

\bibitem{quantumres}
A.~De Simone and A.~Riotto, JCAP {\bf 0708} (2007) 013.


\end{thebibliography}
\end{document}